\RequirePackage{fix-cm}
\documentclass[twocolumn]{svjour3}          % twocolumn
\smartqed  % flush right qed marks, e.g. at end of proof
\usepackage{graphicx}
\usepackage{moreverb}
\usepackage{array}
\usepackage{float}
\usepackage{subfig}
\usepackage{hyperref}
\usepackage{amsmath}
\usepackage{mathtools}
\usepackage{bm}
\usepackage{xcolor}

\usepackage{tikzexternal}
\newlength{\figureheight}
\newlength{\figurewidth}
\newlength{\mriheight}
\newcommand{\beq}{\begin{equation*}}
\newcommand{\eeq}{\end{equation*}}
\newcommand{\beqq}{\begin{equation}}
\newcommand{\eeqq}{\end{equation}}
\newcommand{\dd}{~\text{d}}

\newcommand{\defeq}{\vcentcolon=}

\newcommand{\vecrm}[1]{\boldsymbol{\mathrm{#1}}}

\newcommand{\rs}{\vecrm{R}^\mathrm{S}}
\newcommand{\rd}{\vecrm{R}^\mathrm{0D}}
\newcommand{\ds}{\vecrm{d}}
\newcommand{\p}{\vecrm{p}}
\newcommand{\norm}[2]{\left\| #2 \right\|_{#1}}

\def\vec   #1{\mbox{\boldmath $#1$}{}}
\def\ten   #1{\mbox{\boldmath $#1$}{}}

\newcommand{\fifty}[1]{$\pm50^\circ$}
\newcommand{\sixty}[1]{$\pm60^\circ$}
\newcommand{\seventy}[1]{$\pm70^\circ$}

\usepackage{color}
\definecolor{myred}{rgb}{0.9098,0.1294,0.2078}

\newcommand{\revs}[1]{\textcolor{black}{#1}}
\newcommand{\reva}[1]{\textcolor{black}{#1}}
\newcommand{\revb}[1]{\textcolor{black}{#1}}
\newcommand{\revc}[1]{\textcolor{black}{#1}}

\newcommand{\apex}[1]{\revs{\textit{apex}}}
\newcommand{\free}[1]{\textit{free}}
\newcommand{\normal}[1]{\textit{pericardium}}
\newcommand{\contact}[1]{\textit{pseudo-contact}}

\journalname{Biomechanics and Modeling in Mechanobiology}
\begin{document}

\title{The importance of the pericardium for cardiac biomechanics
}
\subtitle{From physiology to computational modeling}


\author{Martin R. Pfaller$^1$ \and
        Julia M. H\"ormann$^1$ \and
        Martina Weigl$^1$ \and
        Andreas Nagler$^1$ \and
        Radomir Chabiniok$^{2,3,4}$ \and
        Crist\'obal Bertoglio$^{5*}$ \and
        Wolfgang A. Wall$^{1*}$
}

\institute{
M. R. Pfaller\\
martin.pfaller@tum.de\\[0.2cm]
$^1$Institute for Computational Mechanics, Technical University of Munich, Boltzmannstr. 15, 85748 Garching b. M\"unchen, Germany \\[0.2cm]
$^2$ Inria, Paris-Saclay University, Palaiseau, France\\[0.2cm]
$^3$ LMS, Ecole Polytechnique, CNRS, Paris-Saclay University, Palaiseau, France\\[0.2cm]
$^4$ School of Biomedical Engineering \& Imaging Sciences (BMEIS), St Thomas' Hospital, King's College London, UK\\[0.2cm]
$^5$ Bernoulli Institute for Mathematics, Computer Science and Artificial Intelligence, University of Groningen, Nijenborgh 9, 9747 AG Groningen, The Netherlands\\[0.2cm]
$*$ joint last authors
}

\date{Received: date / Accepted: date}

\maketitle

\begin{abstract}
The human heart is enclosed in the pericardial cavity. The pericardium consists of a layered thin sac and is separated from the myocardium by a thin film of fluid. It provides a fixture in space and frictionless sliding of the myocardium. The influence of the pericardium is essential for predictive mechanical simulations of the heart. However, there is no consensus on physiologically correct and computationally tractable pericardial boundary conditions. Here we propose to model the pericardial influence as a parallel spring and dashpot acting in normal direction to the epicardium. \revs{Using a four-chamber geometry, we compare a model with pericardial boundary conditions to a model with fixated apex. The influence of pericardial stiffness is demonstrated in a parametric study.} Comparing simulation results to measurements from cine magnetic resonance imaging reveals that adding pericardial boundary conditions yields a better approximation with respect to atrioventricular plane displacement, atrial filling, and overall spatial approximation error. We demonstrate that this simple model of pericardial-myocardial interaction can correctly predict the pumping mechanisms of the heart as previously assessed in clinical studies. Utilizing a pericardial model can not only provide much more realistic cardiac mechanics simulations but also allows new insights into pericardial-myocardial interaction which cannot be assessed in clinical measurements yet.
\end{abstract}

\section{Introduction}
\subsection{Motivation \label{sec_motivation}}

%
% explanation of layers
\begin{figure*}
\centering
\scriptsize
\subfloat[Dissected mediastinum with cut pericardium and heart surface. Image by G.~M.~Gruber, Medical University of Vienna, Austria. \label{pericardium_cut}]{
\includegraphics[height=5.5cm]{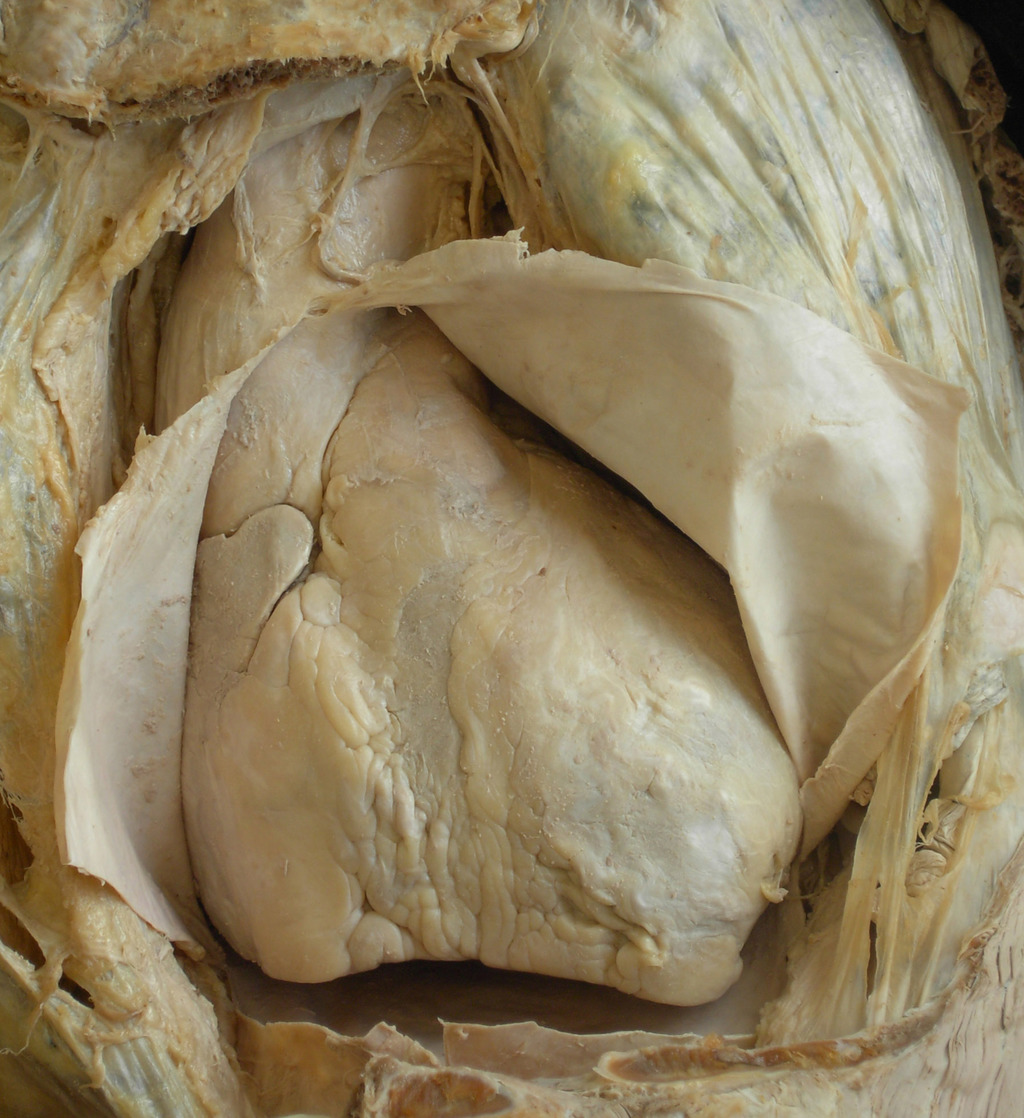}}\quad
\subfloat[Location of the heart with respect to serous and fibrous pericardium. Inspired by \cite{iaizzo15}.\label{pericardium_4ch}]{
\def\svgwidth{.28\textwidth}
\includegraphics[width=.28\textwidth]{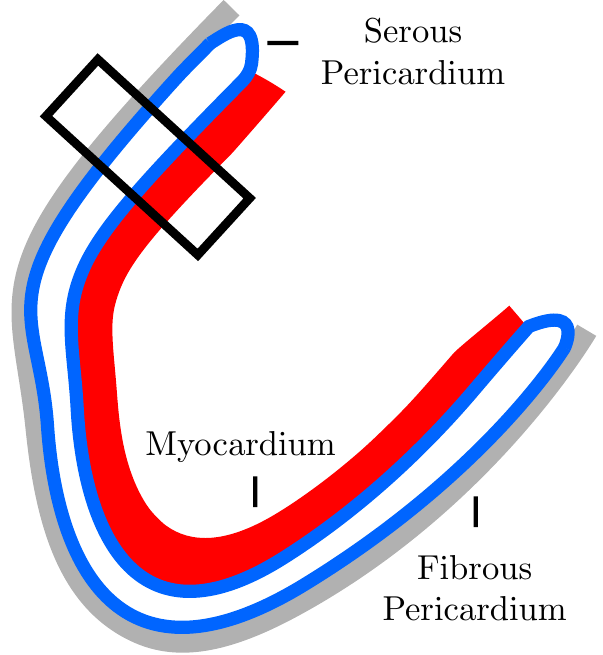}}\quad
\subfloat[Cross-sectional view of transmural layers of heart and pericardium. Inspired by \cite{iaizzo15}.\label{pericardium_layers}]{
\def\svgwidth{.35\textwidth}
\includegraphics[width=.35\textwidth]{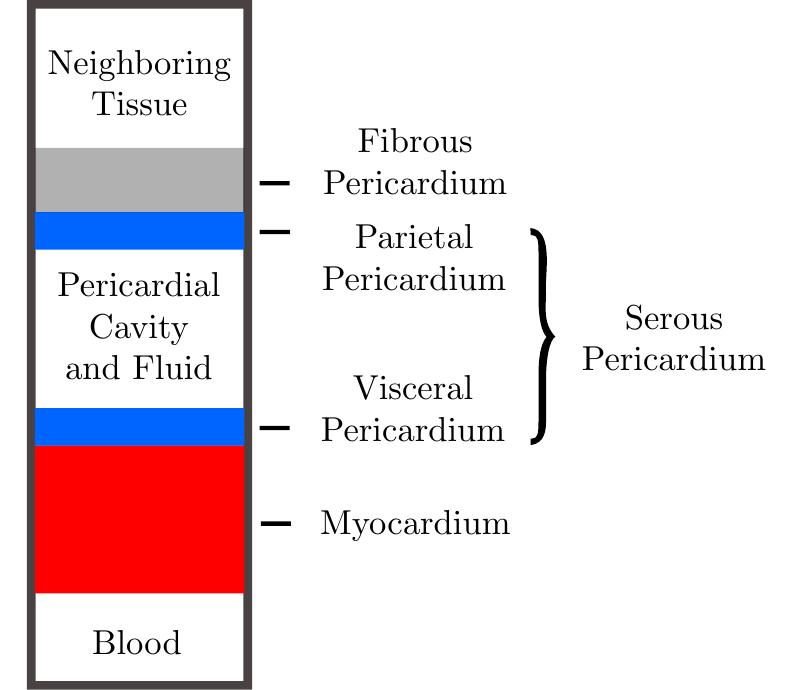}}
\caption{Heart and pericardium.\label{pericardium}}
\end{figure*}
%

% why pericardial bcs are important
Cardiac mechanics simulations consist of solving a nonlinear elastodynamic boundary value problem \cite{sainte-marie06}. Physiological boundary conditions are essential to achieve predictive results for any clinical purposes. The boundary conditions on the structure field of the myocardium are mainly governed by two physiological aspects: Blood flow within the chambers near the inside surface of the myocardium (\emph{endocardium}) and the pericardial sac on the outside surface (\emph{epicardium}), see figure~\ref{pericardium_cut}. \revc{There are many applications for simulating heart blood flow \cite{doost16}. However, for many relevant questions the exact fluid dynamics of blood within the heart or a resolved fluid-solid interaction simulation are often not needed for simulating the myocardium. Instead, a realistic pressure-flow relationship stemming from the circulatory system is sufficient, which is commonly represented by lumped-parameter fluid models that provide the correct normal pressure to the endocardial wall \cite{kerckhoffs07}.}

% objective
However, there is no consensus on boundary conditions to represent the effects of the pericardial sac. The goal of this work is twofold: (a) to provide a detailed literature review of pericardial biomechanics, hence justifying its modeling using a computationally inexpensive viscoelastic model, and (b) to highlight the relevance of such boundary conditions through a  detailed quantitative analysis using a subject-specific cine MRI data set. We employ a four-chamber geometry including parts of the great vessels, as it \reva{provides us with additional options to asses the physiological correctness of our boundary condition, e.g. through the interplay between ventricles and atria during ventricular systole. Note, however, that the pericardial boundary condition is independent of the geometry and is meant to be applied to any kind of cardiac mechanics simulation that includes the epicardial surface.}

% structure of this paper
This work is structured as follows. Following a review of the anatomy and physiology of the pericardium in section~\ref{sec_pericardium}, we review pericadial boundary conditions currently used in cardiac mechanics simulations. We propose a model to represent the influence of the pericardium by parallel springs and dashpots acting in normal direction to the epicardium in section~\ref{sec_methods}. Furthermore, we summarize a three-dimensional elastodynamical continuum model for the myocardium which is monolithically coupled to a zero-dimensional reduced-order windkessel model for the circulatory system. We demonstrate the influence of the pericardial boundary condition in section~\ref{sec_results} through a detailed quantitative comparison of simulation results to cine MRI. For that we evaluate ventricular volume, atrioventricular-plane-displacement, atrioventricular interaction, and introduce a quantitative error measurement by calculating a distance error at endo- and epicardial surfaces between simulation results and cine MRI. We close this work with a discussion of the results, the limitations of our study, future perspectives, and some conclusions in section~\ref{sec_discussion}.

\subsection{The pericardium}
\label{sec_pericardium}
In the following, we review the anatomy of the pericardium and its physiology, where we focus on the mechanical interaction between the pericardium and the heart. Based on this review, we evaluate variants of pericardial boundary conditions and propose a model for pericardial-myocardial interaction.

% what is a pericardium
\subsubsection{Anatomy}
% percardium
As shown in figure~\ref{pericardium_cut}, the pericardium is a sac-like structure with a combined thickness of 1-2~mm that contains the heart and parts of the great vessels \cite{holt70}. Figures~\ref{pericardium_4ch} and \ref{pericardium_layers} show a cross-sectional view of the myocardium and the layers of the pericardium. A common analogy for the location of the heart within the pericardium is that of a fist pushed into an inflated balloon \cite{martini15}.

\begin{figure*}
\centering
\subfloat[Coronal plane \label{pericardium_mri_front.JPG}]{
\includegraphics[height=4cm]{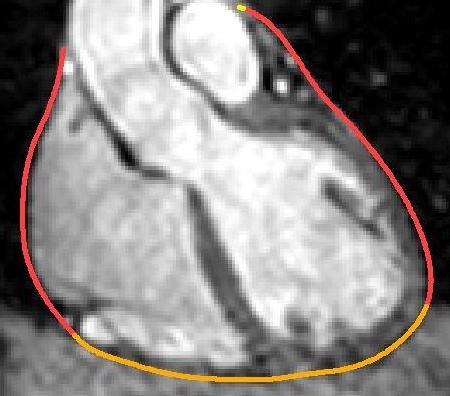}}
\subfloat[Transverse plane \label{pericardium_mri_top.JPG}]{
\includegraphics[height=4cm]{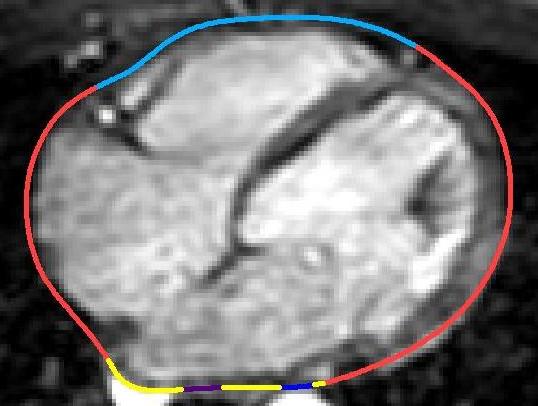}}
\subfloat[Sagittal plane \label{pericardium_mri_side.JPG}]{
\includegraphics[height=4cm]{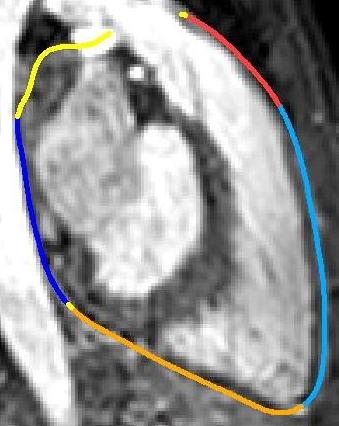}}
\caption{Position of the pericardium indicated in 3D MRI taken during diastasis. The neighboring tissue is color-coded: lungs (red), diaphragm (orange), sternum and ribs (light blue), aorta (dark blue), esophagus (purple), other (yellow). MRI courtesy of R.~Chabiniok, J.~Harmer, E.~Sammut, King's College London, UK. \label{pericardium_mri}}
\end{figure*}

% fibrous pericardium
The \emph{fibrous pericardium} consists of a fibrous layer that forms a flask-like sac with a wavy collagenous structure of three interwoven main layers that are oriented $120^{\circ}$ to each other \cite{standring15}. It has a higher tensile stiffness than the myocardium and is dominated by the viscoelastic behavior of extracellular collagen matrix and elastin fibers \cite{lee85}. The fibrous pericardium is fixed in space by a "three point cardiac seat belt" via the pericardial ligaments to the sternum. Furthermore, it is thoroughly attached to the central tendon of the thoracic diaphragm and additionally supported by the coats of the great vessels \cite{spodick96}. The various tissues, the fibrous pericardium is in contact with, can be seen in figure~\ref{pericardium_mri}.

% serous pericardium
The fibrous pericardium contains a serous membrane, the \emph{serous pericardium}, forming a closed sac.  The serous pericardium is connected to the myocardium (\emph{visceral pericardium}) and the fibrous pericardium (\emph{parietal pericardium}). The composite of fibrous and parietal pericardium is commonly referred to as pericardium, whereas the visceral pericardium in contact with the myocardium is referred to as epicardium \cite{spodick96}. The space between the visceral and parietal pericardium is the pericardial cavity, which is filled by a thin film of fluid with an average volume of 20-25~ml \cite{holt70}. Beneath the visceral pericardium the heart is covered by a layer of adipose tissue, accumulated especially in the interventricular and atrioventricular grooves and around the coronary vessels, constituting about 20\% of the heart weight \cite{rabkin07}.

%
% overview of surrounding tissue
%
% what does it do
\subsubsection{Mechanical physiology \label{sec_physio}}

% general, inluding non-mechanical
The pericardium serves multiple purposes \cite{spodick83} that can be grouped in: (a) membranous, it  serves as a barrier against the spread of infection \cite{standring15} and (b) mechanical, it secures cardiac stability via its attachments within the thorax \cite{shabetai03}, as will be explained in the following. The mechanical properties of the pericardium itself can be found in \cite{sacks03}.

% Evidence that the pericardium has a mechanical importance
There is clear empirical evidence that the pericardium has a direct mechanical impact on the acute and chronic biomechanics of the heart. For example, in \cite{glantz78} it was discovered that the correlation of left and right ventricular pressure is higher with intact pericardium than after its complete removal. Maximal cardiac output during exercise can be increased acutely by the complete removal of the pericardium (\emph{pericardiectomy}) through utilizing the Frank-Starling mechanism \cite{hammond92}. However, removing the pericardial restraint chronically promotes eccentric hypertrophy, i.e.\ an increase in dimension and mass of the heart and a change in shape from elliptical to spherical. The pericardium thus acts as a diastolic constraint for the heart by exercising a radial compression stress. This was confirmed in \cite{joebsis07} where it was observed that the opening angle of the myocardium with intact visceral pericardium is much higher than after its removal. The visceral pericardium is thus important for residual stress and passive stiffness.

% fluid
It is widely accepted that the mechanism of the myocardium-pericardium interaction is through the pericardial fluid. In \cite{holt60} it was found that while increasing the volume of fluid within the pericardial cavity, the pericardial liquid pressure remains constant until it suddenly rises sharply. This led to the conclusion that most of the fluid is contained in pericardial sinuses and groves. This mostly empty space forms the so-called  pericardial reserve volume, acting as a buffer against increasing pericardial liquid pressure. Only a small portion of the pericardial fluid remains as a thin film on the interface between parietal and visceral pericardium. In \cite{santamore90}, a dye was injected into the pericardial cavity near the apex. Fifteen minutes after injection the dye was almost exclusively found in the interventricular and atrioventricular grooves. This suggests that there is no significant fluid movement on the large surface areas of the ventricular free walls, leaving just a very thin film of fluid with an estimated thickness of less than 0.5~mm.

% diastolic constraint
The mechanical constraint of the pericardium on diastolic cardiac function can be quantified by pericardial pressure. Here it is important to distinguish between liquid pressure and contact pressure \cite{smiseth85,tyberg86}. Liquid pressure describes the hydrostatic pressure inside the pericardial fluid and is measured by an open-ended catheter. However, liquid pressure does not describe the constraining effect of the pericardium on the myocardium. The constraint is assessed by contact pressure, which can be measured by a thin, flat, air-filled balloon catheter. In \cite{smiseth85} it was found that liquid pressure is substantially below contact pressure unless the pericardium contains a significant amount of pericardial fluid, which happens e.g. due to pericardial effusion. Furthermore, contact stress and thus ventricular restraint was maintained even though pericardial fluid was completely removed and liquid pressure at the epicardial surface was zero. Pericardial fluid therefore acts as lubrication 
rather than a load balancing mechanism, providing low-friction sliding between pericardium and epicardium \cite{hills85}.

% systolic constraint
There is less information available on the influence of the pericardium during systole. A pericardial restraining effect during systole would require a tension force to be transmitted by the myocardial-pericardial interface. The restraining effect of the pericardium during systole can be well observed in fish, where the parietal pericardium is almost rigid \cite{holt70}. It was observed in \cite{sudak65} that pericardial liquid pressure in smooth dogfish is always negative and decreases further during cardiac contraction. In man, \cite{sutton77} found that pericardial liquid pressure also drops during ventricular systole but remains positive throughout the cardiac cycle. However, to the best of the authors' knowledge there is no study on the change of contact pressure during systole. It can be observed from mammal cine MRI that surrounding tissue moves toward the heart during systole, indicating attachment of pericardium and epicardium. We hypothesize that during systole, through the effect of adhesion, the pericardium remains in contact with the epicardium. This is analogous to the simple experiment of ``gluing'' two glass plates together with a drop of water. The glass plates can hardly be separated in normal direction but can be easily moved relatively to each other in tangential direction.

\subsubsection{Current pericardial boundary conditions}
\label{sec_review_bcs}

% what others did 2ch
For biventricular geometries, the constraining effect of the pericardium in diastole is accounted for in \cite{mansi10,chabiniok12,marchesseau13}, where a no penetration condition is enforced on the epicardium by a unidirectional penalty contact with a rigid pericardial reference surface. However, this neglects any constraining in systole by not allowing the pericardial interface to transmit any tension forces. \revs{Recently, it was proposed in \cite{santiago18a} to completely prohibit movement normal to the epicardial surface, neglecting any elastic effects.} The bi-directional elastic constraining effect of the pericardium is accounted for in spring-type boundary conditions, where a spring-dashpot boundary condition is  enforced either on the base \cite{sainte-marie06} or on apex and valve annuli \cite{sermesant12} with homogeneous Neumann conditions applied to the rest of the epicardium. These boundary conditions are analogous to the external tissue support of the aorta in \cite{moireau12,moireau13}. However, they do not cover the whole epicardial surface thus representing pericardial-myocardial interaction only partially.

% what others did 4ch
Fewer references exist for four chamber geometries. A common combination of boundary conditions for four chamber geometries are homogeneous Dirichlet conditions on vessel cut-offs  and a soft material connected to the apex \cite{chabiniok12,augustin16}, or springs on the outside of great vessels \cite{land17}.  In those cases, however, homogeneous Neumann conditions are applied on the remaining epicardial surface, neglecting any influence of the pericardium as in the biventricular case. In \cite{baillargeon14} "omni-directional" springs acting in all directions are applied to the epicardium, artificially constraining any sliding movement along the pericardial-epicardial interface. To the authors' best knowledge, the most detailed and physiologically correct representation of the pericardium so far was implemented in \cite{fritz13}. The pericardial-myocardial interaction was here modeled by a frictionless sliding, bi-directional penalty contact interaction in normal direction between the epicardium and a solid pericardial reference body. However, this condition is computationally very expensive as it requires solving an adhesial contact interaction problem. \revb{It also requires an additional solid body to be created, representing the surrounding tissue}. Furthermore, no boundary conditions could be enforced at the great vessels since they were not included in the geometry. Thus, a fixation of the apex was necessary. All models based on four chamber geometries reviewed here lack a quantitative validation through comparison of simulation results to measurements, e.g. medical images like magnetic resonance imaging (MRI).

\section{Models}
\label{sec_methods}

\revc{
In this work, we use a patient-specific four-chamber geometry from high-resolution static 3D MRI, including ventricles, atria, adipose tissue, and great vessels. Our cardiac model is formulated in a large displacement, constitutive nonlinear framework with nonlinear boundary conditions. It features high-resolution quadratic tetrahedral finite elements for structural dynamics with implicit time integration. Blood pressure is incorporated through monolithic coupling of the left and right ventricle to windkessel models which include each the atrioventricular and semilunar valves. The reference configuration is prestressed in all four cardiac chambers. The passive myocardial material features a state of the art orthotropic exponential material law proposed in \cite{holzapfel09}. Myofiber contraction in atria and ventricles is modeled with an active stress approach. Passive and active material behavior is based on local fiber orientations.
}

We follow the classic approach of nonlinear large deformation continuum mechanics to model the elastodynamic problem of 3D cardiac contraction. We define the reference configuration $\vec{X}$ and current configuration $\vec{x}$ which are connected by the displacements $\vec{u} = \vec{x} - \vec{X}$. We calculate the deformation gradient $\ten{F}$, the Green-Lagrange strain tensor $\ten{E}$, and the right Cauchy-Green tensor $\ten{C}$
\begin{equation}
\ten{F} = \frac{\partial \vec{x}}{\partial \vec{X}}, \quad \ten{E} = \frac{1}{2}(\ten{C} - \ten{I}), \quad \ten{C}=\ten{F}^{\text{T}}\ten{F}.
\end{equation}

\subsection{Modeling the pericardium \label{sec_review}}
\begin{figure}
\centering
\footnotesize
\def\svgwidth{.15\textwidth}
\includegraphics[width=.45\textwidth]{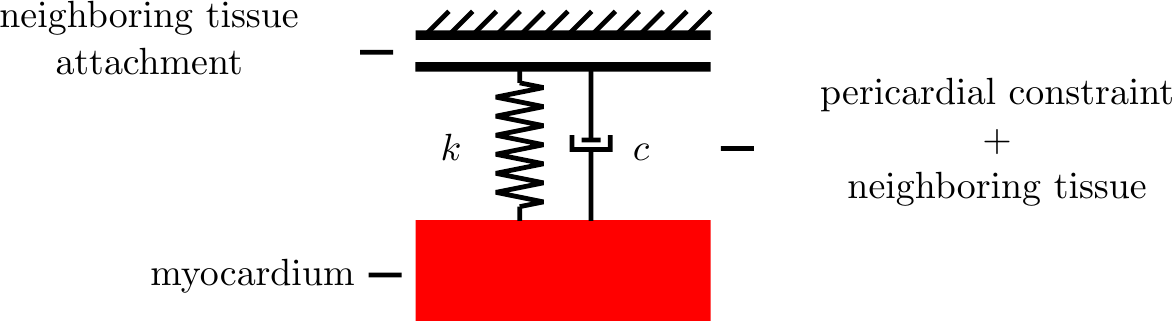}
\caption{Serous pericardium, fibrous pericardium, and neighbouring tissue modeled by a spring (stiffness $k$) and a dashpot (viscosity $c$) in parallel. Spring and dashpot act in normal direction of the epicardial surface. \label{springs}}
\end{figure}

Our aim in this work is to propose and justify a pericardial boundary condition that is both realistic and computationally inexpensive. The pericardial model we propose is based on \cite{rabkin75} and is sketched in figure~\ref{springs}. Using our code it was also already applied to a two-chamber geometry in \cite{hirschvogel16}. It consists of a spring and a dashpot in parallel acting in normal direction to the epicardial surface. Within the tangential plane we allow frictionless sliding to account for the lubricating effect of the pericardial fluid. A spring stiffness~$k$ and dashpot viscosity~$c$ contain the combined effects of serous pericardium, fibrous pericardium, and neighboring tissue. \revc{Generalizing the effect on the ventricles, spring compression models the pericadium's constraining effect during passive ventricular filling, whereas spring expansion models the pericadium's support during ventricular systole.} 

\reva{Note that only in the limit case of $k\to\infty$, we would obtain a boundary condition that penalizes and therefore prohibits any movement in normal direction to the epicardium, as it was recently proposed in \cite{santiago18a}. However, our pericardial boundary condition is meant to be used with finite values for $k$ and $v$, more realistically representing the visco-elastic support of the pericardium and its surrounding tissue and permitting movement normal to the epicardial surface. Furthermore, our parametric study in section~\ref{sec_params} shows that small values of $k$ lead to physiological results.}

\revb{
In the following, we will derive a simple mathematical formulation for the pericardial boundary condition depicted in figure~\ref{springs}. This derivation will be carried out in two steps, where different assumptions are introduced in each step. Only the spring component will be considered during the derivation. However, all conclusions hold equivalently for the dashpot component.}

\begin{figure*}
\centering
\footnotesize
\def\svgwidth{.3\textwidth}
\subfloat[Adhesial contact interaction from \cite{fritz13} between myocardium and fixed surrounding tissue (blue). \label{fig_contact}]{
\includegraphics[width=.3\textwidth]{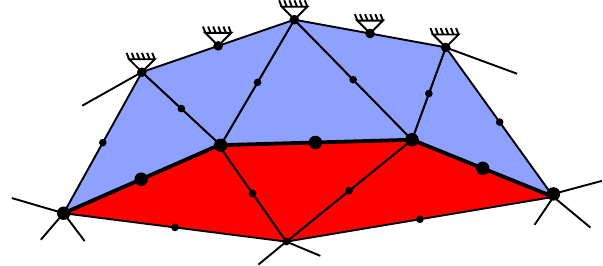}}~
\def\svgwidth{.3\textwidth}
\subfloat[Current normal spring from \eqref{eq_curnormal}.\label{fig_curnormal}]{
\includegraphics[width=.3\textwidth]{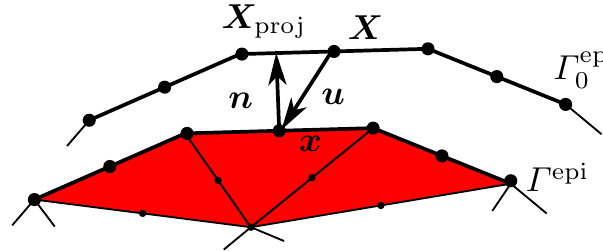}}~
\def\svgwidth{.3\textwidth}
\subfloat[Reference normal spring from \eqref{eq_refnormal}.\label{fig_refnormal}]{
\includegraphics[width=.3\textwidth]{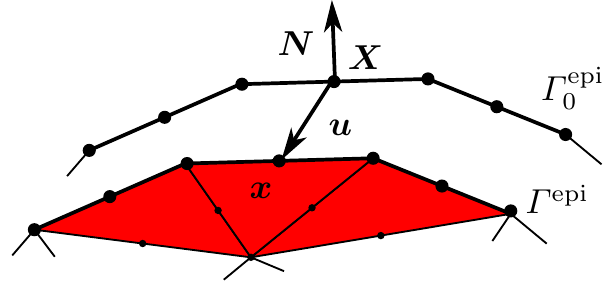}}
\caption{\revb{Different formulations of interaction between myocardium (red) and pericardium.}}
\end{figure*}

\revb{
Our goal is to preserve the features of the detailed pericardial boundary condition in \cite{fritz13} but arrive at a much simpler and cheaper formulation. As reviewed in section~\ref{sec_review_bcs}, pericardial-myocardial interaction is modeled in \cite{fritz13} by adhesial contact between the epicardium and an elastic reference body that is fixed in space and representing the surrounding tissue, see figure~\ref{fig_contact}.}

\revb{
In the first step, we replace the elastic body representing the surrounding tissue in \cite{fritz13} with springs acting in normal direction to the epicardium. Here, we assume that the elasticity of the surrounding tissue is linear with respect to the small movements of the epicardium in its normal direction.}
Note that we enforce the boundary conditions on the epicardial side of the myocardial-pericardial interface, as this does not require a representation of the actual pericardial surface. We therefore do not model the pericardium itself but the forces acting on the myocardium because of its presence.
\revb{
The elastic potential of a linear spring distributed on the epicardial surface $\varGamma^{\text{epi}}$ in current configuration surface is
\begin{equation}
W = \frac{1}{2} \int_{\varGamma^{\text{epi}}} kg^2 \dd a
\end{equation}
with spring stiffness $k$, gap $g$, and surface integral in current configuration $\dd a$. The calculation of the gap is illustrated in figure~\ref{fig_curnormal}. We project a point $\bm x \in \varGamma^{\text{epi}}$ on the current epicardial surface onto the point $\bm X_\text{proj} \in \varGamma_0^{\text{epi}}$ on the reference epicardial surface. The distance between both points projected in the direction of the current outward normal vector $\bm n$ yields the gap function
\begin{equation}
g = (\bm x - \bm X_\text{proj}) \cdot \bm n.
\label{eq_curnormal}
\end{equation}
Though reducing algorithmic and computational demands compared to contact interaction, this boundary condition still requires updates of the normal vector and its linearization with respect to the displacements as well as a projection of each evaluation point onto $\varGamma_0^{\text{epi}}$ in each Newton iteration at each time step.}

\revb{
In a second step, we introduce two further simplifications. Instead of calculating the spring deformation from a projection, we directly use the spatial displacements $\bm u$.
}
Furthermore, we use the epicardial normal vector in reference configuration (i.e. $\vec{N}$ instead of $\vec{n}$), neglecting any change in normal direction throughout the simulation.
\revb{
The formulation of the gap in \eqref{eq_curnormal} is then simplified to
\begin{equation}
g = \bm u \cdot \bm N
\label{eq_refnormal}
\end{equation}
The simplifications leading to \eqref{eq_refnormal} are valid for small rotations of the epicardium, an assumption that is not valid for all parts of the epicardium. However, the performance of both formulations \eqref{eq_curnormal} and \eqref{eq_refnormal} is reviewed in appendix~\ref{sec_ellipsoid}.
}

\revb{
We then arrive at the final expression for the pericardial boundary traction $\vec{t}_{epi}$ acting on the epicardial surface
\begin{equation}
\vec{t}_{epi} = \vec{N} \left(k_p \vec{u} \cdot \vec{N} + c_p \dot{\vec{u}} \cdot \vec{N}\right).
\label{eq_bc_final}
\end{equation}
}
For the sake of simplicity, we use here constant boundary condition parameters $k_p$ and $c_p$ on the whole epicardial surface. As it will be shown in the numerical examples, this simple approach already leads do greatly improved results. But of course, a regional distribution based on neighboring organs as visualized in figure~\ref{pericardium_mri} is also possible.

\subsection{Geometrical model\label{sec_geom}}
\begin{figure*}
\centering
\subfloat[Computational mesh. \label{mesh}]{
\includegraphics[height=.25\textheight]{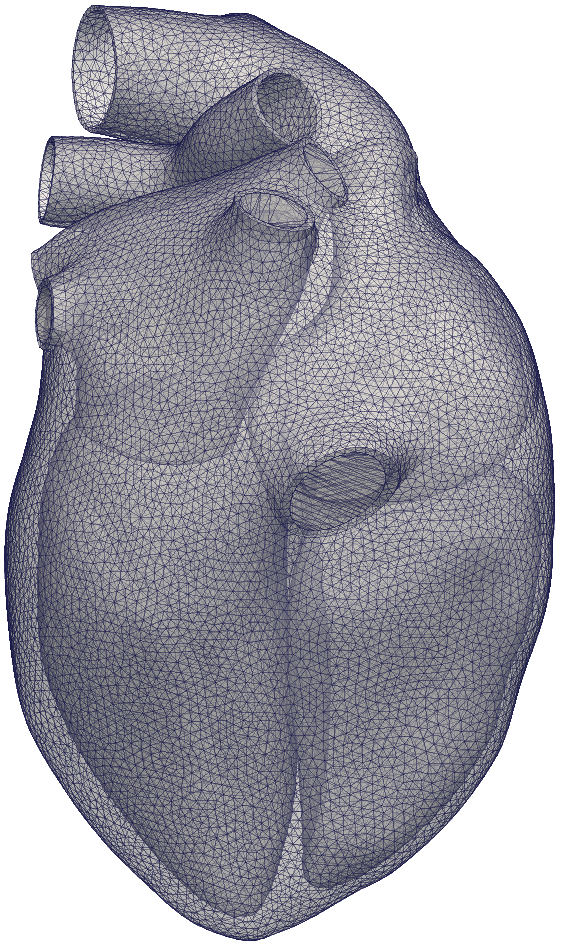}}~
\hspace{.5cm}
\subfloat[\reva{Atrial fibers and ventricular \sixty{} fibers at endocardium. Color shows helix angle with respect to long axis.} \label{fibers}]{
\includegraphics[height=.25\textheight]{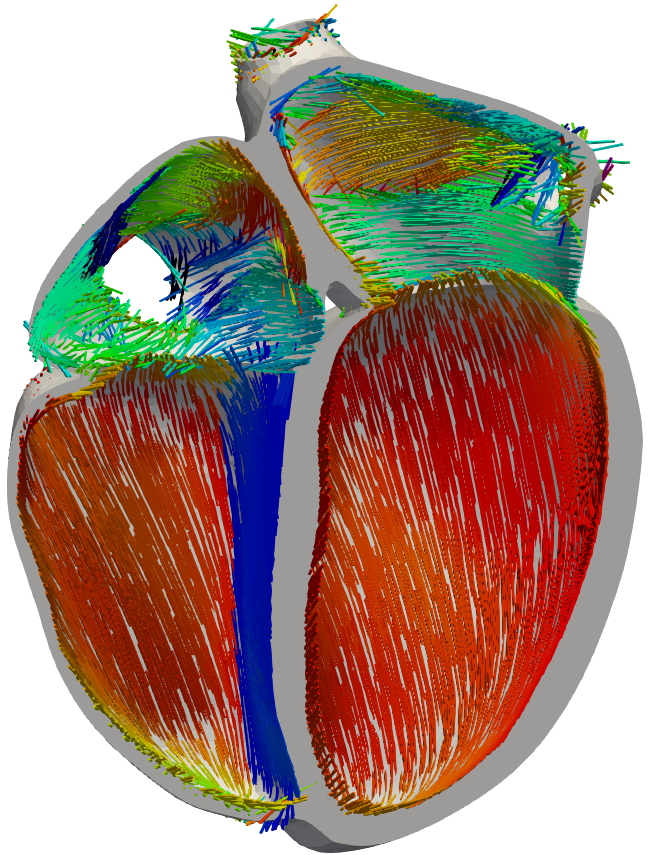}~
\includegraphics[height=.2\textheight]{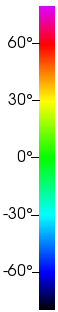}}~
\hspace{.5cm}
\subfloat[Materials. \label{materials}]{
\includegraphics[height=.25\textheight]{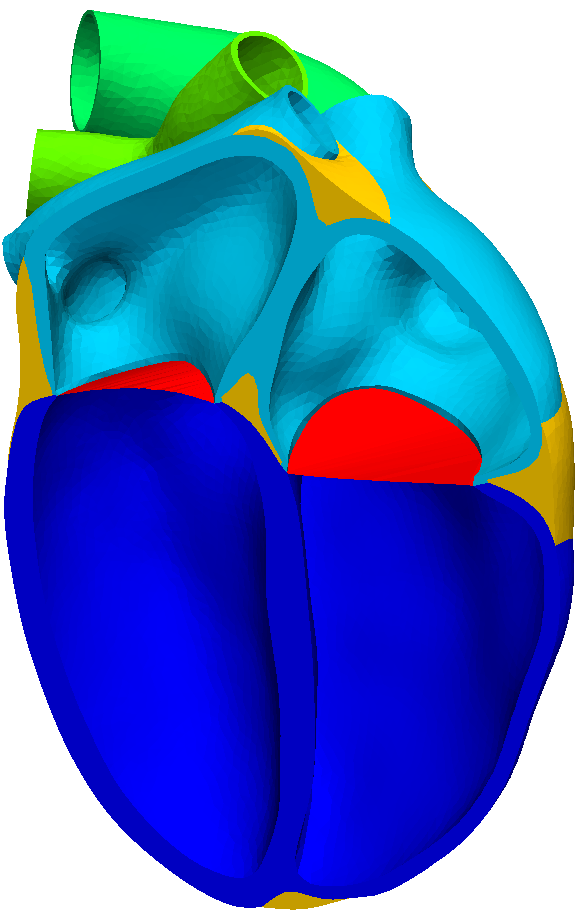}~
\includegraphics[height=.2\textheight]{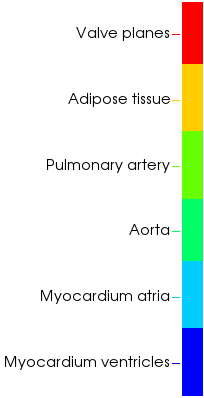}}
\caption{Four chamber patient-specific cardiac geometry. \label{geometry}}
\end{figure*}

To illustrate the effects of our pericardial model, we employ a four chamber geometry obtained in vivo from a 33 year old healthy female volunteer. The imaging data was acquired at King's College London, UK using a Philips~Achieva~1.5T magnetic resonance imaging (MRI) scanner with a dual-phase whole-heart 3D b-SSFP sequence \cite{uribe08}, acquisition matrix $212\times209\times200$, acquired voxel size $2\times2\times2$~mm, repetition time 4.5~ms, echo time 2.2~ms, echo train length 26 and flip angle $90^\circ$. The diastolic rest period (diastasis) was used to generate the computational mesh. The geometry was meshed using Gmsh \cite{geuzaine09} with a resolution of $2\text{ mm}$, yielding $282\,288$ nodes and $167\,232$ quadratic tetrahedral elements, totalling a $846\,864$ structural degrees of freedom. Additionally, our geometry contains triangular surface elements with no additional degrees of freedom to track the movement of the planes of cardiac valves, allowing us to monitor the volumes of all four cardiac cavities. \reva{All four cardiac cavities are closed with surface elements with no additional degrees of freedom at the valve planes depicted in red in figure~\ref{materials} at the left and right atrioventricular plane, respectively.  The atria are additionally closed at their respective connections to the vasculature. We can thus monitor the volumes of all four cardiac cavities and track the movement of cardiac valve planes.} The meshed geometry is shown in figure~\ref{mesh}. The different materials are depicted in figure~\ref{materials}.

Additionally, we prescribe the local angles of cardiac myofibers at epi- and endocardium of the ventricles. Using harmonic lifting, the fiber vectors $\vec{f}_0$ are interpolated to the interior of the domain by solving a Laplace problem \cite{nagler16}. \revs{We interpolate the fiber orientation smoothly at each integration point}. It is well known that the fiber orientation has a strong impact on \revb{passive and active} cardiac mechanics \cite{ubbink06,wong10,gil13,eriksson13a,asner16,nikou16}. In order to make a more clear statement about the pericardial boundary conditions independently of the fiber orientation and to show the interplay between boundary conditions and fiber orientations, we compare in this work three different fiber distributions: \fifty{}, \sixty{}, and \seventy{}. The first and second angle describe the fiber helix angle at the endo- and epicardial surface, respectively, with respect to the local circumferential axis. \revs{The transverse angle is zero throughout the myocardium.} \revb{The sheet normal vector $\bm n_0$ is perpendicular to the epi- and endocardial surfaces. The sheet vector $\bm s_0$ is then obtained from $\bm s_0 = \bm n_0 \times \bm f_0$.} \reva{The atrial fiber architecture is obtained using a semi-automatic registration method based on the fiber definition in atlas atria \cite{hoermann17,hoermann18b}. See figure~\ref{fibers} for atrial fibers and ventricular \sixty{} fibers visualized at the endocardium.}

\subsection{Modeling cardiac contraction}

Balance of momentum, a Neumann windkessel coupling condition with left ventricular pressure~$p_v$ acting on the left endocardium $\varGamma_0^{\text{endo}}$, omni-directional spring-dashpot boundary conditions, and pericardial boundary conditions yield the energy $\delta\Pi$ of the boundary value problem
\begin{equation}
\begin{aligned}
\delta\Pi&=\int_{\varOmega_0} \rho_0 \, \ddot{\vec{u}} \cdot \delta\vec{u}  \dd V 
+ \int_{\varOmega_0} \ten{S} : \delta \ten{E}  \dd V \\
&\revs{+ \sum_{\nu\in\{l,r\}}\int_{\varGamma_0^{\text{endo},\nu}} p_v^\nu \, \ten{F}^{-\mathrm{T}} \cdot \ten{N} \cdot \delta\vec{u} \dd A}\\
&+ \int_{\varGamma^{\text{vess}}} \left[k_v \vec{u} + c_v \dot{\vec{u}}\right] \cdot\delta\vec{u} \dd A
\label{eq_weak}
\end{aligned}
\end{equation}
\revs{with density $\rho_0$, accelerations $\ddot{\vec{u}}$, the second Piola-Kirch\-hoff stress tensor $\ten{S}$, and spring stiffnesses $k_v,k_a$ and viscosities $c_v,c_a$ for vessel and apical surfaces $\varGamma^{\text{vess}}$ and $\varGamma^{\text{apex}}$, respectively. Furthermore, we have virtual displacements $\delta\vec{u}$ and virtual strains $\delta\vec{E}$. Omni-directional springs and dashpots are placed on the outsides of the great vessels $\varGamma^{\text{vess}}$ and apical surface $\varGamma^{\text{apex}}$. The energy $\delta\Pi$ is complemented in section~\ref{sec_results} by the energy of different boundary conditions.}
%
%\begin{figure*}
%\centering
%\includegraphics[height=.25\textheight]{boundary_conditions.png}
%\caption{Surfaces for boundary conditions: cut-offs of great vessels $\varGamma^{\text{cut}}$ (red), epicardium $\varGamma^{\text{epi}}$ (blue), and the outside of the great vessels $\varGamma^{\text{vess}}$ (yellow). \label{bcs}}
%\end{figure*}

We define different materials with the volumes defined as in figure~\ref{geometry} for adipose tissue~\eqref{mat_fat}, \revb{ventricular and atrial myocardium~\eqref{mat_myocard}, and aorta and pulmonary artery~\eqref{mat_vessel}}:
\begin{align}
\label{mat_fat} \ten{S} = \frac{\partial}{\partial \ten{E}} \left( \psi_{\text{NH}} + \psi_{\text{vol}} \right) &+ \frac{\partial}{\partial \dot{\ten{E}}} \, \psi_{\text{visco}},\\
\label{mat_myocard} \revb{\ten{S} =} \revb{\frac{\partial}{\partial \ten{E}} \left( \psi_{\text{exp}} + \psi_{\text{vol}} \right)} & \revb{+\frac{\partial}{\partial \dot{\ten{E}}} \, \psi_{\text{visco}} + \ten{S}_{\text{act}}},\\
\label{mat_vessel} \ten{S} = \frac{\partial}{\partial \ten{E}} \left( \psi_{\text{MR}} + \psi_{\text{vol}} \right) &+ \frac{\partial}{\partial \dot{\ten{E}}} \, \psi_{\text{visco}}.
\end{align}
Each material is composed of a hyperelastic and a viscous contribution depending on the rate of Green-Lagrange strains $\dot{\ten{E}}$. The viscous behavior of myocardial tissue is modeled with a viscous pseudo-potential. Only the ventricular tissue in \eqref{mat_myocard} has an additional active stress component $\ten{S}_{\text{act}}$. The strain energy density functions for \revb{exponential orthotropic solid $\psi_{\text{exp}}$} \cite{holzapfel09}, Mooney-Rivlin solid \cite{sainte-marie06,chabiniok12} $\psi_{\text{MR}}$, Neo-Hookean solid $\psi_{\text{NH}}$, penalty function $\psi_{\text{vol}}$, and viscous pseudo-potential \cite{chapelle12} $\psi_{\text{visco}}$ are given as
\begin{equation}
\begin{aligned}
\revb{\psi_{\text{exp}}} &\revb{= \frac{a}{2b} \left( e^{b(\bar{I}_1-3)} - 1 \right)} \revb{+ \frac{a_{fs}}{2b_{fs}} \left( e^{b_{fs}{I}_{8,fs}^2} - 1 \right)}\\
&\revb{+ \sum_{i\in\{f,s\}} \frac{a_i}{2b_i} \left( e^{b_i({I}_{4,i}-3)} - 1 \right),}\\
\psi_{\text{MR}} &= C_1 (\bar{I}_1 - 3) + C_2 (\bar{I}_2 - 3), \quad 
\psi_{\text{NH}} = \frac{\mu}{2} (\bar{I}_1-3),\\
\revb{\psi_{\text{vol}}} &\revb{= \frac{\kappa}{2} \left( 1-J \right)^2,} \quad
\psi_{\text{visco}} = \frac{\eta}{2} \text{tr}\left(\dot{\ten{E}}^2\right),
\end{aligned}
\end{equation}
with \revb{the Jacobian $J=\det \bm F$ of the deformation gradient and} material parameters $a_i$, $b_i$, $\mu, C_1$, and $C_2$, penalty parameter $\kappa$, viscosity $\eta$, and invariants 
\begin{equation}
\begin{aligned}
\bar{I}_1 &= J^{-2/3}I_1, \quad
\bar{I}_2=J^{-4/3}I_2,\\
I_1 &= \text{tr }\ten{C}, \quad
I_2 = \frac{1}{2}\left[\text{tr}^2\left(\ten{C}\right) - \text{tr}\left(\ten{C}^2\right)\right],\\
\revb{
{I}_{4,f}} &\revb{= \bm f_0 \cdot \bm C \bm f_0, \quad
{I}_{4,s} = \bm s_0 \cdot \bm C \bm s_0, \quad
{I}_{8,fs} = \bm f_0 \cdot \bm C \bm s_0.}
\end{aligned}
\end{equation}
\begin{figure}
\centering
\setlength\figureheight{2.5cm}
\setlength\figurewidth{0.8\columnwidth}
\subfloat[\reva{Prescribed indicator functions $f(t)$ for atria (blue) and ventricles (red)}. \label{activation}]{
\includegraphics{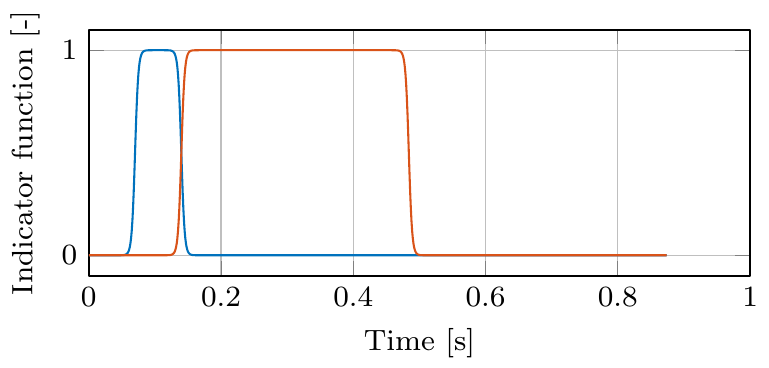}}\\
\subfloat[\reva{Active stress $\tau(t)$ for atria (blue) and ventricles (red) with maximum values $\sigma_\text{a}$ and $\sigma_\text{v}$, respectively}. \label{active_stress}]{
\includegraphics{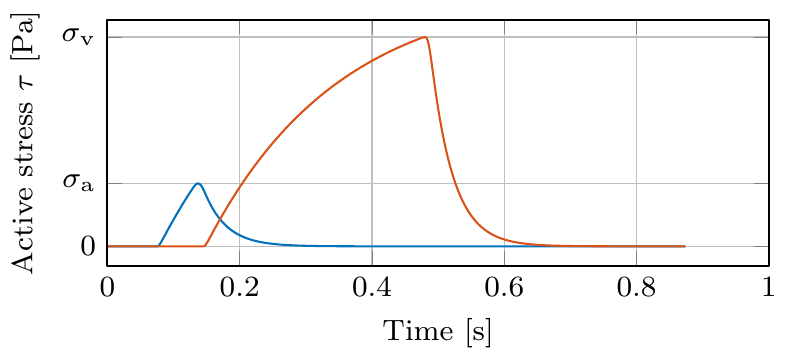}}
\caption{Active stress. \label{wk_time}}
\end{figure}

\reva{We use the same active stress approach for both atrial and ventricular myocardium, though with different parameters. However, for simplicity of notation, we do not distinguish between atria or ventricles in the following equations.} The active stress is computed as
\begin{equation}
\ten{S}_{\text{act}} = \tau(t) \cdot \vec{f}_0\otimes\vec{f}_0
\end{equation}
with fiber stress $\tau$ in local reference fiber direction $\vec{f}_0$. Based on \cite{bestel01}, we model fiber stress by the evolution equation \revc{
\beqq
\dot{\tau}(t) = - |a(t)|\tau(t) + \sigma_0 |a(t)|_+
\label{eq_sigma}
\eeqq
with activation function $a$, contractility $\sigma_0$, and the function $|a(t)|_+ = \max (a(t);0)$. The activation function $a(t)$ is modeled by
\begin{equation}
a(t) = \alpha_{\text{max}} \cdot f(t) + \alpha_{\text{min}} \cdot [1-f(t)]
\end{equation}
}
\reva{with maximum and minimum activation rates $\alpha_{\text{max}}$ and $\alpha_{\text{min}}$, respectively, and functions
\begin{align}
f(t) &= S^+(t-t_{\text{sys}}) \cdot S^-(t-t_{\text{dias}}),\\
S^\pm(\Delta t) &= \frac{1}{2} \left[ 1 \pm \text{tanh} \left( \frac{\Delta t}{\gamma} \right) \right]
\end{align}
with steepness $\gamma=0.005$~s and descending and ascending sigmoid functions $S^+$ and $S^-$, respectively. The indicator function $f\in~]0,1[$ indicates systole. The times $t_{\text{sys}}$ and $t_{\text{dias}}$ model the onset of systole and diastole, respectively.} Note that the active stress tensor $\ten{S}_{\text{act}}$ is the only input of our solid model we prescribe over time. Our structural model can be coupled to a model of electric signal propagation as shown in \cite{hoermann18}. However, as the focus in this work is on pericardial boundary conditions,  a coupled electro-mechanical model would only introduce unnecessary complexity and variability. Using the parameters in table~\ref{mat_solid}, we obtain the active stress curve depicted in figure~\ref{active_stress}. \reva{The values $\sigma_\text{a}$ and $\sigma_\text{v}$ denote the maximum of the active stress $\tau$ for atria and ventricles, respectively.}

\revc{Using the finite element method, } we discretize displacements $\vec{u}$ and virtual displacements $\delta\vec{u}$ arising in the weak form \eqref{eq_weak} by
\begin{equation}
\begin{aligned}
\vec{u}^{(e)} &= \ten{\varphi}^{(e)}(\vec{X}) ~ \ds^{(e)}\\
\delta\vec{u}^{(e)} &= \ten{\varphi}^{(e)}(\vec{X}) ~ \delta\ds^{(e)}
\end{aligned}
\end{equation}
with quadratic basis functions $\ten{\varphi}$ and nodal displacements $\ds$ on each tetrahedral element $(e)$. Assembly of the discretized problem leads to the matrix notation
\begin{equation}
\vecrm{M} \ddot{\ds} + \vecrm{F}(\ds,\dot{\ds},p_v) = \vecrm{0},
\end{equation}
with mass matrix $\vecrm{M}$, force vector $\vecrm{F}$, and discrete displacements, velocities, and accelerations $\ds,\dot{\ds}$, and $\ddot{\ds}$, respectively. We discretize the boundary value problem in time with Newmark's method \cite{newmark59}
\begin{equation}
\begin{aligned}
\dot{\ds}_{n+1} &= \frac{\gamma}{\beta \Delta t} (\ds_{n+1} - \ds_n) - \frac{\gamma-\beta}{\beta} \dot{\ds}_n - \frac{\gamma-2\beta}{2\beta} \Delta t \ddot{\ds}_n,\\
\ddot{\ds}_{n+1} &= \frac{1}{\beta \Delta t^2} (\ds_{n+1} - \ds_n) - \frac{1}{\beta \Delta t} \dot{\ds}_n - \frac{1-2\beta}{2\beta} \ddot{\ds}_n,
\end{aligned}
\end{equation}
with parameters $\gamma\in[0,1]$ and $\beta\in[0,0.5]$, and time step size $\Delta t = t_{n+1} - t_{n}$. Additionally, we apply the generalized-$\alpha$ method \cite{chung93}, yielding quantities at a generalized time step $n+1-\alpha_i$
\begin{equation}
\begin{aligned}
(\bullet)_{n+1-\alpha_i} = (1 - \alpha_i) (\bullet)_{n+1} + \alpha_i (\bullet)_n, \\ \alpha_i\in[0,1], \quad i\in\{f,m\}
\end{aligned}
\end{equation}
depending on the weights $\alpha_f$ and $\alpha_m$ for force vector and mass matrix respectively. \revc{Newmark's method in combination with the generalized $\alpha$-scheme is a common technique for implicit one-step time integration for finite elements in nonlinear solid dynamics.} Finally, we obtain the time and space discrete structural residual 
\beqq
\rs = \vecrm{M} \ddot{\ds}_{n+1-\alpha_m} + \vecrm{F}_{n+1-\alpha_f}.
\eeqq
All parameters used for the elastodynamical model are summarized in table~\ref{mat_solid}. %For $\rho_0$ we use the density of water. The material parameters of the myocardial isotropic Mooney Rivlin solid material $C_1$ and $C_2$ are taken from . \todo{Warum wurden die so gewaehlt?}

\begin{figure}
\centering
\setlength\figureheight{2.5cm}
\setlength\figurewidth{0.8\columnwidth}
\subfloat[Schematic. \label{windkessel_schematic}]{
\includegraphics[width=.45\textwidth]{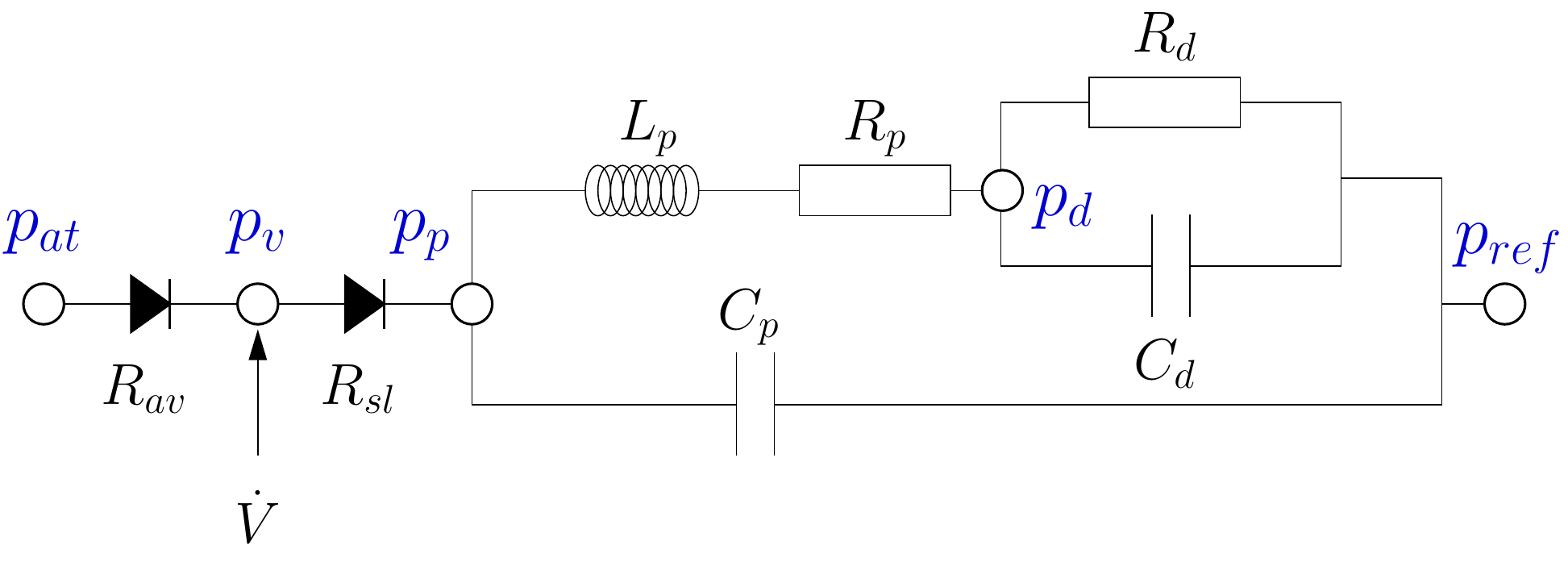}}\\
\subfloat[\revs{Prescribed left (blue) and right (red) atrial pressure $p_{at}(t)$.} \label{p_at}]{
\includegraphics{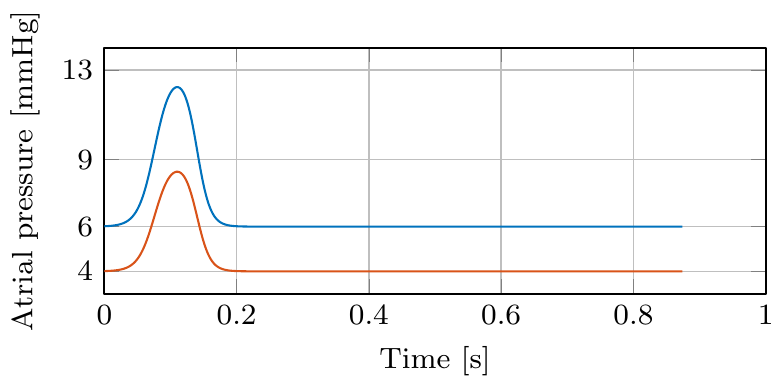}}
\caption{Windkessel model. \label{windkessel}}
\end{figure}

\subsection{Modeling the circulatory system}
\label{sec_windkessel}
\revs{We use the same windkessel model for each ventricle with different parameters. For simplicity of notation, we drop in this section the index of the ventricle $\nu$ in all windkessel parameters and variables. Note that this cardiovascular model does not represent a closed-loop system since the total blood volume is not conserved, i.e. blood exiting the right ventricle into the lungs does not enter the left atrium. However, using a windkessel model for each ventricle provides us with a reasonable approximation of ventricular pressures.}

In this work we use a four element windkessel model based on the ideas in~\cite{westerhof08} and~\cite{kerckhoffs07}. A comprehensive review of different windkessel models is given in \cite{shi11}. The schematic of our windkessel model is given in figure~\ref{windkessel_schematic} using resistances~$R$, compliances~$C$, and an inertance~$L_p$. Pressures at different parts of the model are denoted by~$p$. We distinguish between a proximal (index~$p$) and a distal part (index~$d$) of the outlets, i.e. lung and aorta for the right and left ventricle, respectively. The atrial pressure $p_{at}$ is prescribed to simulate atrial systole, see figure~\ref{p_at}. The reference pressure~$p_{\text{ref}}$ is kept constant.

\revc{
We model the atrioventricular and semilunar valves with a smooth diode-like behavior by non-linear resistances $R_{av} \defeq R(p_v-p_{at})$ and $R_{sl} \defeq R(p_p-p_v)$, respectively, depending on the sigmoid function
\begin{equation}
\begin{aligned}
R(\Delta p) =  R_{\text{min}} + (R_{\text{max}} - R_{\text{min}}) \cdot S^+(\Delta p)
\end{aligned}
\end{equation}
with the steepness $k_p$ of the sigmoid function $S^+$ and the minimal and maximal valve resistance $R_{\text{min}}\to0$ and $R_{\text{max}}\to\infty$, respectively.} This yields the set of differential equations
\begin{equation}
\begin{aligned}
\frac{p_v - p_{at}}{R_{av}} + \frac{p_v - p_p}{R_{sl}} + \dot{V}(\vec{u}) &= 0,\\
q_p - \frac{p_v - p_p}{R_{sl}} + C_p \dot{p}_p &= 0, \\
q_p + \frac{p_d - p_p}{R_p} + \frac{L_p}{R_p} \dot{q}_p &= 0,\\
\frac{p_d - p_{\text{ref}}}{R_d} - q_p + C_d \dot{p}_d &= 0.
\label{eq_wk}
\end{aligned}
\end{equation}
\revc{The 0D windkessel model is strongly coupled to the 3D structural model. The 0D model depends on the structural displacements $\vec{u}$ of the 3D model via the change in ventricular volume $\dot{V}$. On the other hand, the 3D model depends on left and right ventricular pressure from the 0D model. The coupling between both models is described in section~\ref{sec_coupling}.} The vector of primary variables yields $\p=[p_v, p_p, p_d, q_p]^\mathrm{T}$, including the flux $q_p$ through the inertance $L_p$. We discretize the set of windkessel equations~\eqref{eq_wk} in time with the one-step-$\theta$ scheme
\begin{equation}
\begin{aligned}
\dot{(\bullet)}_{n+1} &= \frac{(\bullet)_{n+1} - (\bullet)_n}{\Delta t}, \quad \theta\in[0,1],\\
(\bullet)_{n+\theta} &= \theta (\bullet)_{n+1} + (1-\theta) (\bullet)_n.
\end{aligned}
\end{equation}
This yields the discrete windkessel residual $\rd$ evaluated at time step $n+\theta$. The parameters and initial conditions of the cardiovascular model are summarized in table~\ref{mat_wk} and table~\ref{init} respectively.  \revc{Windkessel parameters are motivated by values from literature and adapted for this heart to yield physiological pressures as well as approximately a periodic state of the windkessel systems.}

\begin{table*}
\centering
\footnotesize
\setlength{\tabcolsep}{.3em}
\renewcommand{\arraystretch}{1.8}
\begin{minipage}{0.45\textwidth}
\subfloat[Parameters of the elastodynamical model. \label{mat_solid}]{
\begin{tabular}{l l l c}
\textbf{Name} & \textbf{Par.} & \textbf{Value} & \textbf{Unit} \\
\hline
\multicolumn{4}{l}{\textit{All tissues}}\\
\hline
Tissue density & $\rho_0$ & $10^3$ & $\left[\frac{\text{kg}}{\text{m}^3}\right]$ \\
Viscosity & $\eta$ & 0.1 & $\left[\text{kPa}\cdot\text{s}\right]$ \\
Volumetric penalty & $\kappa$  & $10^3$ & $\left[\text{kPa}\right]$ \\
\hline
\multicolumn{4}{l}{\textit{Active myocardial tissue}}\\
\hline
Atrial contractility & $\sigma_\text{a}$ & 9.72 & kPa \\
Ventricular contractility & $\sigma_\text{v}$ & \multicolumn{2}{c}{see table~\ref{sigma}} \\
Activation rate & $\alpha_{\text{max}}$ & $+5$  & $\left[\frac{1}{\text{s}}\right]$ \\
Deactivation rate & $\alpha_{\text{min}}$ & $-30$ & $\left[\frac{1}{\text{s}}\right]$ \\
Atrial systole & $t_{\text{sys}}$ & 70 & $\left[\text{ms}\right]$ \\
Atrial diastole & $t_{\text{dias}}$ & 140 & $\left[\text{ms}\right]$ \\
Ventricular systole & $t_{\text{sys}}$ & \multicolumn{2}{c}{see table~\ref{sigma}} \\
Ventricular diastole & $t_{\text{dias}}$ & 484 & $\left[\text{ms}\right]$ \\
\hline
\multicolumn{4}{l}{\textit{Passive myocardial tissue} (\cite{holzapfel09} table 1, shear, figure 7)}\\
\hline
Matrix & $a$ & 0.059 & $\left[\text{kPa}\right]$ \\
 & $b$ & 8.023 & $[-]$ \\
Fiber & $a_f$ & 18.472 & $\left[\text{kPa}\right]$ \\
 & $b_f$ & 16.026 & $[-]$ \\
Sheet & $a_s$ & 2.481 & $\left[\text{kPa}\right]$ \\
 & $b_s$ & 11.120 & $[-]$ \\
Fiber-sheet & $a_{fs}$ & 0.216 & $\left[\text{kPa}\right]$ \\
 & $b_{fs}$ & 11.436 & $[-]$\\
\hline
\multicolumn{4}{l}{\textit{Great vessels}}\\
\hline
Mooney-Rivlin & $C_1$ & 5.0 & $\left[\text{kPa}\right]$ \\
Mooney-Rivlin & $C_2$ & 0.04 & $\left[\text{kPa}\right]$  \\
Spring stiffness & $k_v$ & $2.0 \cdot 10^3$ & $\left[\frac{\text{kPa}}{\text{mm}}\right]$ \\
Dashpot viscosity & $c_v$ & $1.0 \cdot 10^{-2}$ & $\left[\frac{\text{kPa}\cdot\text{s}}{\text{mm}}\right]$\\
\hline
\multicolumn{4}{l}{\textit{Adipose tissue}}\\
\hline
Neo-Hooke & $\mu$ & 1.0 & $\left[\text{kPa}\right]$ \\
\hline
\multicolumn{4}{l}{\textit{Pericardial boundary condition}}\\
\hline
\multicolumn{4}{c}{see table~\ref{tab_bcs}} \\
\hline
\end{tabular}}
\end{minipage}\qquad
 \begin{minipage}{0.45\textwidth}
\subfloat[Parameters of the reduced order cardiovascular model (identical for left and right ventricle). \label{mat_wk}]{
\begin{tabular}{l l l c}
\textbf{Name} & \textbf{Par.} & \textbf{Value} & \textbf{Unit} \\
\hline
Proximal inertance & $L_p$ & $1.3 \cdot 10^5$ & $\left[\frac{\text{kg}}{\text{m}^4}\right]$ \\
Proximal capacity & $C_p$ & $7.7 \cdot 10^{-9}$ & $\left[\frac{\text{m}^4 \cdot \text{s}^2}{\text{kg}}\right]$ \\
Distal capacity & $C_d$ & $8.7 \cdot 10^{-9}$ & $\left[\frac{\text{m}^4 \cdot \text{s}^2}{\text{kg}}\right]$ \\
Proximal resistance & $R_p$ & $7.3 \cdot 10^6$ & $\left[\frac{\text{kg}}{\text{m}^4\cdot\text{s}}\right]$ \\
Distal resistance & $R_d$ & $1.0 \cdot 10^8$ & $\left[\frac{\text{kg}}{\text{m}^4\cdot\text{s}}\right]$ \\
Reference pressure & $p_{\text{ref}}$ & 0 & $\left[\text{Pa}\right]$ \\
Closed valve resistance & $R_{\text{max}}$ & $1.0 \cdot 10^{13}$ & $\left[\frac{\text{kg}}{\text{m}^4\cdot\text{s}}\right]$ \\
Open valve resistance & $R_{\text{min}}$ & $1.0 \cdot 10^6$ & $\left[\frac{\text{kg}}{\text{m}^4\cdot\text{s}}\right]$ \\
Valve steepness & $k_p$ & $1.0 \cdot 10^{-3}$ & $\left[\text{Pa}\right]$ \\
\hline
\end{tabular}} \\
\subfloat[Initial conditions of the reduced order cardiovascular model. \label{init}]{
\begin{tabular}{l l l l c}
\textbf{Name} & \textbf{Par.} & \multicolumn{2}{c}{\textbf{Value}}& \textbf{Unit} \\[-1em]
& & \textbf{Left} & \textbf{Right }&\\
\hline
Atrial pressure & $p_{at}(0)$ & 6.0 & 4.0 & $\left[\text{mmHg}\right]$ \\
Ventricular pressure & $p_v(0)$ & 8.0 & 6.0 & $\left[\text{mmHg}\right]$ \\
Proximal pressure & $p_p(0)$ & 61.8 & 24.0 & $\left[\text{mmHg}\right]$ \\
Distal pressure & $p_d(0)$ & 59.7 & 23.2 & $\left[\text{mmHg}\right]$ \\
Proximal flow & $q_p(0)$ & 38.3 & 14.9 & $\left[\frac{\text{cm}^3}{\text{s}}\right]$ \\
\hline
\end{tabular}} \\
\subfloat[Numerical time integration parameters. \label{tab_numerical}]{
\begin{tabular}{l l}
\renewcommand{\arraystretch}{1}
\textbf{Par.} & \textbf{Value} \\
\hline
\multicolumn{2}{l}{\textit{Generalized-$\alpha$}}\\
\hline
$\gamma$, $\alpha_f$, $\alpha_m$ & 0.5 \\
$\beta$ & 0.25 \\
\hline
\multicolumn{2}{l}{\textit{One-step-$\theta$}}\\
\hline
$\theta$ & 1.0 \\
\hline
\end{tabular}}
\end{minipage}
\caption{\revs{Overview of parameters in four-chamber cardiac model.}\label{tab_parameters}}
\end{table*}

\subsection{Prestress \label{prestress}}
For our reference configuration we use a patient-specific geometry segmented from static 3D MRI at diastolic rest period (diastasis), see section~\ref{sec_geom}. \revc{Diastasis is very suitable for the reference configuration, since both ventricular and atrial myofibers are relaxed, the heart is not accelerated, and blood pressures are minimal and constant. This simplifies the task of obtaining the stress state of the reference configuration, which in this case is determined by the static blood pressures within the cardiac cavities.} We therefore have to prestress our geometry with the initial ventricular pressure from table~\ref{init}. In this work, we use the Modified Updated Lagrangian Formulation as proposed in \cite{gee09a,gee10}. \revc{This method incrementally calculates a deformation gradient with respect to an unknown stress-free reference configuration. From this deformation gradient, a stress field is calculated so that the segmented geometry of the heart is in balance with the prestressed pressure state.} This yields a prestress within the myocardium as well as in the pericardial boundary condition in case \normal{}. Note that while this technique allows to model prestress, we do not account for the residual stress inherent in myocardial tissue \cite{joebsis07}.

\subsection{Solving the 0D-3D coupled problem}
\label{sec_coupling}
We solve the coupled 0D-3D model with the structural and windkessel residuals $\rs$ and $\rd$, respectively, at time step $n+1$ with the Newton-Raphson method
\begin{equation}
\renewcommand{\arraystretch}{1.8}
\left[
\begin{matrix}
\frac{\partial \rs}{\partial \ds} & \frac{\partial \rs}{\partial \p} \\
\frac{\partial \rd}{\partial \ds} & \frac{\partial \rd}{\partial \p} \\
\end{matrix}
\right]^i_{n+1}
\cdot
\left[
\begin{matrix}
\Delta \ds \\
\Delta \p
\end{matrix}
\right]^{i+1}_{n+1}
= -
\left[
\begin{matrix}
\rs \\
\rd
\end{matrix}
\right]^i_{n+1},
\end{equation}
\revc{in a monolithic fashion} for increments in displacements and windkessel variables $\Delta\ds_{n+1}$ and $\Delta\p_{n+1}$, respectively, at iteration $i+1$ until convergence. We build and solve this coupled system using our in-house code BACI \cite{wall10}. The numerical parameters for the time integration of our model are listed in table~\ref{tab_numerical}.

\section{Results} 
\label{sec_results}

\begin{figure}
\centering
\subfloat[Case \apex{} with omni-directional spring-dashpots on $\varGamma^{\text{apex}}$ (green). \label{case_apex}]{
\includegraphics[height=.24\textheight]{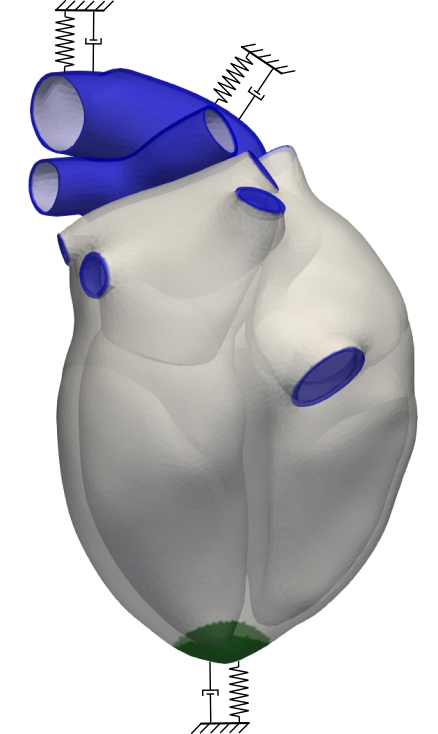}}\quad
%\hspace{.5cm}
\subfloat[Case \normal{} with normal spring-dashpots on $\varGamma^{\text{epi}}$ (red). \label{case_pericardium}]{
\includegraphics[height=.24\textheight]{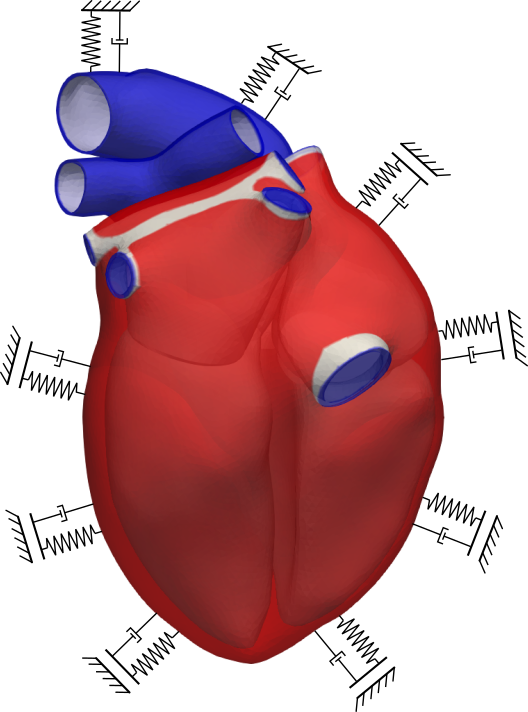}}
\caption{Surface definitions for boundary conditions with omni-directional spring-dashpots on  $\varGamma^{\text{vess}}$ (blue) and homogeneous Neumann boundary conditions (white). \label{cases}}
\end{figure}

In order to investigate the influence of pericardial boundary conditions, we compare simulations \revc{with and without pericardial boundary condition} on the epicardial surface $\varGamma_0^{\text{epi}}$. The simulations will be denoted by \apex{} and \normal{} in the following. See table~\ref{tab_bcs} for an overview of used parameters.

\revs{Case \apex{} depicted in figure~\ref{case_apex} yields the boundary-vale problem
\begin{align}
0 = \delta\Pi+ \int_{\varGamma^{\text{apex}}} \left[k_a \vec{u} + c_a \dot{\vec{u}}\right] \cdot\delta\vec{u} \dd A
\end{align}
adding the energy for omni-directional spring dashpots to the energy \eqref{eq_weak}, where $\varGamma^{\text{apex}}$ is the apical surface. The apical surface is defined as the epicardial surface within 10\,mm of the apex, see figure~\ref{cases}. It resembles homogeneous Neumann boundary conditions on $\varGamma_0^{\text{epi}} \setminus \varGamma_0^{\text{apex}}$, i.e. the absence of any pericardial boundary conditions as frequently found in literature \cite{sermesant12,chabiniok12,augustin16}.}

\revs{Case \normal{} depicted in figure~\ref{case_pericardium} yields the boundary-vale problem 
\begin{align}
0 = \delta\Pi+ \int_{\varGamma^{\text{epi}}} \vec{N} \left[k_p \vec{u} \cdot \vec{N} + c_p \dot{\vec{u}} \cdot \vec{N}\right] \cdot\delta\vec{u} \dd A
\end{align}
adding the energy for the pericardial boundary condition to the energy \eqref{eq_weak}, where $\varGamma^{\text{epi}}$ is the pericardial surface. The choice of parameters $k_v$ and $c_v$ is detailed in appendix~\ref{sec_params}.}

\revs{
The remainder of this section is structured as follows. We first give an overview of all methods used in this work to quantify the difference of simulation and MRI in section~\ref{sec_assessment}. Next, we calibrate model parameters for both cases in section~\ref{sec_params}.} In the following sections, we validate various outputs of both simulation cases \apex{} and \normal{} with measurements from cine MRI. We begin with scalar windkessel outputs and compare the simulation volume curves to MRI in section~\ref{sec_res_scalar}. A qualitative evaluation of displacement results is given in section~\ref{sec_ref_es} by comparing end-systolic simulation results to cine MRI frames at multiple views. We quantify the differences in pumping motion for simulation cases \apex{} and \normal{} by comparing the displacements of the left and right atrioventricular plane to MRI in section~\ref{sec_avpd}. The interplay between ventricles and atria with and without the presence of pericardial boundary conditions is investigated in section~\ref{sec_inter}. In section~\ref{sec_prr} we calculate a spatial error for the left and right endocardium to quantify the overall approximation quality. Finally, we evaluate the contact stress of our pericardial boundary condition of case \normal{} in section~\ref{sec_stress}.

\revs{
\subsection{Assessment of cardiac function}
\label{sec_assessment}
In this section, we briefly describe the various methods we use throughout this work to quantify cardiac function of different simulations.
}

\begin{figure}
\centering
\setlength\figureheight{5cm}
\subfloat[Four chamber view. \label{4ch}]{
\includegraphics[height=\figureheight]{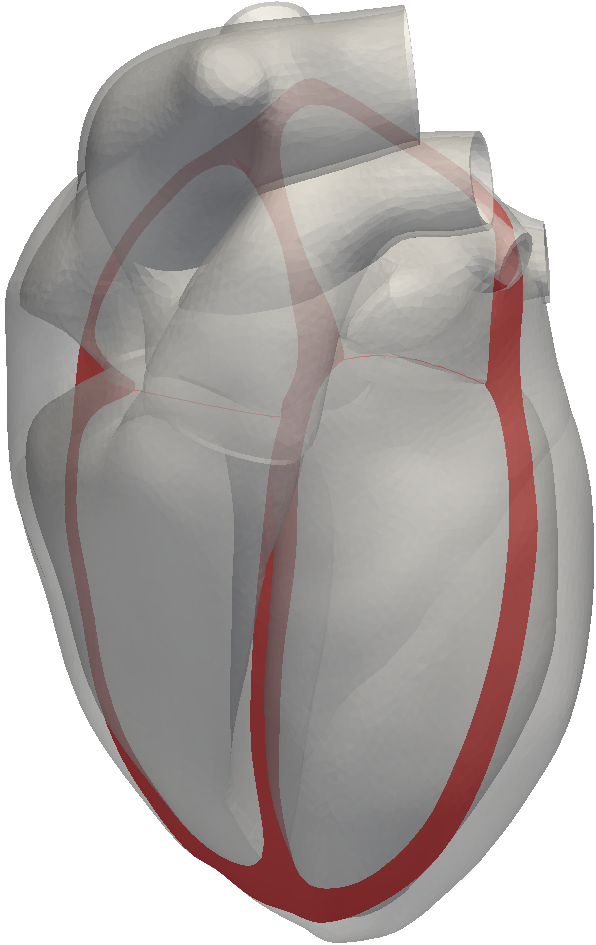}}~
\hspace{0.25cm}
\subfloat[Short axis endocardial contours used for error calculation. \label{contours}]{
\includegraphics[height=\figureheight]{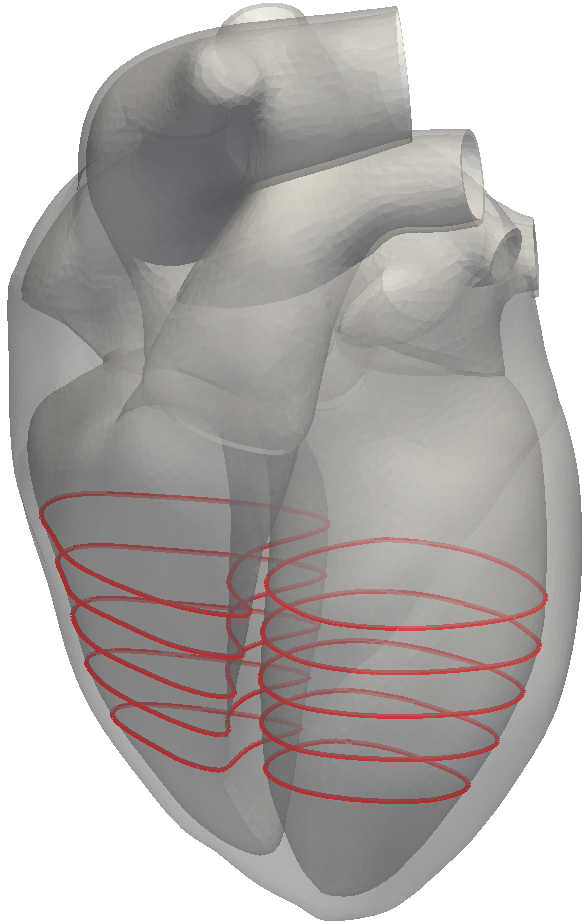}}\\
%
%\hspace{0.25cm}
\subfloat[Short axis slice 9. \label{sax9}]{
\includegraphics[height=\figureheight]{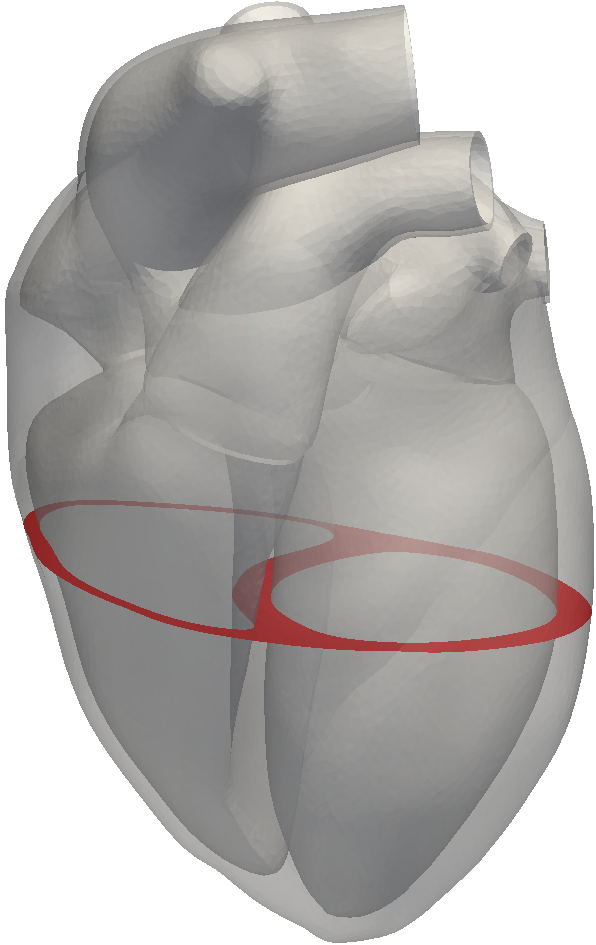}}
\hspace{0.25cm}
\subfloat[Short axis slice 6. \label{sax6}]{
\includegraphics[height=\figureheight]{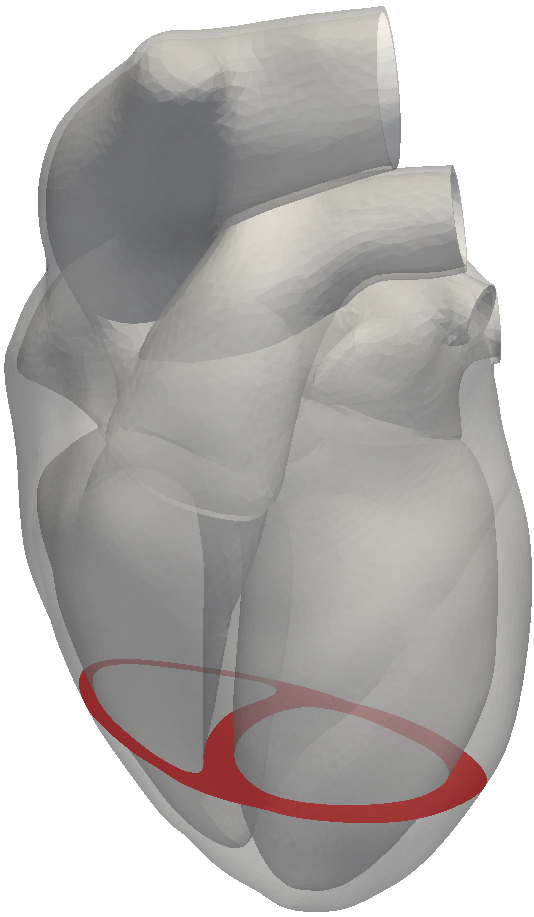}}~
\caption{Post processing planes for simulations and cine MRI.\label{mri_views}}
\end{figure}

\paragraph{Cine MRI}
We use cine MRI with a temporal resolution of $\sim$30~ms in four- (figure~\ref{4ch}), three-, and two-chamber views and short axis planes with a slice distance of 8~mm (figures~\ref{sax9}, \ref{sax6}, \ref{contours}). All cine MRI data used in this work is rigidly registered to the static 3D image \revc{taken during diastasis and} used for geometry creation to account for any movement of the subject during image acquisition.

It is important to note that the reference configuration of our simulation is obtained from static 3D imaging with a fine isotropic resolution and acquired in free-breathing, as explained in section~\ref{sec_geom}. For the comparison of simulation results to cine MRI however, we have to rely on sparsely distributed images acquired in expiration breath-hold. The used image types rely on different MRI acquisition parameters and pulse sequences. Therefore, it is impossible for our simulation to match the cine MRI data perfectly, even in reference configuration. \revc{This error however is usually smaller than the approximation error of the cardiac model.}

\paragraph{Left ventricular volume}
We obtain a reference for left ventricular volume by manually segmenting the left endocardial surface obtained from the short axis cine MRI stack at all time steps. We add the sum of areas in each short axis slice multiplied by the slice thickness. We cut the volume at the top and bottom according to the limits of the left ventricle at each time step, as observed in two chamber and four chamber views.

In order to be fair and not introduce a bias towards our more realistic model, for each simulation, we calibrate the contractility $\sigma_0$. It is a key parameter describing cardiac elastodynamics, resembling the asymptotical active fiber stress in \eqref{eq_sigma}. It controls maximum deformation during systole. In order to make simulations comparable, we adapt $\sigma_0$ for each combination of boundary condition and fiber orientation to match the left ventricular volume at end-systole as segmented from cine MRI of $V_{\text{min}}=57~\text{ml}$. The heart thus yields a stroke volume of 75~ml and an ejection fraction (EF) of 57\%.

\paragraph{Atrioventricular plane displacement (AVPD)}
The movement of the left or right plane of the valve separating atrium and ventricle in long axis direction during the cardiac cycle is described by AVPD. For left and right ventricle those valves are termed mitral and tricuspital valve, respectively. As a scalar parameter, AVPD at end-systole is an important clinical parameter to describe and predict cardiac vitality \cite{carlsson07,willenheimer97}.

We evaluate AVPD in this work as it gives us a quantitative measurement of the displacements in long axis direction. We semi-automatically extract the displacement of the left and right atrioventricular plane from two, three, and four chamber cine MRI using the freely available software Segment version 2.0 R5585 \cite{heiberg10}. In our simulations, we average the displacements on all nodes on the valve plane (see the red planes in figure~\ref{materials}) and project them onto the long axis direction. A positive sign indicates a movement of the base towards the apex.
% visualized in the four chamber view in figures~\ref{4ch_free} and \ref{4ch_normal}

\paragraph{Spatial error}
We validate displacement in long axis direction using AVPD as measurement. To validate radial displacement we compare the movements of cardiac surfaces in simulations to the ones from short axis cine MRI. For comparison, we select the left and right endocardium, as it shows how pericardial boundary conditions prescribed on the epicardium act on the interior of the domain.

For each MRI time step (temporal resolution $\sim$30~ms) we select the closest simulation time step (temporal resolution 1~ms). \revc{Spatially, we extract the simulations' displacement results at the same positions where the cine MRI slices were acquired. This is possible since we use the MRI scanner's global coordinate system for all images and the simulation. This method can be thought of as taking a virtual cine MRI of the simulation. This yields an Eulerian description of motion, as the observer is fixed in space. The difference of simulated displacements to cine MRI data was used previously, e.g. in \cite{chabiniok12} to estimate local tissue contractility. Note that this technique does not allow us to track rotations of the left ventricle due to its rotational symmetry.}

We manually segment the contours of left and right endocardium from short axis cine MRI for slices 5 to 9 at all MRI time steps, see figure~\ref{contours}. These slices are selected because the myocardium is recognizable for all MRI time steps and not disturbed by either apex or AVP. \revc{The function $A$ converts both MRI and simulated endocardial contours $\vec{d}^s_{\text{MRI}}$ and $\vec{d}^s$, respectively, to binary images with a resolution of $1 \times 1 \text{ mm}^2$ for every slice $s$. We use the Dice metric to compare both binary images
\begin{equation}
\renewcommand{\arraystretch}{1.5}
\epsilon = 1 - \frac{1}{5} \sum_{s=5}^{9} ~ \frac{2 \, \left| A\left(\vec{d}^s_{\text{MRI}} \right) \cap A \left( \vec{d}^s \right)\right|}{\left|A\left(\vec{d}^s_{\text{MRI}}\right)\right| + \left|A\left(\vec{d}^s\right)\right|} \in [0,1]
\end{equation}
where $|\bullet|$ denotes the area of the binary image.% The error is displayed in figure~\ref{mri_err}. 
}

\paragraph{Ventricular-atrial interaction}
Utilizing a four chamber geometry allows us to investigate the interaction between ventricles and atria. Specifically, we want to study the influence of ventricular contraction on atrial filling. We therefore analyze atrial volumes over time. Furthermore, we segmented left and right atrial volumes at ventricular diastasis and end-systole from isotropic 3D MRI.

\paragraph{Pericardial contact stress}

\revs{
We evaluate the stresses transmitted between the epicardial boundary conditions and the myocardium for both cases \apex{} and \normal{}. We use different averaged stresses for both cases to quantify the constraining effect of each boundary condition. In case \apex{} the stresses are concentrated on the small apical area and acting in any direction. We thus integrate the stress vectors of the apical boundary condition over the apical surface and normalize by the apical area to obtain the mean apical stress
\begin{align}
\bar{\vec{t}}_{\text{apex}}(t) = \frac{\int_{\varGamma^{\text{apex}}} \vec{t}_{\text{apex}} \dd a}{\int_{\varGamma^{\text{apex}}} 1 \dd a}, \quad \vec{t}_{\text{apex}} = k_a \vec{u} + c_a \dot{\vec{u}}.
\end{align}
In case \normal{} the boundary stresses are distributed over the whole epicardial surface and acting only in normal direction. Therefore, we extract the (signed) normal component $t_\text{epi}$ and integrate it over the epicardial surface to obtain the mean pericardial stress
\begin{align}
\bar{t}_{\text{epi}}(t) = \frac{\int_{\varGamma^{\text{epi}}} t_\text{epi} \dd a }{\int_{\varGamma^{\text{epi}}} 1 \dd a}, \quad t_\text{epi} = k_p \vec{u} \cdot \vec{N} + c_p \dot{\vec{u}} \cdot \vec{N},
\end{align}
normalized by the epicardial area.
}

\revc{
\subsection{Selection of pericardial paramaters}\label{sec_params}
}

\revc{
Since in case \apex{} the purpose of the apical boundary condition is fixing the apex throughout cardiac contraction, we chose a high spring stiffness permitting only little motion. For case \normal{}, the parameters $k_p$ and $c_v$ describing pericardial stiffness and viscosity, respectively, need to be calibrated. The chosen value for pericardial viscosity has on its own, i.e. without parallel spring, only little influence on cardiac dynamics. However, in combination with the spring, it prevents unphysiological oscillations of the heart. Pericardial stiffness controls the amount of displacement perpendicular to the epicardial surface and thus the radial motion of the myocardium.}

\revc{
We investigate in the following the influence of the parameter $k_p$ on the contraction of the heart. For this study, we limit ourselves to the \sixty{} fiber distribution, as it is commonly used in cardiac simulations, see e.g. \cite{chabiniok12,lee15a,hirschvogel16}. We tested the following parameter values:
\begin{align}
k_p \in \{ 0.1,0.2,\dots,1.0,1.5,\dots,5.0 \} \left[\frac{\text{kPa}}{\text{mm}}\right]
\end{align}
For each choice of $k_p$, we calibrated active stress to yield the same end-systolic volume as measured from MRI. All parameters except $k_p$ are kept constant throughout this study. Specifically, we did not adjust the timing of ventricular systole to match the volume curve from MRI. However, since all simulations reach the end-systolic state it will be used in this section for quantitative comparisons.}

\begin{figure}
\centering
\setlength\figureheight{3cm}
\setlength\figurewidth{0.35\textwidth}
\includegraphics{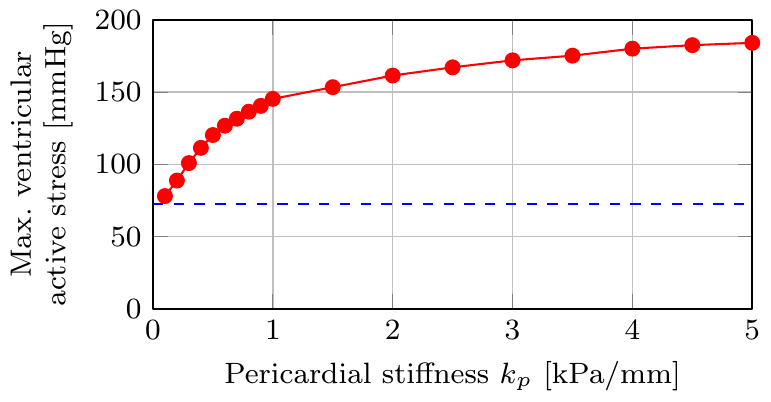}\\
\includegraphics{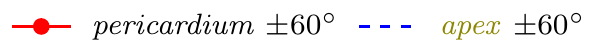}
\caption{\revc{Maximum ventricular active stress $\sigma_v$ calibrated to yield identical end-systolic volume. Shown for case \normal{} with varying pericardial stiffness compared to case \apex{}, both with \sixty{} fiber distributions.} \label{contractility_parametric}}
\end{figure}

\begin{figure}
\centering
\setlength\figureheight{3cm}
\setlength\figurewidth{0.35\textwidth}
\includegraphics{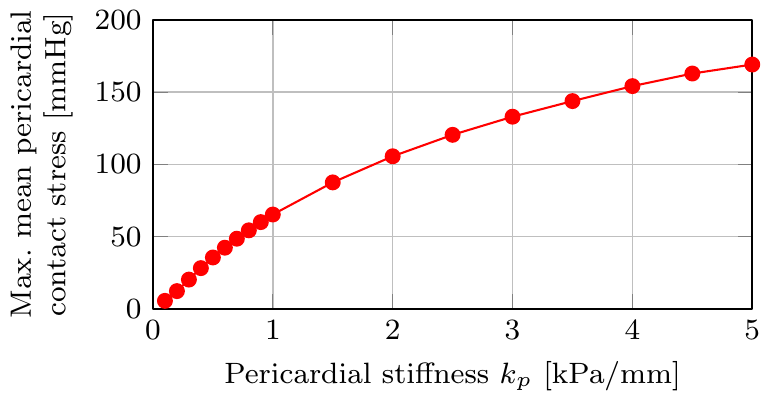}
\caption{\revc{Maximum of mean pericardial contact stress $\bar{t}_\text{epi}$ for case \normal{} with varying pericardial stiffness and \sixty{} fiber distributions.} \label{springstress_parametric}}
\end{figure}

\revc{
The results of the calibration are shown in figure~\ref{contractility_parametric}, where maximum active stress is plotted against pericardial stiffness. For comparison, the result for case \apex{} with \sixty{} fibers is included. Active stress required to yield identical end-systolic volume rises strongly with increasing pericardial stiffness. The temporal maximum of pericardial contact stress averaged over the epicardium also increases strongly with $k_p$, as shown in figure~\ref{springstress_parametric}. For high $k_p$, contact stress has the same order of magnitude as active stress and exceeds maximum left ventricular pressure. For small $k_p$, it has the same order of magnitude as atrial pressure, which experimentally shown to be a good predictor for pericardial contact stress \cite{tyberg86}.}

\begin{figure}
\centering
\setlength\figureheight{3cm}
\setlength\figurewidth{0.35\textwidth}
\includegraphics{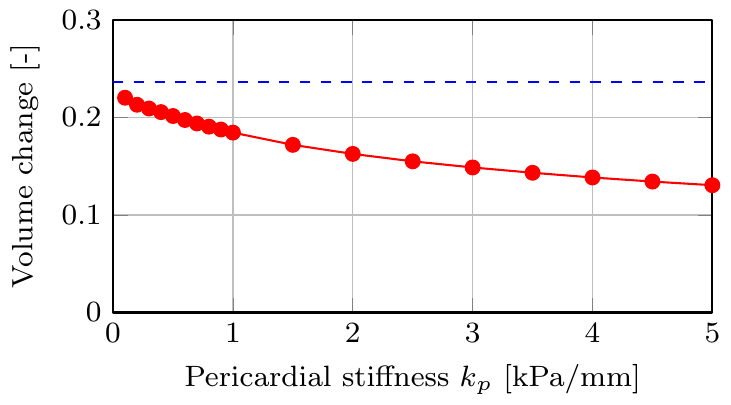}\\
\includegraphics{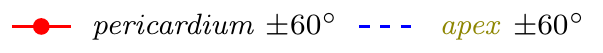}
\caption{\revc{Volume change for case \normal{} with varying pericardial stiffness compared to case \apex{}, both with \sixty{} fiber distributions.} \label{volume_change}}
\end{figure}

\revc{
Figure~\ref{volume_change} shows the volume within the pericardial cavity, calculated as the combined volume of all tissue inside the pericardium and the volume within the four cardiac cavities. Case \normal{} yields a lower volume change than case \apex{} and decreases further with increasing $k_p$.}

\begin{figure*}
\centering
\setlength\mriheight{0.29\textheight}
\subfloat[Four chamber view \label{4ch_parametric}]{
\includegraphics[height=\mriheight]{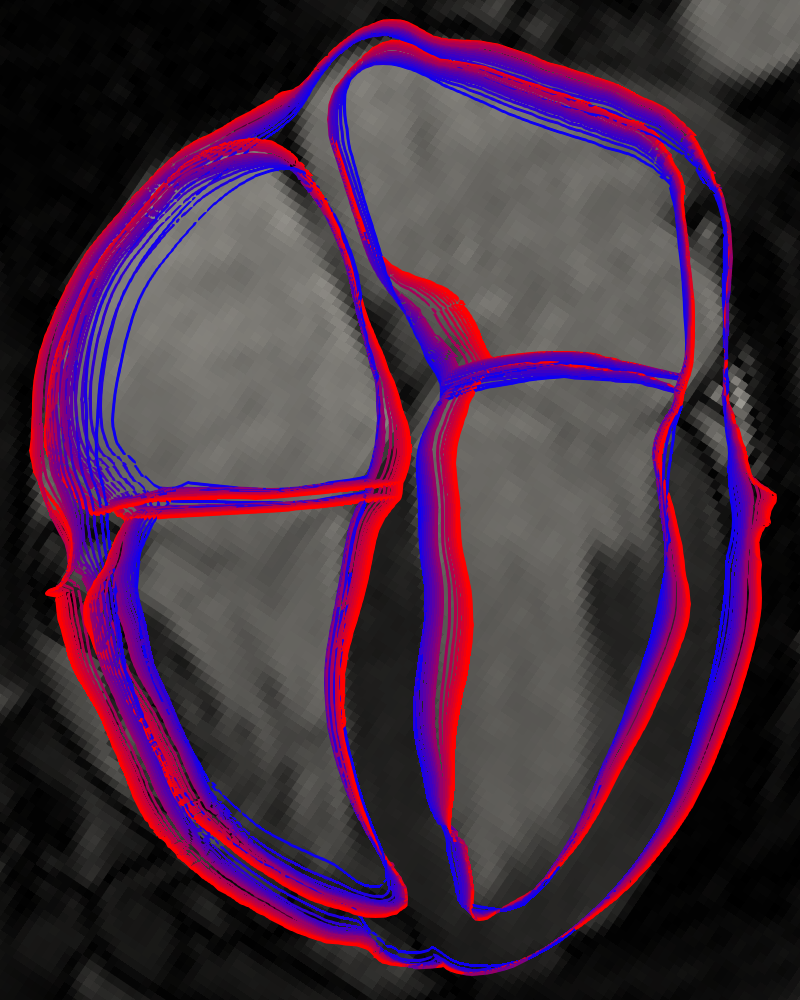}}~
\subfloat[Short axis view slice 9 \label{sax9_parametric}]{
\reflectbox{\rotatebox[origin=c]{180}{\includegraphics[height=\mriheight]{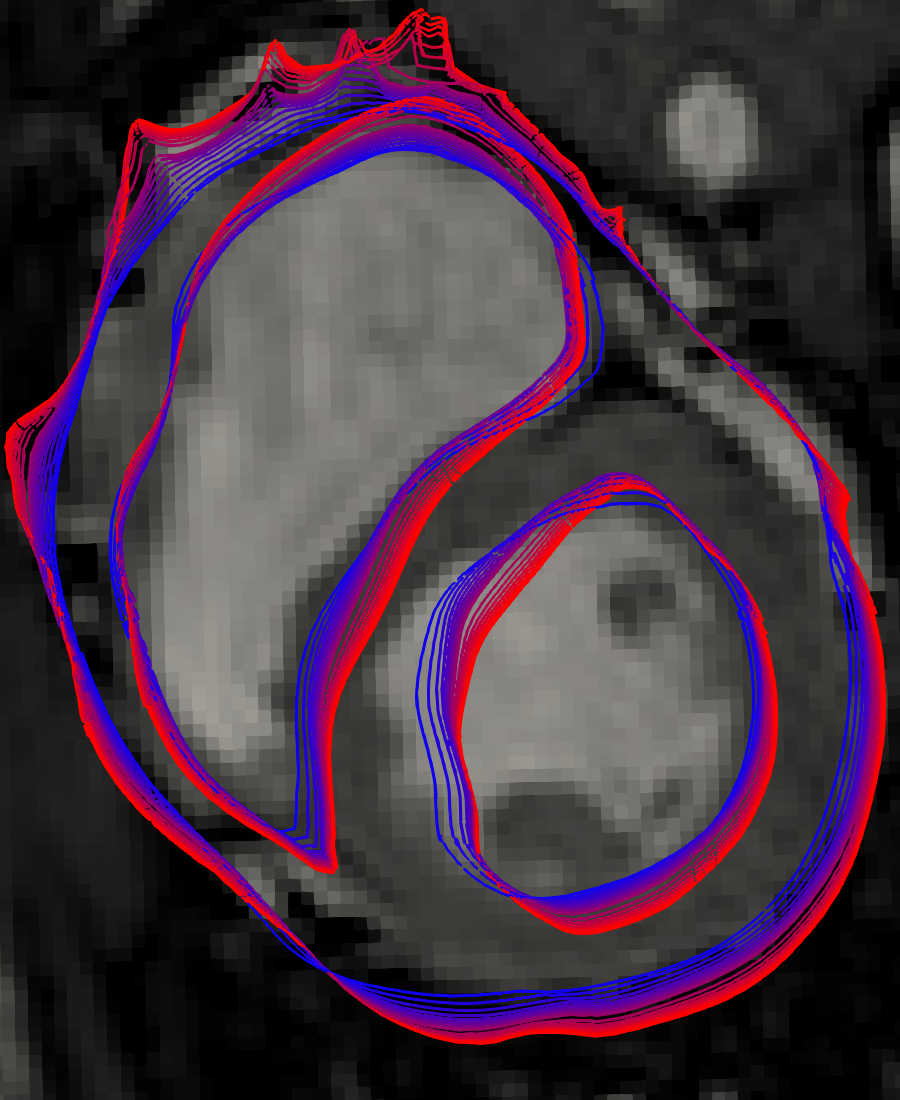}}}}~
\subfloat[Short axis view slice 6\label{sax6_parametric}]{
\reflectbox{\rotatebox[origin=c]{180}{\includegraphics[height=\mriheight]{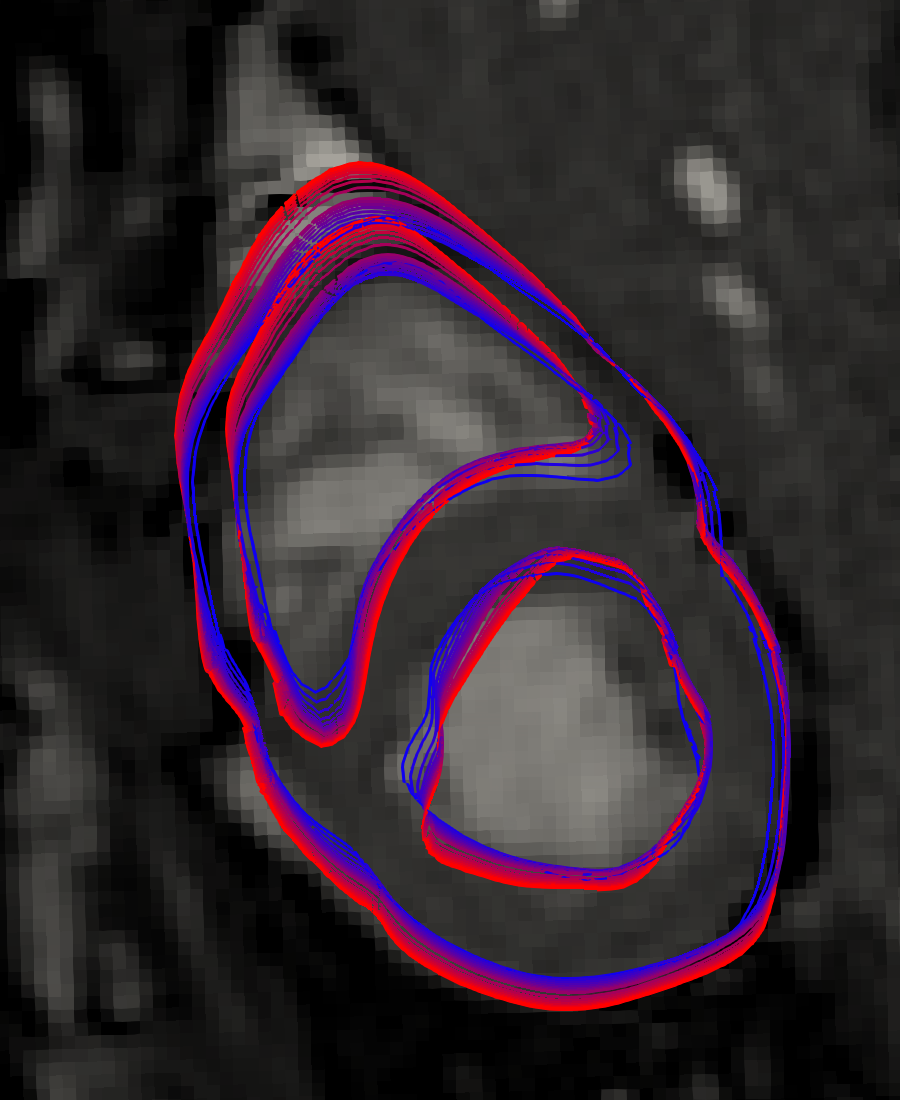}}}}
\caption{\revc{Cine MRI at end-systole for case \normal{} with \sixty{} fiber distribution from $k_p=0.1$ (blue) to $k_p=5.0$ (red). Views as defined in figure~\ref{mri_views}. MRI courtesy of R.~Chabiniok, J.~Harmer, E.~Sammut, King's College London, UK.}\label{mri_es_parametric}}
\end{figure*}

\revc{
The end-systolic state of the simulations is shown in figure~\ref{mri_es_parametric} compared to MRI. The images contain all simulated variants for $k_p$, where the color changes continuously from $k=0.1$\,kPa/mm (blue) to $k=5.0$\,kPa/mm (red). All MRI views in figure~\ref{mri_es_parametric} show clearly that pericardial stiffness controls radial displacement of the epicardium. High stiffness values result in less radial inward motion during ventricular systole than visible in cine MRI and vice versa. This is also well observable in figure~\ref{4ch_parametric} for the atria in four-chamber view. The short axis views in figures~\ref{sax9_parametric} and \ref{sax6_parametric} additionally show that the interventricular septum is stretched and rotated as compared to MRI for high $k_p$.}

\begin{figure*}
\centering
\setlength\figureheight{3cm}
\setlength\figurewidth{0.35\textwidth}
\subfloat[Left ventricular endocardial error. \label{err_lv_parametric}]{
\includegraphics{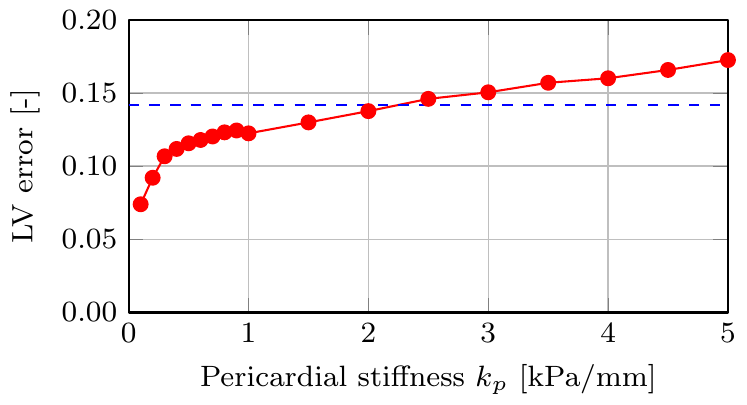}}~
\subfloat[Right ventricular endocardial error. \label{err_rv_parametric}]{
\includegraphics{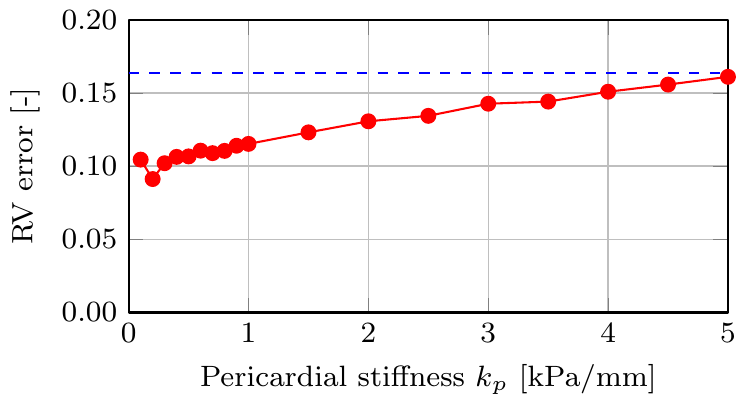}}\\
\subfloat[Left atrioventricular plane displacement. \label{lvpd_parametric}]{
\includegraphics{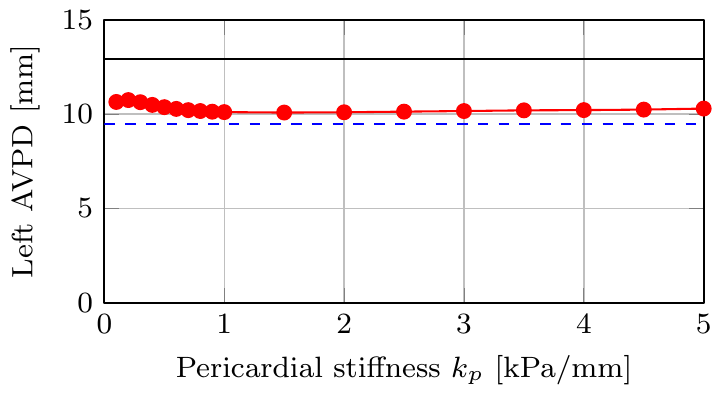}}~
\subfloat[Right atrioventricular plane displacement. \label{rvpd_parametric}]{
\includegraphics{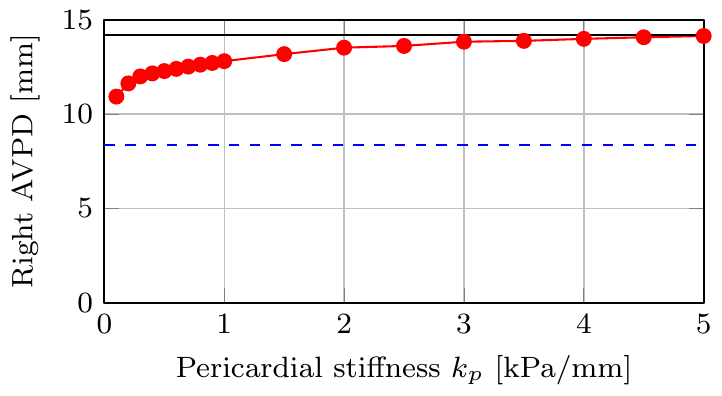}}\\
\subfloat[Left atrial volume. \label{volume_la_parametric}]{
\includegraphics{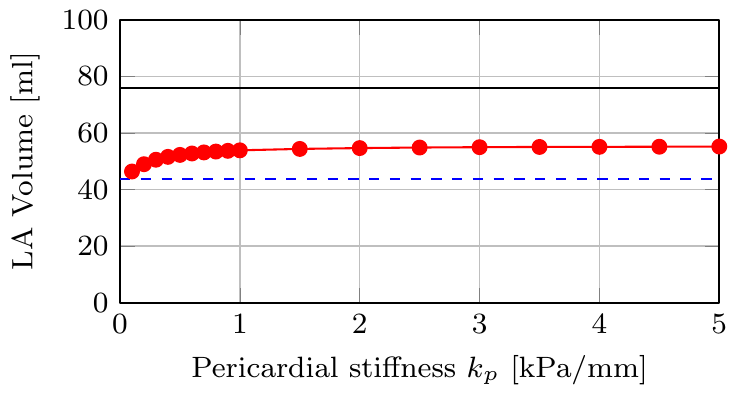}}~
\subfloat[Right atrial volume. \label{volume_ra_parametric}]{
\includegraphics{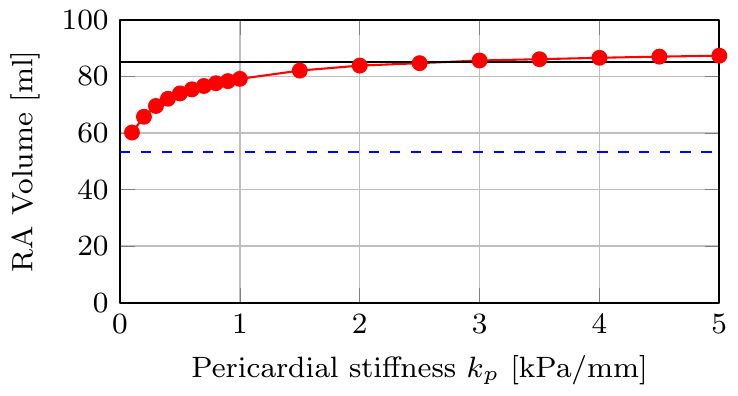}}\\[0.3cm]
\includegraphics{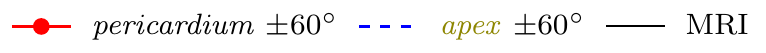}
\caption{\revc{Kinematic scalar cardiac quantities at end-systole for case \normal{} with varying pericardial stiffness $k_p\in[0.1,5.0]$ compared to case \apex{} both with \sixty{} fiber distributions and MRI.} \label{parametric_kinematic}}
\end{figure*}

\revc{
The spatial error at left and right ventricular endocardium is shown in figures~\ref{err_lv_parametric} and \ref{err_rv_parametric}, respectively. The increasing mismatch between simulations and MRI for increasing $k_p$ as visible in figures~\ref{sax9_parametric} and \ref{sax6_parametric} is quantified as  increasing spatial error.}

\revc{
Left and right AVPD is displayed in figures~\ref{lvpd_parametric} and \ref{rvpd_parametric}, respectively. In the left ventricle, i.e. at the mitral valve, AVPD is not very sensitive to the choice of $k_p$. However, it is higher than in case \apex{} but much lower than in MRI. For the right ventricle, i.e. at the tricuspid valve, AVPD is greatly enhanced by increasing $k_p$ towards the value measured in MRI. An identical trend is observable for left and right atrial volume in figures~\ref{volume_la_parametric} and \ref{volume_ra_parametric}, respectively.}

\revc{
To conclude the parametric study for pericardial stiffness, we choose for all following simulations the value $k_p=0.2$\,kPa/mm. It offers a low spatial error at the ventricles but has higher atrial volume and AVPD than the simulation with $k=0.1$\,kPa/mm.}\\

\subsection{Model personalization}

\revc{
All simulations are carried out in the following using three different fiber distributions, i.e. \fifty{}, \sixty{}, and \seventy{}. The results for the calibration of $\sigma_0$ are shown in table~\ref{sigma}. Note that we show here the maximum value $\sigma_\text{v}$ of active stress instead of $\sigma_0$. 
}
It can be observed that $\sigma_\text{v}$ is larger in case \normal{} than in case \apex{}. Furthermore, $\sigma_\text{v}$ increases from \fifty{} to \seventy{} fibers for more vertical fiber distributions. 

\revc{The onset of systole and diastole, $t_\text{sys}$ and $t_\text{dias}$, as well as the myofiber activation and deactivation rates, $\alpha_\text{max}$ and $\alpha_\text{max}$, are adapted to the left ventricular volume curve for ventricles and atria. Here, parameters for the atria are fitted from the interval $t\in[0,0.2\,\text{s}]$ and $t\in[0.2\,\text{s},0.9\,\text{s}]$ for the ventricles. The material parameter $\eta$ controlling the viscosity of the tissue is fitted during ventricular diastole, i.e. $t\in[0.5\,\text{s},0.9\,\text{s}]$. Since active stress is zero during this interval, viscosity controls the relaxation speed of the model. A summary of all calibrated model-specific material parameters is given in table~\ref{tab_sigma0}. For parameters identical in all models see table~\ref{tab_parameters}.}

% what tissue stiffness/length is equivalent to spring stiffness?
%
\begin{table}
\centering
\scriptsize
\renewcommand{\arraystretch}{1.5}
\subfloat[Spring stiffness and dashpot viscosity on apical and epicardial surface. \label{tab_bcs}]{
\revs{
\begin{tabular}{ | l | c | c | c | c |}
\hline
& \multicolumn{2}{c|}{$\varGamma_0^{\text{apex}}$} & \multicolumn{2}{c|}{$\varGamma_0^{\text{epi}}$} \\
& $k_a \left[\frac{\text{kPa}}{\text{mm}}\right]$ & $c_a \left[\frac{\text{kPa}\cdot\text{s}}{\text{mm}}\right]$ & $k_p \left[\frac{\text{kPa}}{\text{mm}}\right]$ & $c_p \left[\frac{\text{kPa}\cdot\text{s}}{\text{mm}}\right]$  \\
\hline
\hline
\apex{} & $1.0 \cdot 10^3$& $1.0 \cdot 10^{-2}$ & 0 & 0 \\
\hline
\normal{} & 0 & 0 & 0.2 & $5.0 \cdot 10^{-3}$ \\
\hline
\end{tabular}}
}\\
\subfloat[Maximal myocardial active stress $\sigma_\text{v}$ and ventricular activation time $t_\text{sys}$. \label{sigma}]{
\revb{
\begin{tabular}{| l | c | c | c | c | c | c |}
\hline
& \multicolumn{3}{c|}{$\sigma_\text{v}$ [kPa]} & \multicolumn{3}{c|}{$t_\text{sys}$ [ms]}\\
& \fifty{} & \sixty{} & \seventy{} & \fifty{} & \sixty{} & \seventy{}\\
\hline
\hline
\apex{} & 63.5 & 72.4 & 91.0 & 143 & 155 & 172 \\
\hline
\normal{} & 79.4 & 90.7 & 129 & 161 & 170 & 193 \\
\hline
\end{tabular}}
}
\caption{Calibrated parameters for simulation cases \apex{} and \normal{} and different fiber orientations.\label{tab_sigma0}}
\end{table}

\subsection{Scalar windkessel results}
\label{sec_res_scalar}
\begin{figure*}
\centering
\setlength\figureheight{3cm}
\setlength\figurewidth{0.35\textwidth}
\subfloat[Case \apex{}: LV volume \label{scalar_v_free}]{
\includegraphics{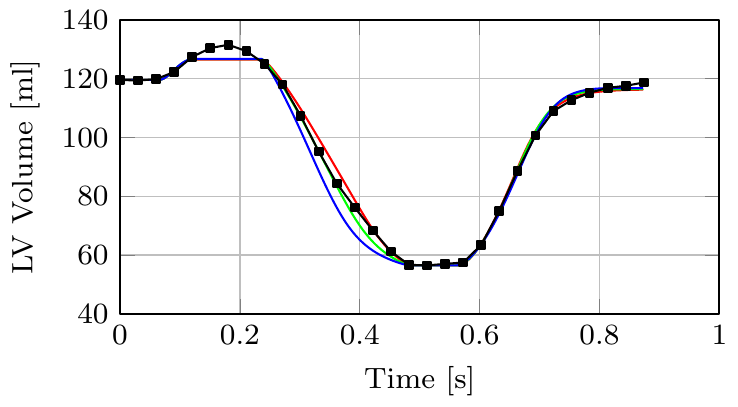}}~
\subfloat[Case \apex{}: LV pressure \label{scalar_p_free}]{
\includegraphics{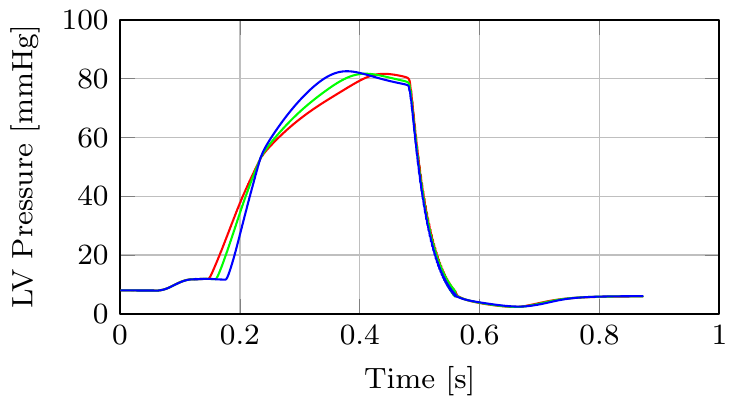}}\\
\subfloat[Case \normal{}: LV volume \label{scalar_v_normal}]{
\includegraphics{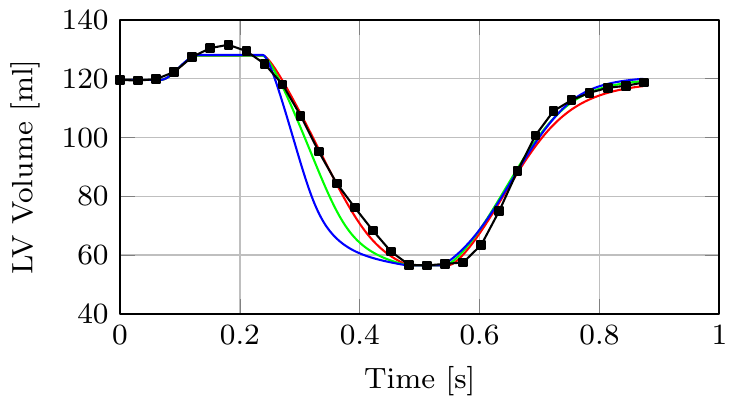}}~
\subfloat[Case \normal{}: LV pressure \label{scalar_p_normal}]{
\includegraphics{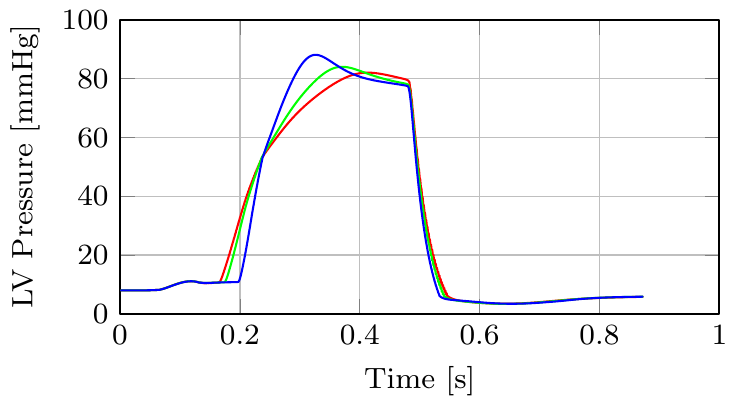}}\\[0.3cm]
\includegraphics{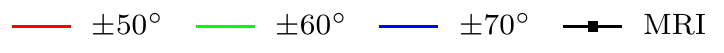}
\caption{Simulation results for volume (left) and pressure (right) of the left ventricle (LV) for boundary condition cases \apex{} (top) and \normal{} (bottom). Volume results are compared to cine MRI. \label{scalar}}
\end{figure*}

Firstly, in figure~\ref{scalar} we compare the scalar outputs volume (left) and pressure (right) of the left ventricle of our windkessel model. As explained in section~\ref{sec_params}, the contractility $\sigma_0$ was calibrated in all simulations to match end-systolic volume as segmented from cine MRI. Therefore, in figures~\ref{scalar_v_free} and \ref{scalar_v_normal} the volumes of MRI and all simulations match at $t=0.51$~s. Furthermore, although they result from simulations with very different boundary conditions and fiber orientations the volume curves are very similar. The maximum volume due to \reva{atrial contraction and} the prescribed atrial pressure in figure~\ref{p_at} is \revs{similar in both cases but lower than in MRI}. As for the volume curves, the pressure curves in figures~\ref{scalar_p_free} and \ref{scalar_p_normal} are remarkably similar despite the different simulation settings. Because case \normal{} exhibits a faster decay in volume during systole than in case \apex{}, the pressure peak during systole is more pronounced.

\subsection{Displacements at end-systole}
\label{sec_ref_es}
\begin{figure*}
\centering
\setlength\mriheight{0.26\textheight}
\subfloat[\revc{Reference configuration (diastasis)} four chamber view \label{4ch_ref}]{
\includegraphics[height=\mriheight]{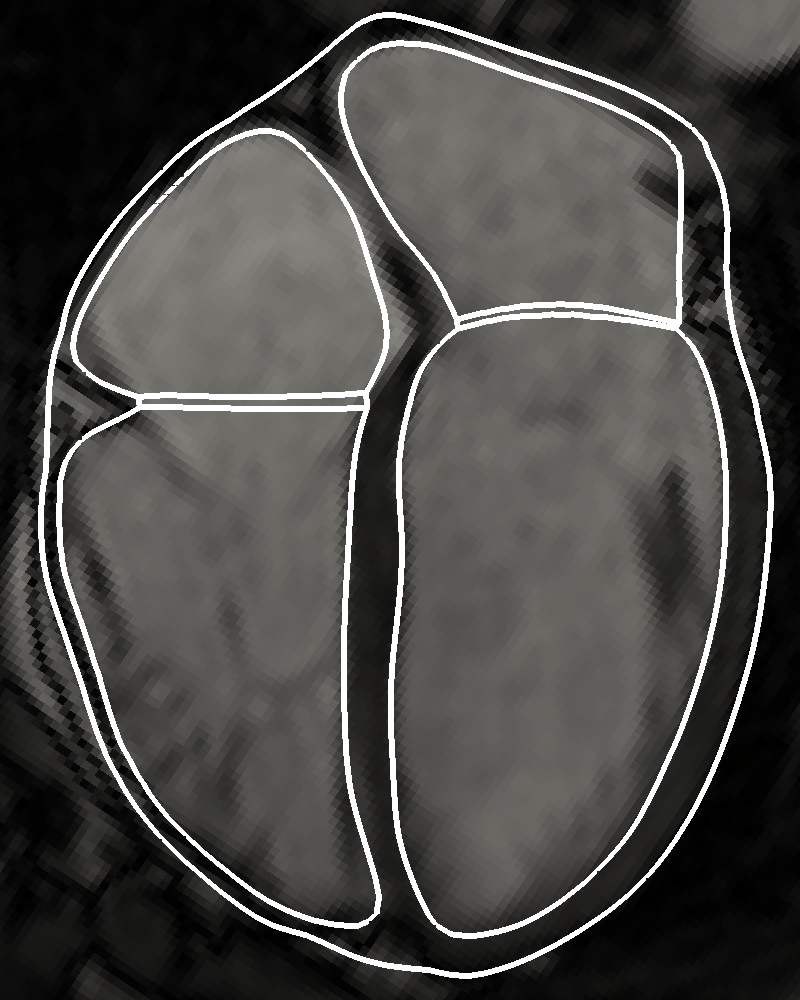}}~
\subfloat[\revc{Reference configuration (diastasis)} short axis view slice 9 \label{sax9_ref}]{
\reflectbox{\rotatebox[origin=c]{180}{\includegraphics[height=\mriheight]{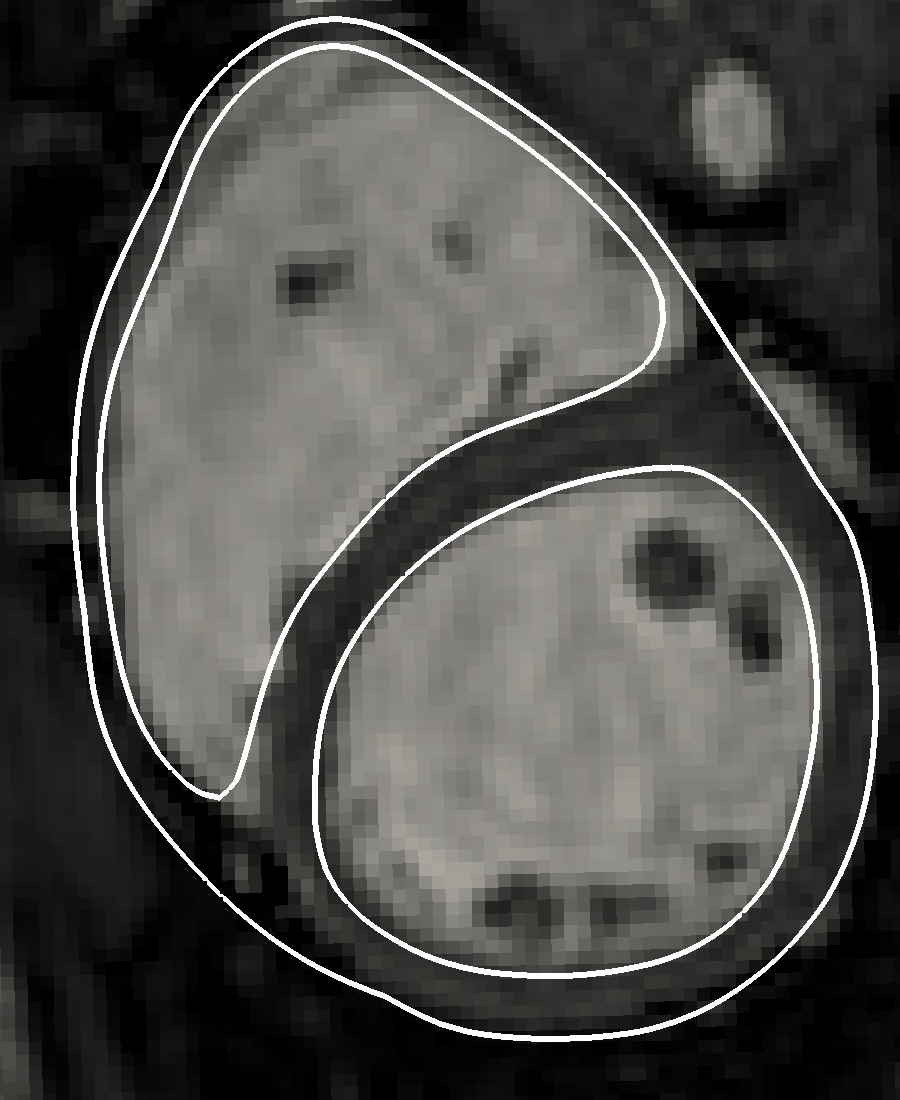}}}}~
\subfloat[\revc{Reference configuration (diastasis)} short axis view slice 6 \label{sax6_ref}]{
\reflectbox{\rotatebox[origin=c]{180}{\includegraphics[height=\mriheight]{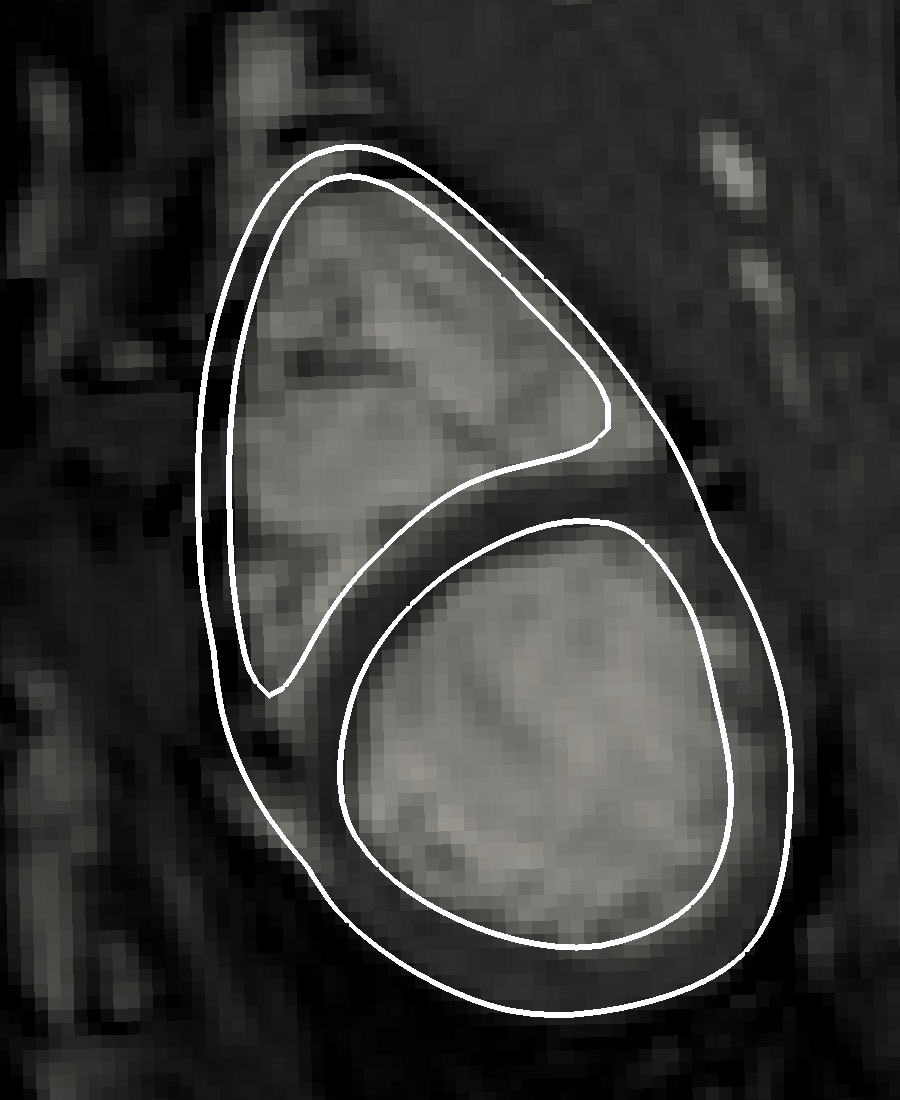}}}}\\
\subfloat[Case \apex{} four chamber view \label{4ch_free}]{
\includegraphics[height=\mriheight]{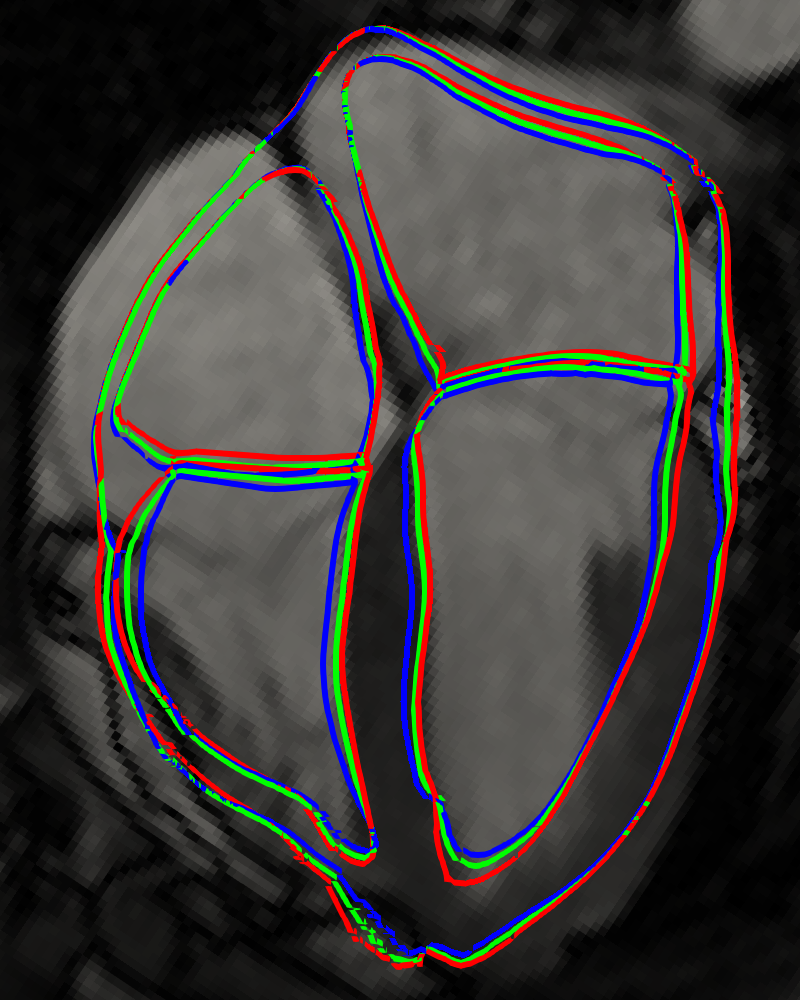}}~
\subfloat[Case \apex{} short axis view slice 9 \label{sax9_free}]{
\reflectbox{\rotatebox[origin=c]{180}{\includegraphics[height=\mriheight]{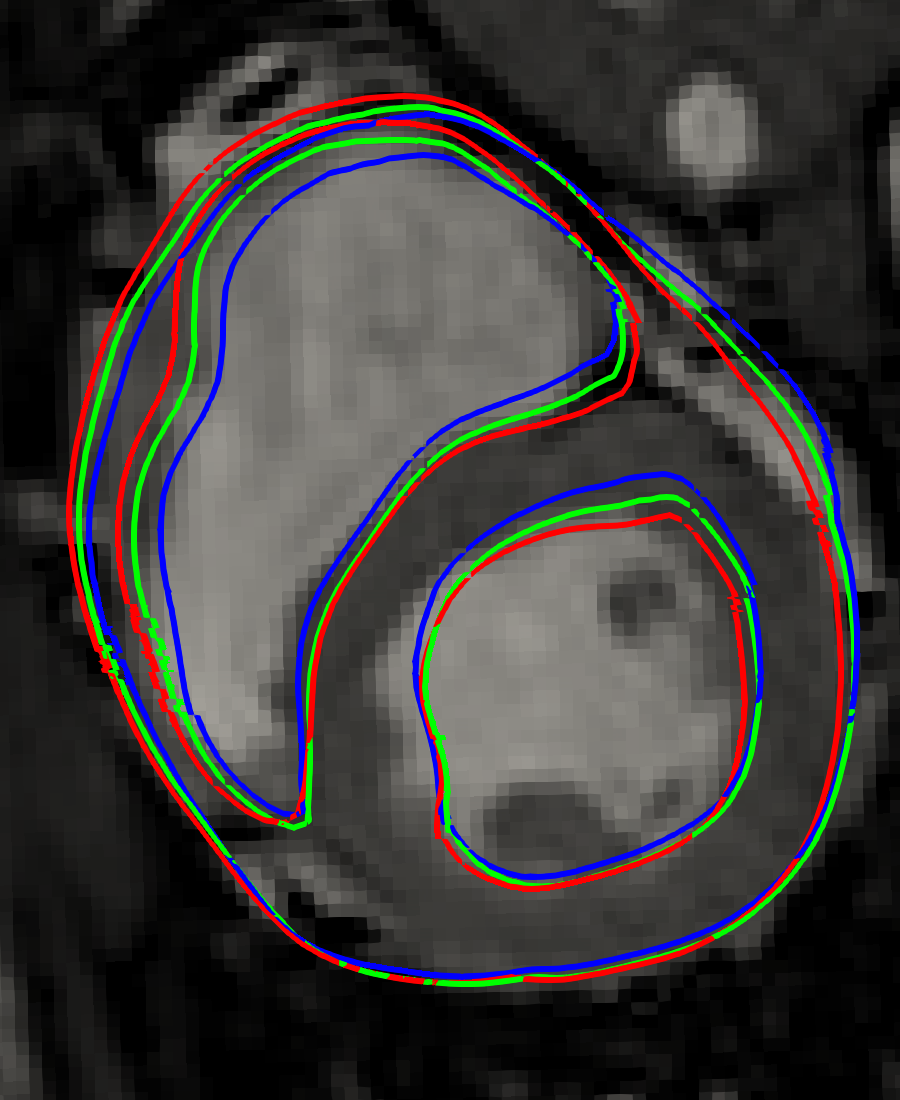}}}}~
\subfloat[Case \apex{} short axis view slice 6 \label{sax6_free}]{
\reflectbox{\rotatebox[origin=c]{180}{\includegraphics[height=\mriheight]{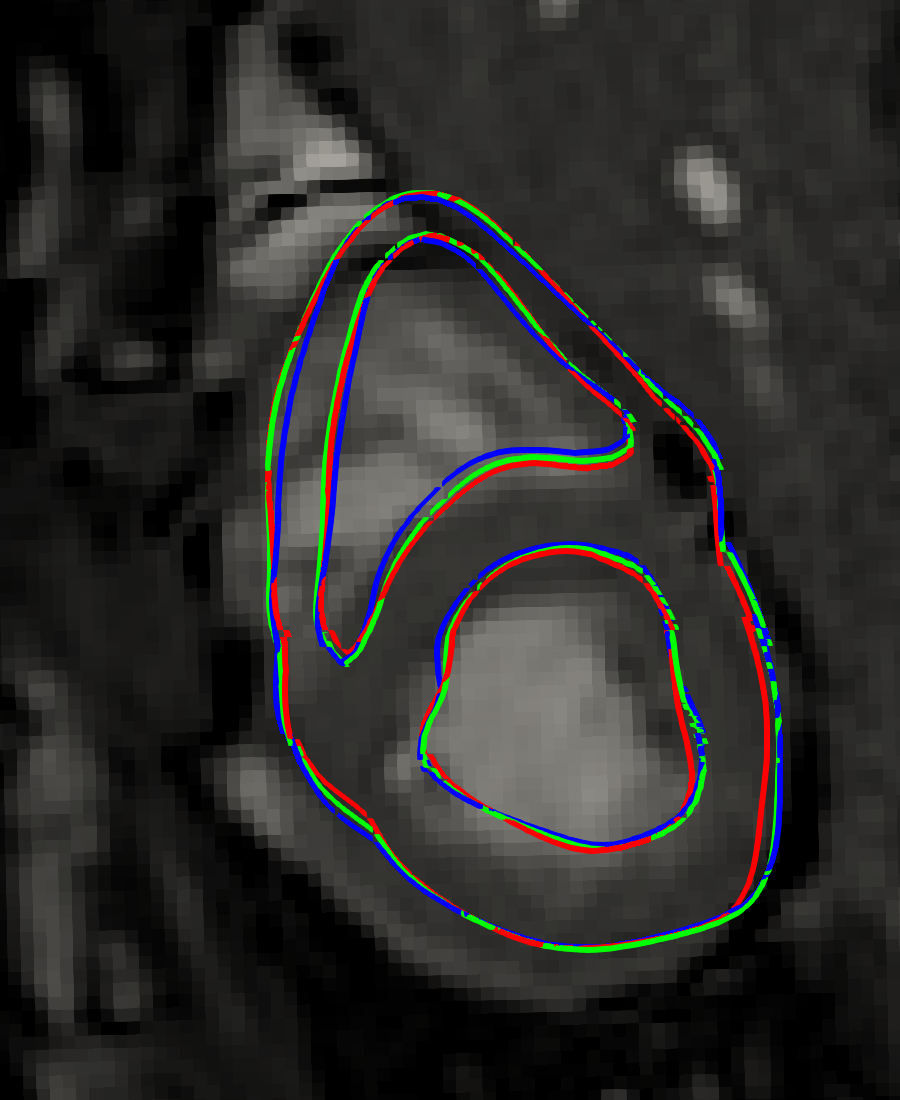}}}}\\
\subfloat[Case \normal{} four chamber view \label{4ch_normal}]{
\includegraphics[height=\mriheight]{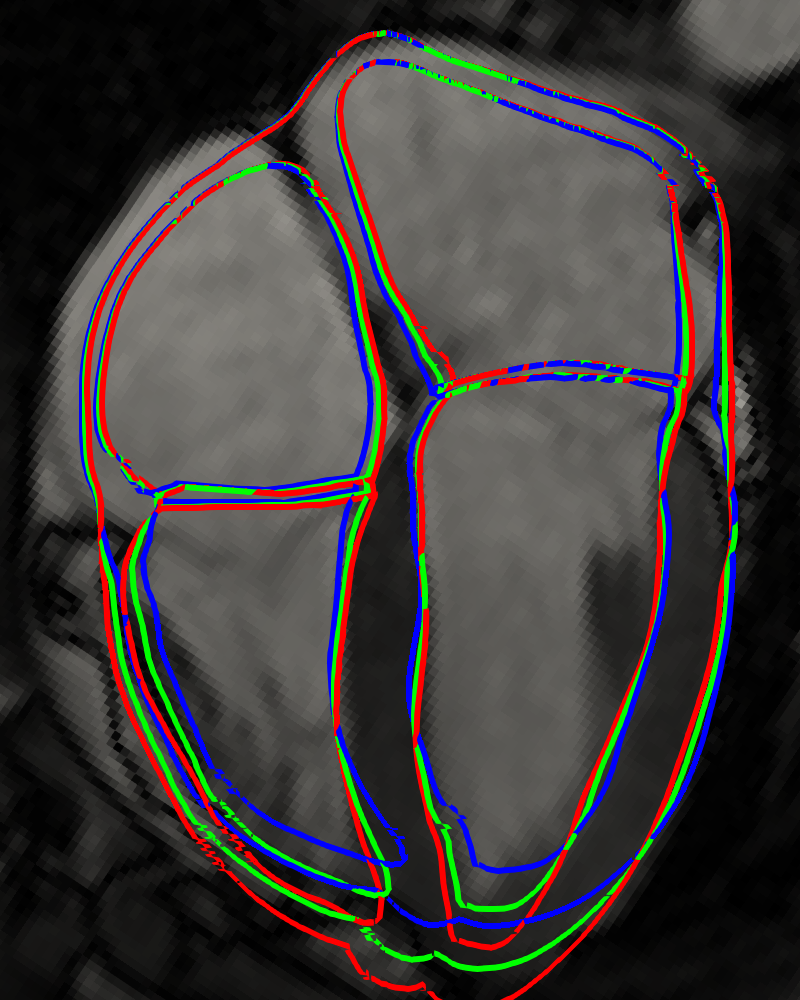}}~
\subfloat[Case \normal{} short axis view slice 9 \label{sax9_normal}]{
\reflectbox{\rotatebox[origin=c]{180}{\includegraphics[height=\mriheight]{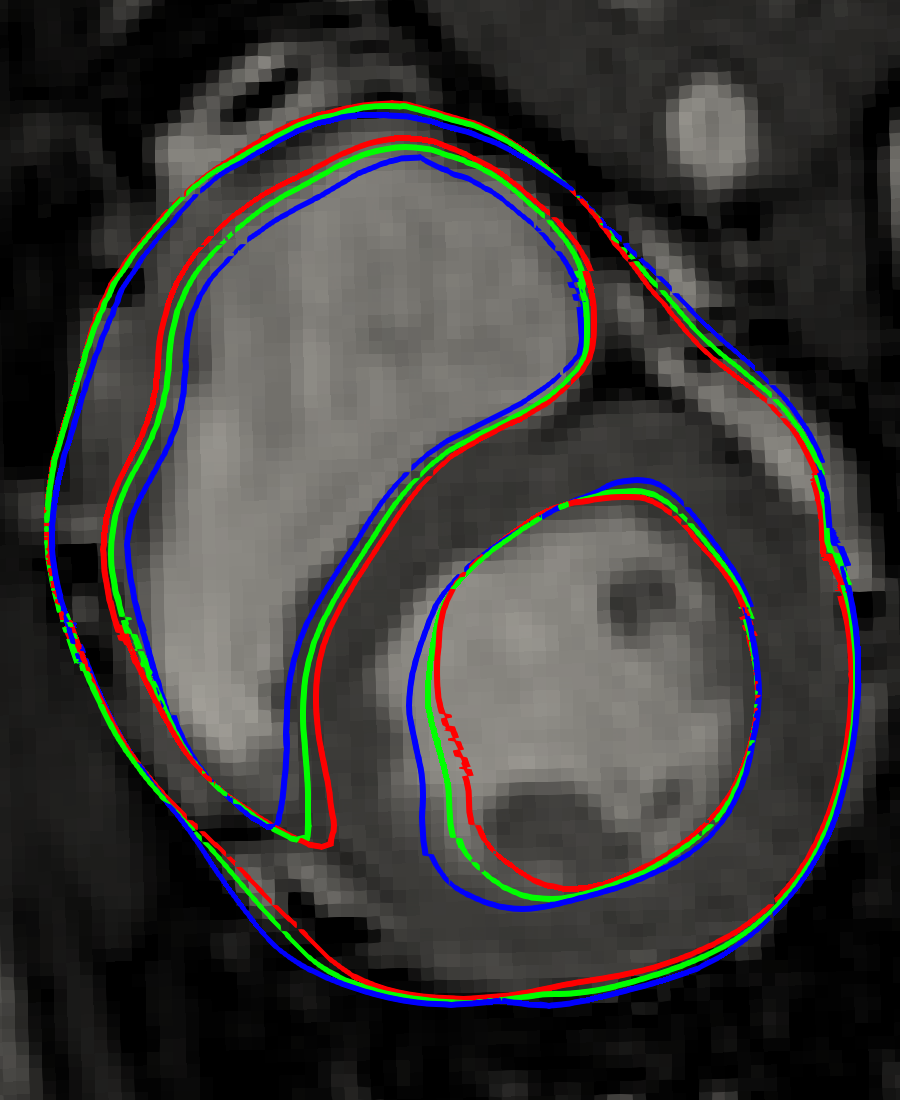}}}}~
\subfloat[Case \normal{} short axis view slice 6\label{sax6_normal}]{
\reflectbox{\rotatebox[origin=c]{180}{\includegraphics[height=\mriheight]{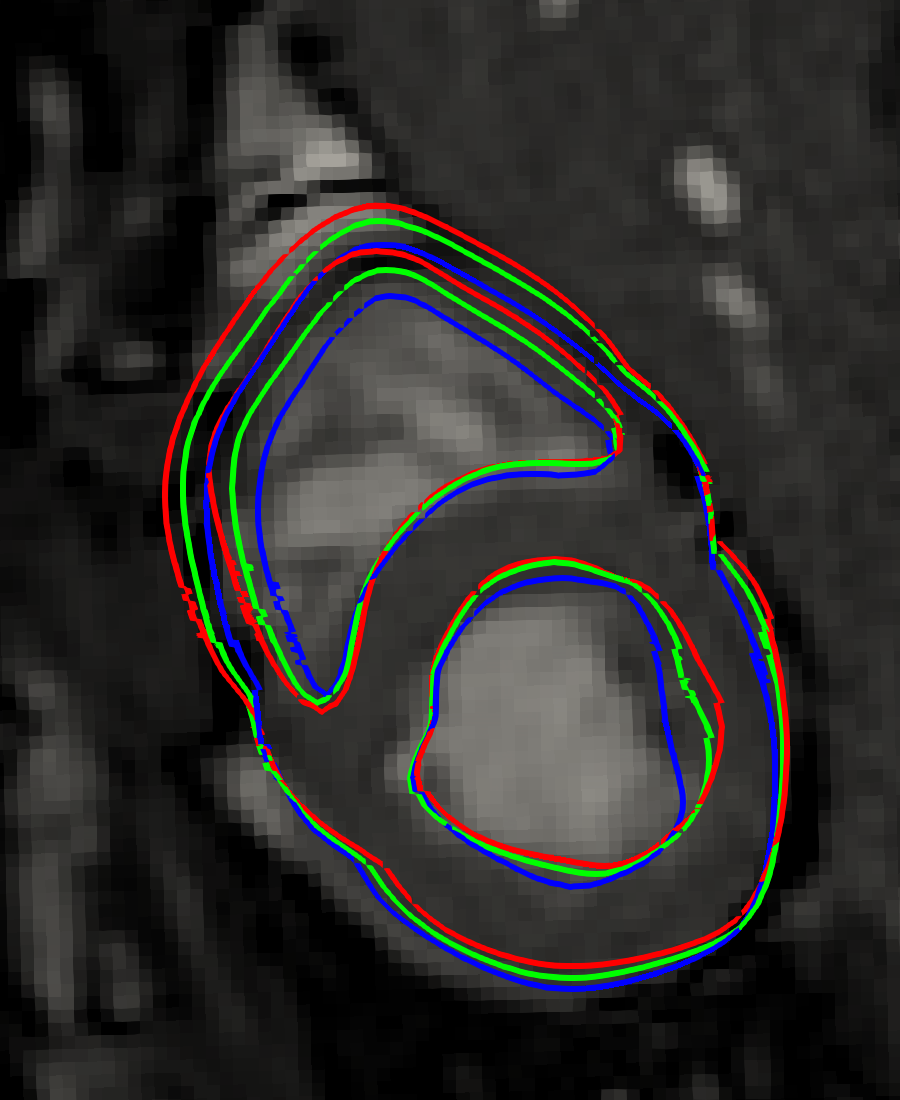}}}}\\[0.1cm]
\includegraphics{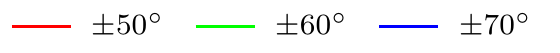}
\caption{\revc{Reference configuration (diastasis) as well as} simulation results and cine MRI at end-systole in four chamber view and short axis views as defined in figure~\ref{mri_views}. MRI courtesy of R.~Chabiniok, J.~Harmer, E.~Sammut, King's College London, UK\label{mri_es}}
\end{figure*}

% intro
As demonstrated in section~\ref{sec_res_scalar}, the results of the scalar output parameters left ventricular volume and pressure are mostly invariant to changes in boundary conditions or fiber orientation. Validating the elastodynamical model of cardiac contraction thus requires a comparison of displacement results to spatially distributed MRI observations, see  figure~\ref{mri_es}. The reference configuration \revc{(diastasis)} of the simulation is shown in figures~\ref{4ch_ref}, \ref{sax9_ref}, \ref{sax6_ref}. We compare the MRI frames at \revc{end-systole} to our simulation results using the four chamber view, see figures~\ref{4ch_free} and \ref{4ch_normal}, and two different short axis views, see figures~\ref{sax9_free}, \ref{sax9_normal}, \ref{sax6_free}, \ref{sax6_normal}. The location of the view planes is visualized in figures~\ref{4ch}, \ref{sax9}, \ref{sax6}.

% free
%
\revs{
For case \apex{}, there is a radial inward movement of the myocardial wall. In figure~\ref{4ch_free}, this is especially visible at right atrial free wall  and at the left and right epicardial free wall. There is a large mismatch between simulation and MRI at the interatrial septum. Due to the radial contraction motion, atrioventricular plane displacement (AVPD) is lower than in MRI. The fixation of the apex in case \apex{} causes a mismatch between simulations and MRI at the apex, as the apex slightly moves during cardiac contraction. The interventricular septum's matches well with MRI in figures~\ref{sax9_free} and \ref{sax6_free}. However, the mismatch of epicardial contours is clearly visible and sensitive to fiber orientation.
}

% pericardium
Comparing figures~\ref{4ch_free} and \ref{4ch_normal}, the influence of the pericardial boundary condition becomes clearly visible. It can be observed for case \normal{} in figure~\ref{4ch_normal} that the epicardial contour matches  the MRI much closer than case \apex{} in figure~\ref{4ch_free} for any fiber orientation. The movements of the left and right atrioventricular plane also match well with MRI, for both orientation and displacement in normal direction. The displacements at the apical region are also predicted more accurately than in case \apex{}. Comparing the shape of the right ventricle in figures~\ref{4ch_free} and \ref{4ch_normal}, one can observe that the pumping motion of the right ventricle in case \apex{} is the result of radial movement, whereas in case \normal{} it is the result of a downward movement of the atrioventricular plane. The same observation holds for a less visible degree for the left ventricle. Through the constraining effect of the pericardium, the atria are visibly more stretched than in case \apex{}. There is also an influence of the fiber orientation in case \normal{}, although it is more bound to the endocardial surfaces. The more vertical the fiber orientation, i.e. from \fifty{} (red) to \seventy{} (blue), the larger the displacements of the atrioventricular planes and the smaller the displacement of the apex in anterior direction. There are some mismatches between simulation and MRI at the interatrial and interventricular septum but less pronounced than in case \apex{}. The deviation at the interventricular septum can be observed for short axis slice 9 in figure~\ref{sax9_normal}. For short axis slice 6 in figure~\ref{sax6_normal} there is a good agreement with simulation and MRI at all regions of the left and right myocardium.

\subsection{Atrioventricular plane displacement}
\label{sec_avpd}
\begin{figure*}
\centering
\setlength\figureheight{3cm}
\setlength\figurewidth{0.35\textwidth}
\subfloat[Case \apex{}: left AVPD \label{avpd_free_left}]
{\includegraphics{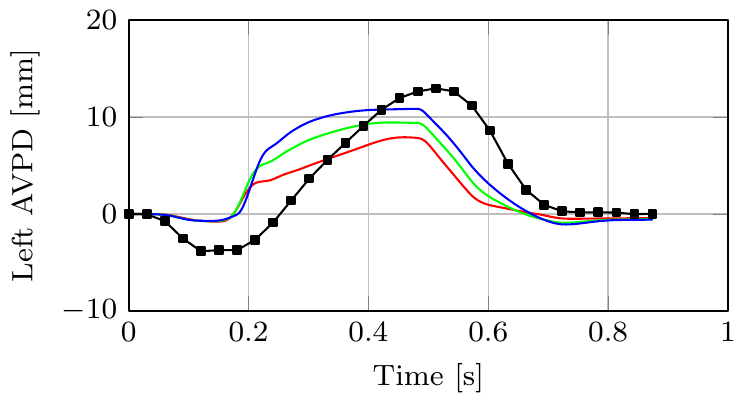}}~
\subfloat[Case \apex{}: right AVPD \label{avpd_free_right}]
{\includegraphics{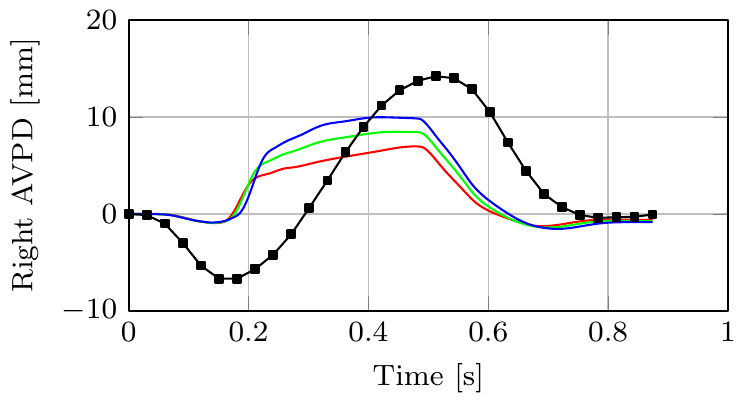}}\\
\subfloat[Case \normal{}: left AVPD \label{avpd_normal_left}]
{\includegraphics{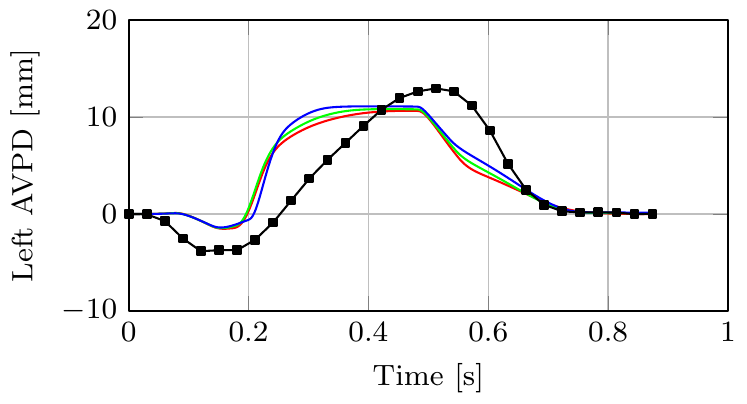}}~
\subfloat[Case \normal{}: right AVPD \label{avpd_normal_right}]
{\includegraphics{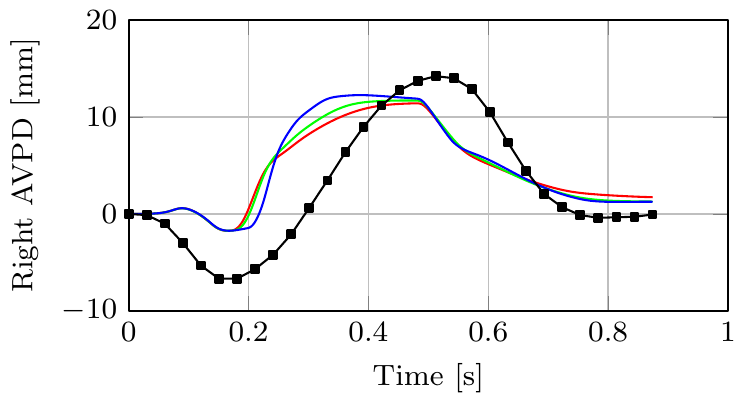}}\\[0.3cm]
\includegraphics{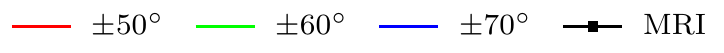}
\caption{Simulated atrioventricular plane displacement for left and right ventricle compared to cine MRI. \label{mri_avpd}}
\end{figure*}

% MRI
The AVPDs of simulations and MRI are compared in figure~\ref{mri_avpd}. The left and right AVPD from MRI (black) is zero at the beginning and at the end of the cardiac cycle. During atrial systole, the left and right atrio\-vencular planes (AVP) move away from the apex and reach both their minimal value at atrial end-systole at $t=0.17s$. Followed by ventricular systole, the AVPs move towards the apex and both reach their maximal value at ventricular end-systole at $t=0.51s$.

% comparison atrial systole
\reva{
During atrial systole for $t\in[0,0.25s]$, negative AVPD, i.e. movement of the AVP towards the atria, is less pronounced and delayed in both cases as compared to MRI. However, extremal AVPD at atrial end-systole is slightly higher in case \normal{} than in case \apex{}.
}

% comparison ventricular systole
\revs{
Comparing AVPD cases \apex{} and \normal{} in figures~\ref{avpd_free_left}, \ref{avpd_normal_left}, \ref{avpd_free_right}, \ref{avpd_normal_right}, one can observe that in both cases maximum AVPD depends on fiber orientation: Maximum AVPD increases from horizontal \fifty{} fibers (red) to vertical \seventy{} fibers (blue). The dependence on fiber orientation is more pronounced in case \apex{} than in case \normal{}. In general, AVPD is slightly higher in case \normal{} than in case \apex{} but still underestimates measurements from MRI.
}

\subsection{Ventricular-atrial interaction}
\label{sec_inter}
\begin{figure*}
\centering
\setlength\figureheight{3.1cm}
\setlength\figurewidth{0.35\textwidth}
\subfloat[Case \apex{}: Left atrium \label{volume_la_free}]{
\includegraphics{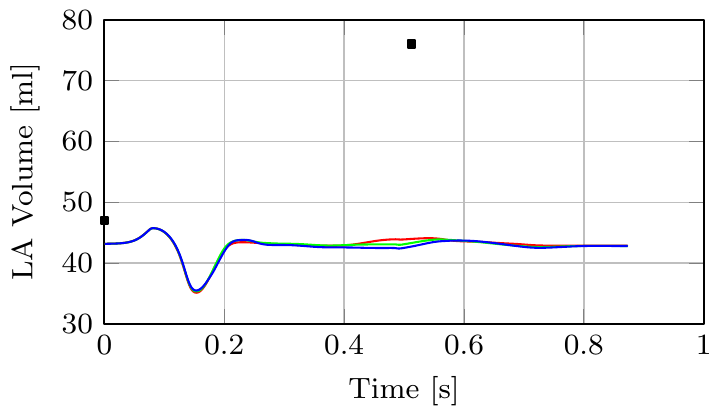}}~
\subfloat[Case \apex{}: Right atrium \label{volume_ra_free}]{
\includegraphics{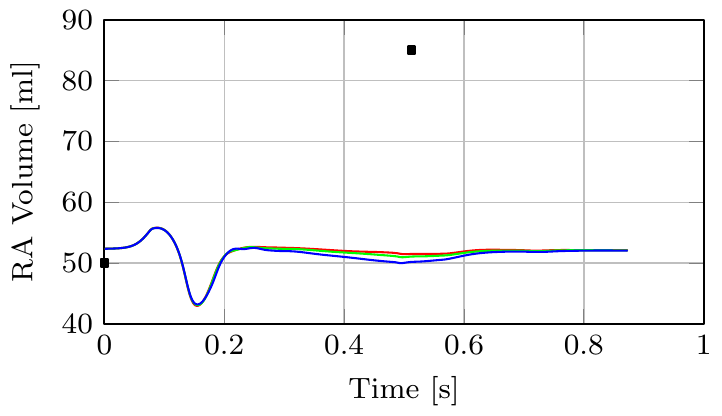}}\\
\subfloat[Case \normal{}: Left atrium \label{volume_la_normal}]{
\includegraphics{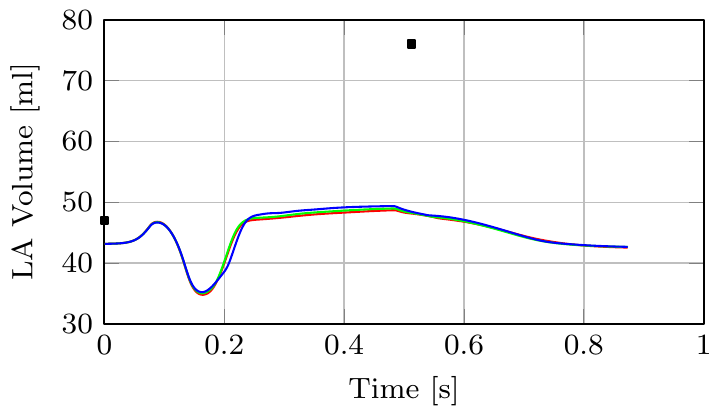}}~
\subfloat[Case \normal{}: Right atrium \label{volume_ra_normal}]{
\includegraphics{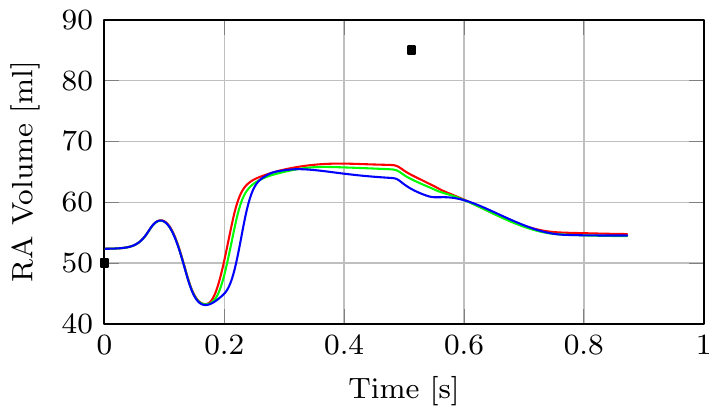}}\\[0.3cm]
\includegraphics{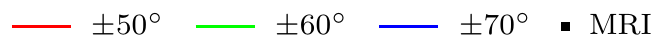}
\caption{Simulated volume curves for left and right atrium compared to 3D MRI at ventricular diastasis and end-diastole. \label{volume_atria}}
\end{figure*}

\reva{Atrial systole is visible by the drop in atrial volume in both cases.}
\revs{Passive atrial filling is non-existent in case \apex{}, as the volume in figures~\ref{volume_la_free} and \ref{volume_ra_free} stay constant during ventricular systole. This is also visible at the end-systolic four-chamber view in figure~\ref{4ch_free}. For \seventy{} fibers, the right atrium is even slightly emptied during ventricular systole, as observed in figure~\ref{volume_ra_free}. Atrial filling can be observed for case \normal{} in figures~\ref{volume_la_normal} and \ref{volume_ra_normal}. Both atria are visibly filled during ventricular systole, although maximum atrial volume remains smaller than in MRI.}

\subsection{Spatial error}
\label{sec_prr}
\begin{figure*}
\centering
\setlength\figureheight{3cm}
\setlength\figurewidth{0.35\textwidth}
\subfloat[Case \apex{}: left endocardium spatial error \label{err_LV_free}]{
\includegraphics{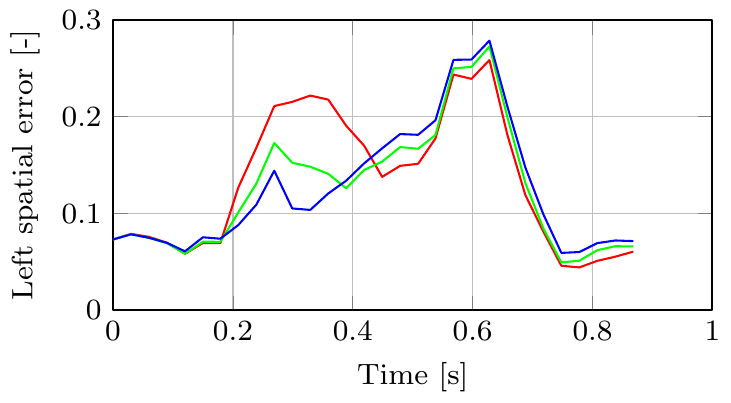}}~
\subfloat[Case \apex{}: right endocardium spatial error \label{err_RV_free}]{
\includegraphics{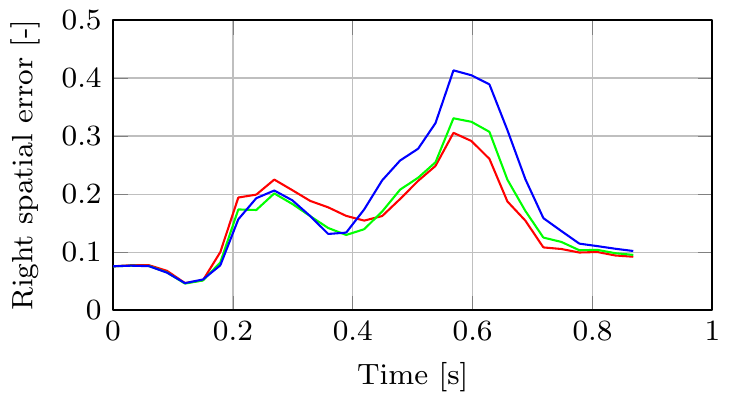}}\\
\subfloat[Case \normal{}: left endocardium spatial error \label{err_LV_normal}]{
\includegraphics{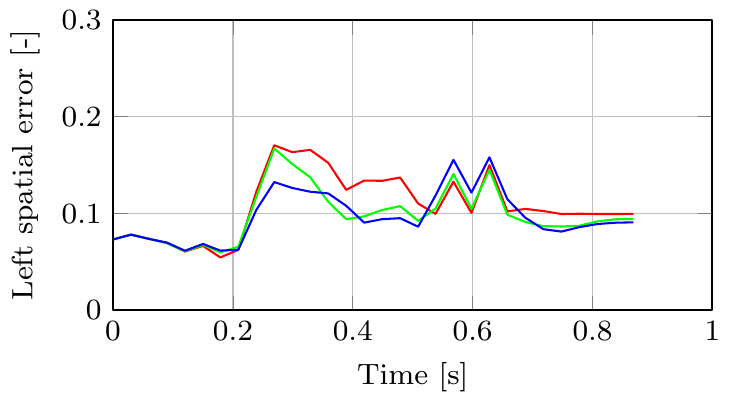}}~
\subfloat[Case \normal{}: right endocardium spatial error \label{err_RV_normal}]{
\includegraphics{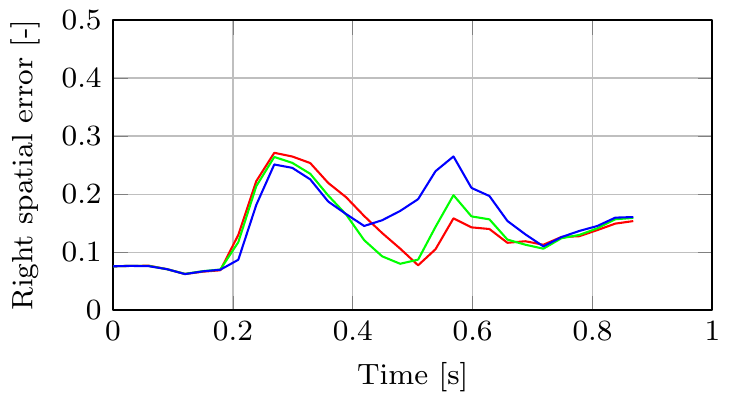}}\\[0.3cm]
\includegraphics{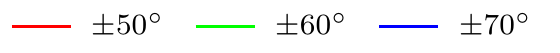}
\caption{Relative spatial error of simulation results and cine MRI at left and right endocardium. \label{mri_err}}
\end{figure*}

\revs{
For case \apex{} in figures~\ref{err_LV_free} and \ref{err_RV_free} the error is lowest in both ventricular endocardia during contraction at end-systole at $t=0.51$. The error rises during ventricular contraction and relaxation. Errors at the end of the simulation higher than the ones at $t=0$ suggest that the state at the end of the simulation differs from the reference configuration. The overall error is much lower in case \normal{} than in case \apex{}.
}
 
\subsection{Boundary stresses}
\label{sec_stress}

\begin{figure*}
\centering
\setlength\figureheight{3cm}
\setlength\figurewidth{0.35\textwidth}
\subfloat[Mean apical contact stress $\bar{t}_{\text{apex}}$ for case \apex{}.\label{pressure_apex}]{\includegraphics{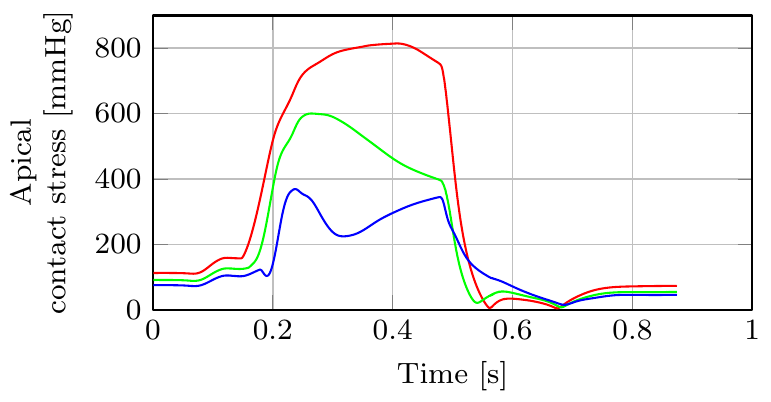}}~
\subfloat[Mean pericardial contact stress $\bar{t}_{\text{epi}}$ for case \normal{}.\label{pressure_refspring}]
{\includegraphics{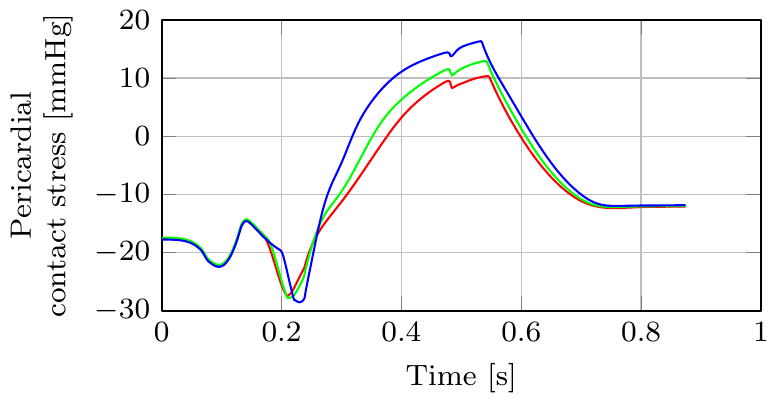}}\\
\includegraphics{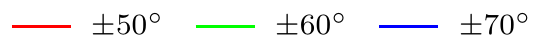}
\caption{Boundary condition stress.\label{boundary_stress}}
\end{figure*}

\revs{
Both scalar boundary stresses $\bar{t}_{\text{apex}} = \norm{2}{\bar{\vec{t}}_{\text{apex}}}$  and $\bar{t}_{\text{epi}}$ are visualized in figure~\ref{boundary_stress} over time for all fiber orientations. It can be observed that apical stress in case \apex{} is orders of magnitude higher than pericardial stress in case \normal{} and more dependent on fiber orientation. Positive values of $\bar{t}_{\text{epi}}$ indicate predominant tensile stresses between epicardium and pericardium. It can be seen that mean pericardial stress in figure~\ref{pressure_refspring} is a compressive stress for most of the cardiac cycle, except at the end of systole and onset of diastole.
}

\begin{figure*}
\centering
\subfloat[Case \apex{} mean apical stress vectors $\bar{\vec{t}}_{\text{apex}}$ (scaled by magnitude) for \fifty{} (red), \sixty{} (green), and \seventy{} (blue) fiber orientations at $t=0.45$.\label{pressure_apex_3D}]{
\hspace{1.5cm}
\includegraphics[height=6cm]{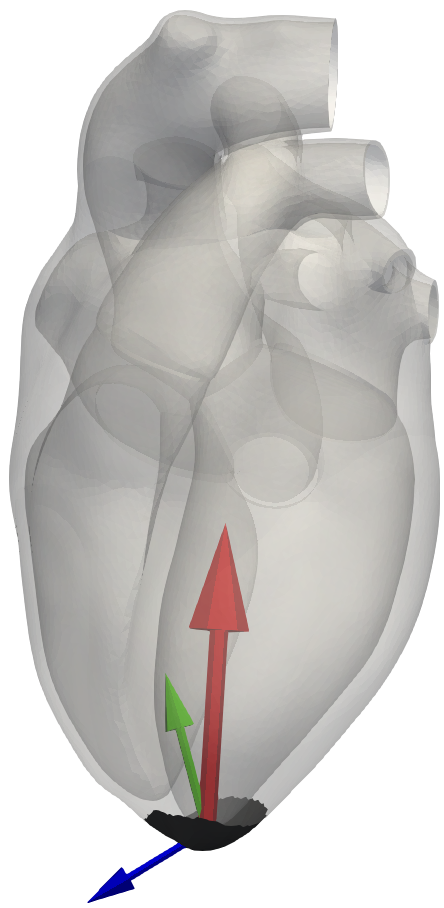}~
\hspace{1.5cm}}~
\hspace{.5cm}
\subfloat[Case \normal{} pericardial contact stress $t_{\text{epi}}$ on epicardial surface with \sixty{} fibers at end-systole $t=0.51$.\label{pressure60_surf}]{
\includegraphics[height=6cm]{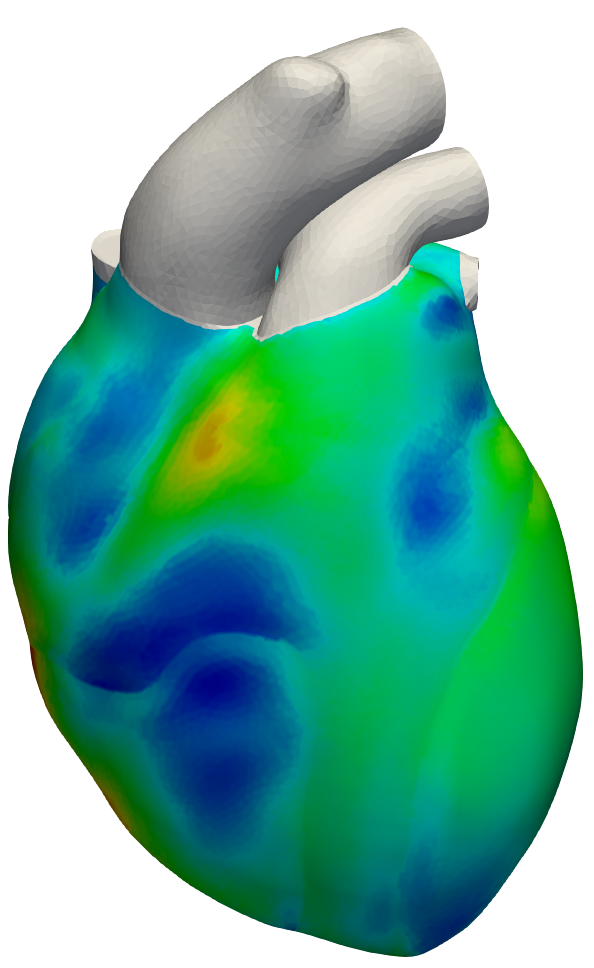}~
\hspace{.5cm}
\includegraphics[height=6cm]{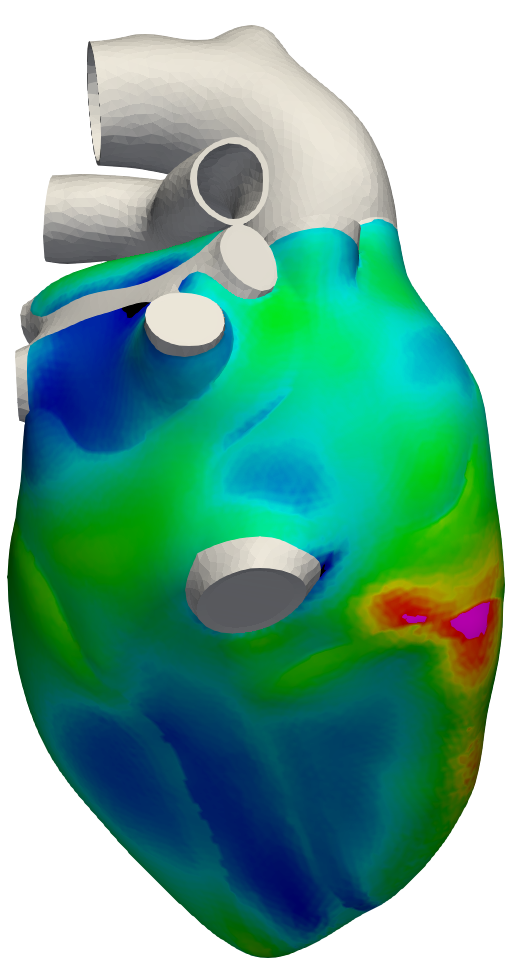}~
\hspace{.5cm}
\raisebox{+0.5\height}{\includegraphics[height=3cm]{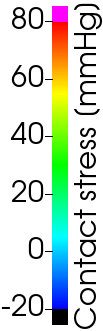}}}
\caption{Visualization of boundary stresses. \label{pressure_3D}}
\end{figure*}

\revs{
Boundary stresses are visualized in figure~\ref{pressure_3D}. For case \apex{}, the mean stress vectors $\bar{\vec{t}}_{\text{apex}}$ for all three fiber distributions are shown in figure~\ref{pressure_apex_3D} at $t=0.45$ and scaled according to their magnitude. Fiber orientation has not only a strong influence on the magnitude but also the direction of mean apical stress.
}

\revs{
The local distribution of pericardial contact stress with \sixty{} fibers at end-systole is shown in figure~\ref{pressure60_surf} in reference configuration. At end-systole, compressive as well as tensile stresses occur. Stresses are centered around a tensile stress of 20~mmHg. Areas of high compressive stresses are at the left atrium, the anterior and posterior right ventricle, the posterior left ventricle, and the anterior left ventricular apex. Areas of high tensile stresses are the right ventricle close to the anterior part of the right ventricular outflow tract and the left and right ventricular free wall. Overall, pericardial contact stress is evenly distributed around the epicardial surface. 
}

\section{Discussion}
\label{sec_discussion}

\revs{Our objective was to analyze the effects of the pericardial boundary condition proposed in section~\ref{sec_review} based on the physiology of the pericardium, comparing simulation cases \normal{} and \apex{}. We first performed a parametric study to explore the influence of pericardial stiffness. Each simulation case was then personalized and evaluated for the fiber orientations \fifty{}, \sixty{}, and \seventy{}. We then compared scalar left ventricular pressure and volume. The displacements at end-systole were qualitatively compared to multi-view cine MRI. Additionally, we quantified the differences of both simulation cases to MRI by atrioventricular plane displacement (AVPD), passive atrial filling, and spatial approximation error at the left and right ventricular endocardium.}

\subsection{Pericardial stiffness}
\revc{
The parametric study for pericardial stiffness in case \normal{} in section~\ref{sec_params} revealed that the ventricles are well approximated by the lowest tested stiffness values, e.g. $k_p=0.1$\,kPa/mm. Here, the error at left and right ventricular endocardium was minimized and much lower than in case \apex{}.}

\revc{
In contrast, right AVPD and right atrial passive filling matched well with measurements from MRI for high stiffness values, e.g. $k_p=3.0$\,kPa/mm. Choosing this value globally for pericardial stiffness lead however to some undesirable consequences, namely unphysiologically high myocardial contractility and pericardial stress as well as bad approximation of the interventricular septum.}

\revc{
In future studies, it might thus be reasonable to select spatially varying pericardial parameters. This hypothesis is supported by the fact that the pericardial tissue is in contact with various organs of different material properties as outlined in section~\ref{sec_physio}. A starting point could be the estimation of regional pericardial parameters based on the surface definitions in figure~\ref{pericardium_mri} with the objective to match MRI measurements in section~\ref{sec_params}.}

\revc{
In case of a biventricular geometry, no atria are present. Thus AVPD is not controlled by the interaction of atria and pericardium. Furthermore, atrial filling is not taken into account. We thus expect that a global value of $k_p=0.1$\,kPa/mm for pericardial stiffness yields good results for a biventricular geometry with \sixty{} fibers. This value was also used in \cite{hirschvogel16}, although it was not really analyzed there, e.g. with respect to MRI.}

\subsection{Pumping mechanism}
\label{sec_pumping}
% pumping mechanism
We calibrated cardiac contractility in all simulations in section~\ref{sec_params} to yield the same \revc{end-systolic} volume. It was shown that in case \normal{}, higher contractilities are required than in case \apex{}. \revc{Therefore, for a given contractility, a heart constrained with the pericardial boundary condition yields less output. This result is in agreement with the experimental observation that cardiac output is greatly increased after the removal of the pericardium \cite{hammond92}. The result further agrees with the numerical experiments performed in \cite{fritz13}. For identical active stress, left ventricular ejection fraction decreased from 71\,\% to 63\,\% when including the pericardium.}

The main pumping mechanism of the heart is shortening in long axis direction, which is quantified by AVPD \cite{arutunyan15,arvidsson15}. In \cite{maksuti15}, the pumping function of the heart was compared to a piston unit with the AVP as a piston. This mechanism could be observed in section~\ref{sec_avpd} for case \normal{}, where left and right AVPD \revs{is higher than in case \apex{} but still lower than in MRI.}

%passive atrial filling
The upper part of the left atrium is fixated by pulmonary veins. Ventricular contraction forces the mitral ring towards the apex and promotes the filling of the left atrium from the pulmonary veins \cite{fujii94}. In section~\ref{sec_inter} we compared atrial filling during ventricular systole with and without pericardium. It was observed that the simulations of case \normal{} which promoted higher AVPD in section~\ref{sec_avpd} contribute more to atrial filling during ventricular systole. \revs{Case \normal{} predicts maximal atrial volume at ventricular end-systole as segmented from isotropic 3D MRI better than case \apex{}. The simulated values are however still lower than in MRI for the chosen pericardial parameters.}

% volume change
Keeping in mind that all simulations yield the same \revc{end-systolic} volume, it was shown that the pumping mechanism of the heart is very different for cases \apex{} and \normal{} although their pressure and volume curves were similar in section~\ref{sec_res_scalar}.  Comparison of four-chamber and short axis slices of the left and right ventricle from simulation results to cine MRI in section~\ref{sec_ref_es} revealed an unphysiological radial pumping motion without pericardial boundary conditions in case \apex{}. In \cite{emilsson01} it was found that the outer diameter of the left ventricle shortens only about 2~mm during systole. Furthermore, the total volume enclosed by the pericardium changes only by about 5-8\,\% during the cardiac cycle \cite{bowman03,arvidsson15}. \revc{We found that for the \sixty{} fiber orientation the total change in pericardial volume is 24\,\% and 21\,\% for cases \normal{} and \apex{}, respectively. This mismatch is mainly due to the unphysiological change in atrial volume during ventricular contraction.}

% pericardial stiffness
\revs{As demonstrated in the parametric study in section~\ref{sec_params}, AVPD and atrial filling could be increased to the values measured in MRI by increasing the global pericardial stiffness. However, this was shown to lead to a worse approximation of the interventricular septum. This motivates the use of a regionally distributed pericardial stiffness. Another reason for underestimating AVPD and atrial filling might be an atrial material model, which is in our case identical to the ventricular one, that is too stiff.}

\subsection{Fiber orientation}
We compared in this work three fiber orientations within the myocardium, namely \fifty{}, \sixty{}, and \seventy{}. We studied the influence of fiber direction for both boundary conditions cases \apex{} and \normal{}. It was shown that fiber orientation has a strong influence on the displacements. In section~\ref{sec_avpd} \seventy{} fiber orientations exhibited larger AVPD for both boundary condition cases. This can be attributed to the fact that the fiber orientation is more vertical, i.e. more aligned with the long axis, than \sixty{} and \fifty{} fiber orientations. Since myofiber contraction is prescribed in fiber direction, more vertical fiber orientations inherently apply a greater force pushing apex and AVP together, thus yielding higher AVPD. Since the AVP is also attached to the atria, AVPD is also linked to atrial filling. In section~\ref{sec_inter} it was shown that the more vertical \seventy{} fiber orientation also yielded the highest atrial filling during ventricular systole. Comparing results to short axis cine MRI slices, it was shown that a more horizontal \fifty{} fiber orientation leads to a more radial contraction of the heart. The maximum pericardial stress at end-systole was highest for \fifty{} fibers. This can be explained by the observation in figure~\ref{sax9_normal} where \fifty{} fibers (red) exhibited the most radial inward movement during systole. Since the myocardial-pericardial interface can only transmit forces in normal direction, a more radial contraction exerts a higher pericardial tensile stress. The overall spatial approximation error was also shown to be dependent on fiber direction. However, the dependence was more pronounced in case \apex{} than in case \normal{}.

\subsection{Pericardial contact stress}
% diastole
\revs{
In \cite{smiseth85} end-diastolic pericardial contact pressure was measured with a flat balloon catheter at the left ventricular anterolateral epicardial surface with around 15\,mmHg. In vivo experiments on humans in \cite{tyberg86} showed pericardial pressures on the left lateral surface of the heart between 0 and 15\,mmHg. The prestressing procedure in our model not only includes the myocardium but also the pericardial boundary condition in case \normal{}. Here, we measure a contact pressure of 20\,mmHg at diastasis on the left ventricular epicardial surface, agreeing well with experimental observations.
}

% apex vs. pericardium
\revs{
The stresses exerted by the boundary conditions on the epicardial surface of the heart were found to be over one order of magnitude higher in case \apex{} than in case \normal{}. The exact stress values in case \apex{} depend on the choice of apical spring stiffness, which was not calibrated in this study. It is nevertheless evident, that unphysiologically high stresses are concentrated in a very small area of the heart. In case \normal{}, all boundary stresses are evenly distributed on the epicardial surface.
}

% systole
The similar maximum values of mean pericardial contact stress for all fiber directions in case \normal{} suggest that pericardial constraint is displacement-controlled. Pericardial constraint is determined by the deviation of the heart throughout the cardiac cycle from its end-diastolic state. However, as outlined in section~\ref{sec_physio}, to the best of the authors' knowledge there are no measurements of pericadial contact pressure during the cardiac cycle to validate the stresses experienced in our computational study. Pericardial contact stress is thus an output of our computational model which is not yet available in clinical practice. \revc{Given that our model was solely calibrated to kinematic data, the pericardial contact stresses predicted by our model should be considered as qualitative results.}

\revc{
\subsection{Numerical performance}
We ran all simulations on two nodes of our Linux cluster. One node features 64 GB of RAM and two Intel Xeon E5-2680 "Haswell" processors, each equipped with 12 cores operating at a frequency of 2.5\,GHz. The computation time of cases \apex{} and \normal{} was almost identical, which was about 18 hours for each of the simulations performed in this work, including prestressing. The pericardial boundary condition requires little effort to evaluate, since the pericardial boundary condition only requires the displacement field, which is computed anyway, and reference surface normals, which are computed once at the initialization of the simulation. Some differences in numerical performance arise since the calculated displacement fields of both cases are different. Taking the \sixty{} fiber distribution, case \apex{} had an average of 7.8 Newton iterations per time step and 28 linear solver iterations per Newton iteration. For case \normal{}, these values were 8.3 and 25, respectively. 
}

\subsection{Limitations and future perspectives}
% instantaneous activation
As mentioned earlier, in this work, we did not account for the propagation of the electrical signal sent from the sinus node. Rather, all myocardial tissue in our simulations was activated simultaneously. We recently demonstrated the ability to couple our mechanical model to an electrophysiological model \cite{hoermann18}, which we can include in further studies. However, since the data came from a healthy volunteer, we do no expect relevant variations.

% material
\revb{Ex-vivo experiments on myocardial tissue in \cite{yin87,dokos02,sommer15} showed anisotropic tissue characteristics, depending on myocardial fiber and sheet orientation. In our model, we used the anisotropic material model proposed in \cite{holzapfel09} for myocardial tissue.} \revc{Due to the lack of sufficient experimental data, we used identical material properties for left and right myocardium, as well as the atria.} \revb{However, no studies have been carried out how material parameters obtained from experiments on ex vivo tissue correlate to in vivo material behavior. Furthermore, it should be noted that vastly different material parameters have been estimated in \cite{holzapfel09} and \cite{gueltekin16} when being fitted to measurements from either biaxial extension tests or shear tests.}

% windkessel
Our structural model was coupled to a lumped-para\-meter windkessel model of hemodynamics of the systemic \revs{and pulmonary} circulation \revs{with prescribed atrial pressures}. The interaction between atria and ventricles should be investigated in further studies using a volume-preserving closed-loop model, including both pulmonary and systemic circulation. \revs{Furthermore, none of our cardiac simulations are perfectly periodic, i.e. the values at the end of the cardiac cycle are not equal to the initial conditions. In future studies, achieving a periodic state should be incorporated into parameter estimation.}

% fiber orientation
\revc{In this work we interpolated the local helix fiber directions at the integration points from three different prescribed cons\-tant-per-surface fiber orientations. Results showed that fiber orientation has a large influence on AVPD. However, we have no knowledge of patient-specific fiber orientation and assumed equal distributions in left and right ventricle.} Patient-specific cardiac fiber orientations can be estimated from diffusion tensor MRI \cite{nagler17} (DTMRI). \revc{However, while applicable to in vivo DTMRI (as shown in \cite{nagler17}), to the best of our knowledge, fiber estimation has not been tested and validated with in vivo DTMRI yet.} Further quantitative studies of cardiac dynamics require a fine resolution of patient-specific fibers.

% pericardial stiffness
We further assumed constant stiffness and viscosity parameters of our pericardial boundary condition over the epicardial surface. \revc{Given reliable material parameters for the myocardium, constant pericardial stiffness and viscosity could be estimated from measured AVPD.} The choice of constant parameters might however be oversimplified, as the pericardium is in contact with various tissues of different mechanical behaviors, as illustrated in figure~\ref{pericardium_mri}. For example, the movement of the apex in anterior direction in case \normal{} as observed in figure~\ref{4ch_normal} suggests a higher pericardial stiffness to model the influence of the sternum and the diaphragm. This will however introduce more parameters to the model, which will need to be calibrated to measurements from e.g. cine \revc{or 3D tagged} MRI. For this study we kept the number of parameters small in order to make evident the general effect of the pericardial boundary condition even by using a simplified modeling approach.

% model personalization
\revc{
From a machine learning perspective, we split our limited available data from cine MRI into a training set and a test set. The training set data is used during model personalization. The rest of the data can then be used in the test set to check how well the model actually predicts data that was not used during personalization. In our case, we used as training set left ventricular volume and ventricular epicardial contours to tune timing, (de-) activation rates, and contractility for atria and ventricles and global material viscosity and pericardial stiffness. We then used as test set AVPD, atrial volume, and ventricular endocardial contours, each left and right, to quantify the simulations' approximation error. Many more parameters of our cardiac model could be personalized for this patient-specific study. However, using the metrics in our test set for model calibration would disqualify using them to test model accuracy and limit our abilities to test the model.}

% comparison of rotation. cine: Euler. tagged: Lagrange
We validated our simulation results solely with cine MRI data. Cine MRI can be interpreted as an Eulerian description of cardiac movement, as the imaging planes stay fixed in space throughout the cardiac cycle. This observation however cannot detect any rotational movement with respect to the long axis, as the left ventricle is almost rotationally symmetric. To properly validate any rotational movement of the myocardium, a comparison to data from 3D tagged MRI is necessary, which can be interpreted as a Lagrangian observation of cardiac motion. Furthermore, pressure measurements from within ventricles and atria are required. Pressure values at end-diastole yield initial values for the stress state of the myocardium, which cannot be assessed from imaging alone. Pressure curves over the cardiac cycle would yield a ground truth to validate the outputs of our windkessel model. \reva{Figure~\ref{volume_atria} demonstrates that the model, while using the pericardial constraint, does predict accurately the atrial volume at ventricular end-systole. However, we have no data available at atrial end-systole. In future studies, if detailed cine data of atria are available (e.g. cine stack in trasverse orientation with respect to the body, and using thin slices of 5mm), we will consider a more detailed analysis of atrial contraction.}

\subsection{Concluding remarks}
In this work we gave an overview of the anatomy and mechanical function of the pericardium and motivated to model its influence on the myocardium as a parallel spring and dashpot acting on the epicardial surface. Following a review of pericardial boundary conditions currently used in mechanical simulations of the heart, we proposed to compare two simulation cases, one with and one without pericardial boundary conditions. Following calibration to stroke volume as measured from short axis cine MRI, we compared several physiological key outputs of our model and validated them using multi-view cine MRI. Although exhibiting similar volume and pressure curves, the displacement results of both simulation cases were radically different. The simulations with pericardial boundary conditions matched MRI measurements much closer than without, especially with respect to atrioventricular plane displacement and atrial filling during ventricular systole, quantities which were not included in the calibration of the model. By establishing an overall spatial approximation error at the left and right endocardium, we showed that the introduction of only two global parameters for the pericardial boundary condition already yields a big gain in model accuracy. Our ultimate goal is to obtain more comprehensive data sets, adding 3D tagged MRI and pressure measurements, to further validate our model of pericardial-myocardial interaction. \revc{Measurements of pericardial contact stress at different locations on the epicardium throughout the cardiac cycle would help to test the qualitative predictions of pericardial contact stresses by our model and will probably lead to further model improvements.}

\section*{Compliance with ethical standards}
\textbf{Conflict of interest} The authors declare that they have no conflict of interest.

\appendix

\begin{figure*}
\centering
\setlength\figureheight{7cm}
\subfloat[Ellipsoid model in reference configuration with cross-section (yellow). \label{fig_ell_geo}]{
\includegraphics[height=\figureheight]{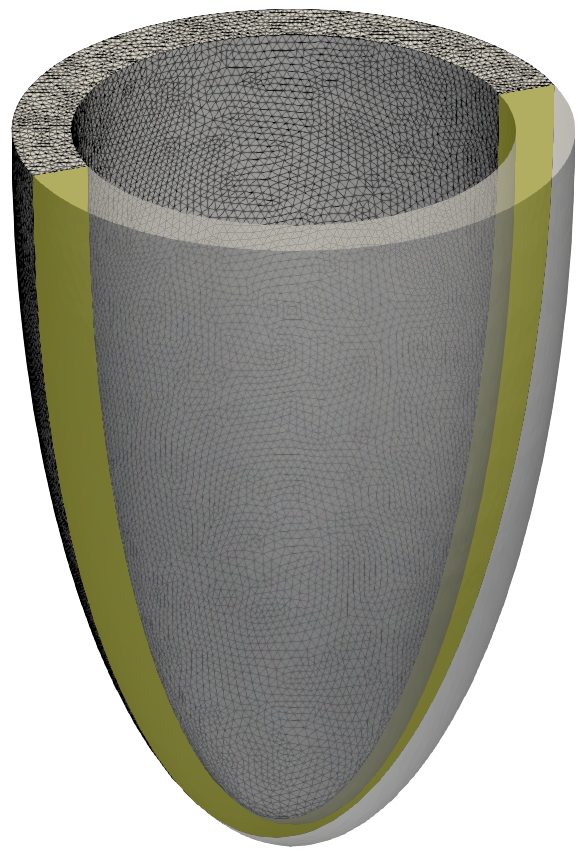}}\qquad
\subfloat[Frontal view of epi- and endocardial cross-section at end-systole. \label{fig_ell_short}]{
\includegraphics[height=\figureheight]{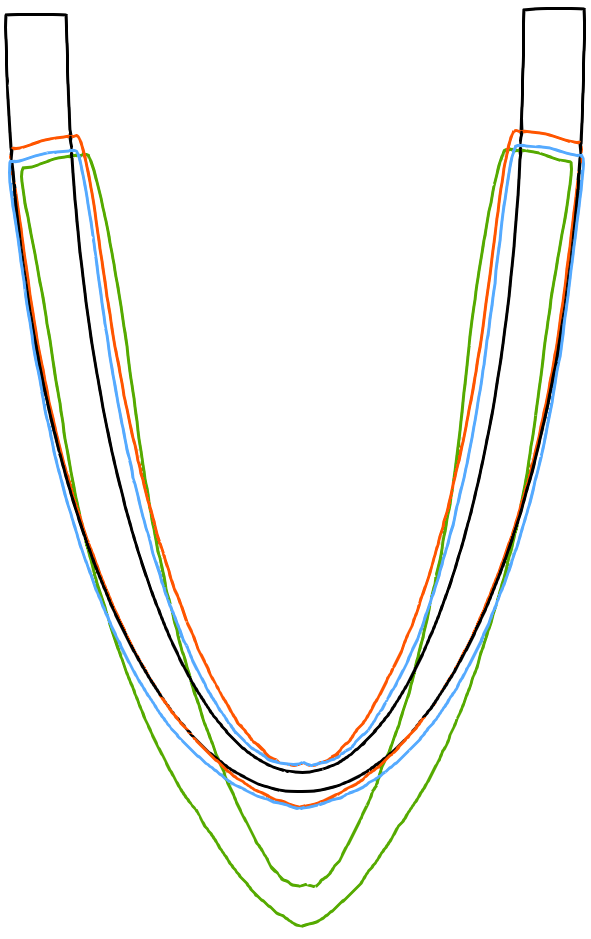}}\qquad
\subfloat[Top-down view of epicardial cross-section at end-systole. \label{fig_ell_twist}]{
\includegraphics[height=\figureheight]{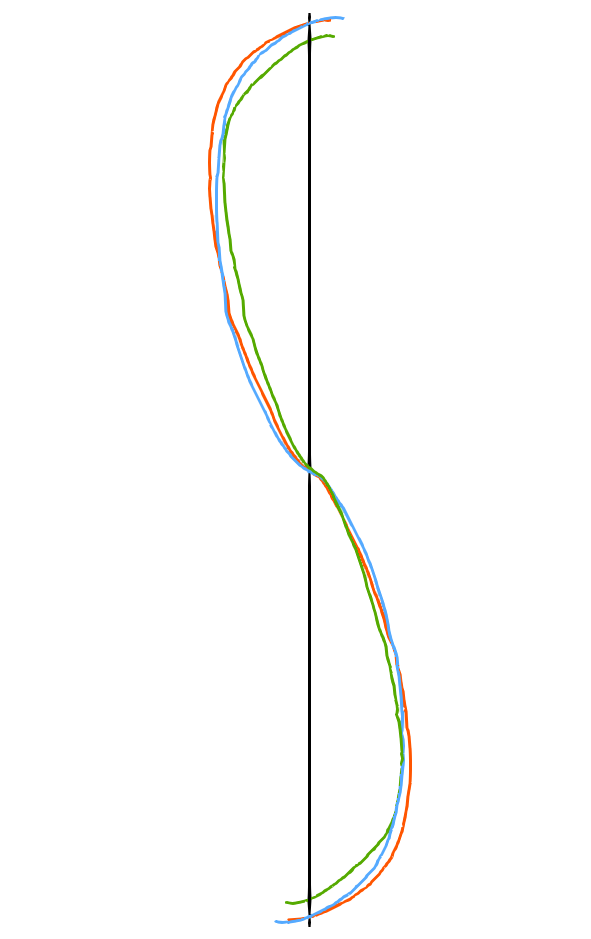}}
\caption{\revb{Ellipsoid model in reference configuration (black) with cases \free{} (green), \normal{} (blue), and \contact{} (orange).} \label{fig_ell}}
\end{figure*}

\revb{
\section{Comparison of spring formulations}
\label{sec_ellipsoid}
We show in section~\ref{sec_review} how the pericardial boundary condition in case \normal{} can be derived from adhesive sliding contact by introducing several simplifications. To justify the simplifications made by our pericardial boundary condition, we use a very simple geometry of a hollow half-ellipsoid with \sixty{} fibers, which roughly represents the shape of the left ventricle, see figure~\ref{fig_ell_geo}. It is able to show the consequences of each approach while being simple enough to isolate the effects of the boundary condition. The parameters of the ellipsoid model are given in table~\ref{tab_parameters_ellipsoid}. We use the same active stress model introduced in \eqref{mat_myocard} to mimic cardiac contraction. All three simulations use the same contractility parameter.}

\revb{
As in section~\ref{sec_results}, case \normal{} utilizes the pericardial boundary condition proposed in \eqref{eq_bc_final} using the gap \eqref{eq_refnormal}. Additionally, we introduce case \contact{}, which uses the definition of the gap in \eqref{eq_curnormal} based on projection and the current normal vector to the epicardium. Case  \free{} has homogeneous zero-Neumann boundary conditions on the whole epicardial surface.}

\revb{
The results of the contraction simulation are shown in figures~\ref{fig_ell_short} and \ref{fig_ell_twist} at end-systole. Displayed is the reference configuration and all three boundary condition cases for a cross-section of the ellipsoid. Figure~\ref{fig_ell_short} shows in a frontal view the shortening of the ellipsoid with visible epi- and endocardial contours. While cases \normal{} and \contact{} are very similar with little differences only in radial direction, case \free{} exhibits much less longitudinal shortening. There is almost no longitudinal shortening but a translational movement of the whole geometry instead.} %This observation is analog to the observation that atrioventricular plane displacement is lower in case \free{} than in case \normal{} in section~\ref{sec_avpd}.

\revb{
Figure~\ref{fig_ell_short} shows the epicardial contour of the ellipsoid in a top-down view to observe the twisting motion of the ellipsoid. All three boundary condition cases are very similar. This confirms that the normal springs in cases \normal{} and \contact{} in fact allow tangential sliding and do not prohibit any rotational movement, as they are very similar to case \free{}. Furthermore, the similarity of cases \normal{} and \contact{} shows that the simplified spring formulation \eqref{eq_refnormal} in case \normal{} is sufficient to represent the effects of the pericardium compared to the more detailed formulation \eqref{eq_curnormal} in case \contact{}.}

\begin{table}[htb!]
\centering
\footnotesize
\setlength{\tabcolsep}{.3em}
\renewcommand{\arraystretch}{1.8}
\begin{minipage}{0.45\textwidth}
\subfloat[Parameters of the elastodynamical model. \label{mat_solid_ell}]{
\begin{tabular}{l l l c}
\textbf{Name} & \textbf{Par.} & \textbf{Value} & \textbf{Unit} \\
\hline
Tissue density & $\rho_0$ & $10^3$ & $\left[\frac{\text{kg}}{\text{m}^3}\right]$ \\
Viscosity & $\eta$ & 10 & $\left[\text{Pa}\cdot\text{s}\right]$ \\
Volumetric penalty & $\kappa$  & $10^4$ & $\left[\text{kPa}\right]$ \\
Ventricular contractility & $\sigma_\text{v}$ & 185 & $\left[\text{kPa}\right]$ \\
Mooney-Rivlin & $C_1$ & 10 & $\left[\text{kPa}\right]$ \\
Mooney-Rivlin & $C_2$ & 40 & $\left[\text{Pa}\right]$  \\
Spring stiffness  base& $k_b$ & 1 & $\left[\text{kPa}\right]$  \\
\hline
\multicolumn{4}{l}{\textit{Pericardial spring stiffness}}\\
\hline
Case \free{} & $k_p$ & 0 & $\left[\frac{\text{kPa}}{\text{mm}}\right]$ \\
Case \normal{} & $k_p$ & 20 & $\left[\frac{\text{kPa}}{\text{mm}}\right]$ \\
Case \contact{} & $k_p$ & 20 & $\left[\frac{\text{kPa}}{\text{mm}}\right]$ \\
\hline
\end{tabular}}
\end{minipage}
\caption{\revb{Overview of material parameters in ellipsoid model. For numerical parameters see table~\ref{tab_numerical}}.\label{tab_parameters_ellipsoid}}
\end{table}

%\begin{acknowledgements}
%\end{acknowledgements}

\bibliographystyle{1_WileyNJD-AMA}
\bibliography{2_references}   % name your BibTeX data base

\end{document}